\documentclass[12pt]{article}

\usepackage{keyval}
\usepackage{dsfont}
\usepackage{natbib}

\usepackage{rotating}
\usepackage{endnotes}
\usepackage{amssymb,amsmath,lscape,bm,amsthm}
\usepackage{graphicx}
\usepackage{color}
\usepackage{multirow}
\usepackage{subfig}
\usepackage{caption}
\usepackage{setspace}
\usepackage{verbatim}
\usepackage{epstopdf}
\usepackage{booktabs}
\usepackage{array}
\usepackage{adjustbox}
\usepackage{dcolumn}
\usepackage{longtable}
\usepackage{mathrsfs}

\newcommand{\possessivecite}[1]{\citeauthor{#1}'s (\citeyear{#1})}

\begin{document}

\renewcommand{\baselinestretch}{1.5}
\small\normalsize

\renewcommand{\thefootnote}{\fnsymbol{footnote}}

\begin{center}
{\bf\Large Quantile-based modeling of scale dynamics in financial returns for Value-at-Risk and Expected Shortfall forecasting}

\bigskip

{\small \emph{This is the author accepted manuscript.}\\
\emph{Accepted for publication in International Journal of Forecasting.}\\
\emph{DOI:} 10.1016/j.ijforecast.2025.12.002}

\end{center}

\renewcommand{\baselinestretch}{1.5}
\normalsize

\begin{center}

Xiaochun Liu\footnotemark\  and   Richard Luger\footnotemark\

\bigskip

\addtocounter{footnote}{-1}\footnotetext{Department of Economics, Finance and Legal Studies, Culverhouse College of Business, University of Alabama, Tuscaloosa, AL 35487, USA. \emph{E-mail address:}  xliu121@ua.edu.} 

\stepcounter{footnote}\footnotetext{Corresponding author.  Department of Finance, Insurance and Real Estate, Laval University, Quebec City, Quebec G1V 0A6, Canada.
  \emph{E-mail address:} richard.luger@fsa.ulaval.ca.}

\end{center}

\thispagestyle{empty}

\normalsize

\bigskip

\noindent \textbf{Abstract:} We introduce a semiparametric approach for forecasting {Value-at-Risk} (VaR) and {Expected Shortfall} (ES) by modeling the conditional scale of financial returns, defined as the difference between two specified quantiles, via restricted quantile regression. Focusing on downside risk, VaR is derived from the left-tail quantile of rescaled returns, and ES is approximated by averaging quantiles below the VaR level. The method delivers robust, distribution-free estimates of extreme losses and captures skewness, heavy tails, and leverage effects. Simulation experiments and empirical analysis show that it often outperforms established models, including GARCH and joint VaR-ES conditional-quantile approaches. An application to daily returns on major international stock indices, spanning the COVID-19 period, highlights its effectiveness in capturing risk dynamics.

\bigskip

\noindent {\bf JEL classification:} C14, C22, G17, G32 

\bigskip

\noindent {\bf Keywords:} Conditional scale dynamics, {CAViaR} models, Multiple quantiles, Robust risk estimation, Financial risk forecasting

\renewcommand{\baselinestretch}{1.75}
\small\normalsize

\thispagestyle{empty}

\newpage

\pagenumbering{arabic}

\renewcommand{\thefootnote}{\arabic{footnote}}
\setcounter{footnote}{0}

\section{Introduction}

Forecasting risk measures such as {Value-at-Risk} (VaR) and {Expected Shortfall} (ES) is crucial for financial institutions, risk managers, and regulators. VaR, which estimates a conditional quantile in the lower tail of the return distribution, has long been a standard tool for assessing potential losses. 
Specifically, the VaR for period $t$ at probability level $\alpha$ is defined as
\begin{equation} 
\text{VaR}_{r_t}(\alpha) = Q_{r_{t}}(\alpha \,| \, \mathcal{I}_{t-1}), 
\label{defVaR} 
\end{equation} 
where $r_{t}$ denotes the asset return at time $t$ and $Q_{r_{t}}(\alpha \,\vert \, \mathcal{I}_{t-1})$ is the $\alpha$th conditional quantile given the information set $\mathcal{I}_{t-1}$.
However, VaR has notable limitations, particularly its inability to account for extreme losses beyond the quantile threshold defined by 
$\alpha$, and its potential failure to capture diversification benefits \citep{Embrechts-McNeil-Straumann:2002}.
This is because VaR is not a coherent risk measure; it does not satisfy the property of subadditivity, meaning that the VaR of a combined portfolio can exceed the sum of the VaRs of its individual components. As a result, VaR may inadequately reflect the benefits of diversification.

In contrast, ES is a coherent risk measure, addressing these shortcomings by estimating the expected loss beyond the VaR threshold  and, in particular, satisfying 
subadditivity  \citep{Artzner-Delbaen-Eber-Heath:1999, Acerbi-Tasche:2002}. Conditional on information $\mathcal{I}_{t-1}$, ES for period $t$ at probability level $\alpha$ is defined as 
\begin{equation} 
\text{ES}_{r_t}(\alpha) = \mathbb{E}[r_t \, | \, r_t \leq \text{VaR}_{r_t}(\alpha), \, \mathcal{I}_{t-1}], 
\label{defES} 
\end{equation} 
providing the average loss when $r_t$ falls below the VaR threshold in~(\ref{defVaR}). This makes ES a more theoretically sound measure that incentivizes diversification and better accounts for extreme risks.
Recognizing these advantages, the {Basel Committee}, through its {Basel III} framework and the {Fundamental Review of the Trading Book (FRTB)}, has replaced VaR with ES as the primary metric for market risk capital requirements \citep{Basel:2019}.
This regulatory shift underscores the need for robust ES forecasting models that can accurately quantify tail risk under changing market conditions.

ES forecasts can often be derived as a natural extension of many VaR forecasting methods. Nonparametric approaches, including historical simulation and kernel density estimation, generate density forecasts from which both VaR and ES predictions can be obtained. Similarly, parametric methods, typically based on models for the conditional variance, such as GARCH specifications combined with a precise assumption about the conditional return distribution, provide concurrent forecasts for both risk measures.

A prominent semiparametric method for VaR forecasting is quantile regression, specifically the {Conditional Autoregressive VaR ({CAViaR})} models of \citet{Engle-Manganelli:2004}, which  model the conditional quantile directly without relying on specific distributional assumptions. 
Let $y_t = r_t - \mu_t$, where $\mu_t$ represents the conditional location of returns, often assumed to be zero, a constant, or a dynamic process such as AR(1), as in \citet{Kuester-Mittnik-Paolella:2006}. A first-order symmetric absolute value (SAV) {CAViaR} specification is expressed as
\begin{equation}
Q_{y_t}(\alpha \,\vert \, \mathcal{I}_{t-1}) = \omega(\alpha) + \beta(\alpha) Q_{y_{t-1}}(\alpha \,\vert \, \mathcal{I}_{t-2}) + \gamma(\alpha) \lvert y_{t-1} \rvert,
\label{SAV}
\end{equation}
with the VaR in (\ref{defVaR}) given by $Q_{r_t}(\alpha \,\vert \, \mathcal{I}_{t-1}) = \mu_t + Q_{y_t}(\alpha \,\vert \, \mathcal{I}_{t-1})$. 
This model treats positive and negative deviations of $y_{t-1}$ symmetrically, relying solely on $\lvert y_{t-1} \rvert$ and thereby implicitly responding to volatility symmetrically. Alternatively, an asymmetric slope (AS)  {CAViaR}  specification is given by 
\begin{equation}
\begin{aligned}
Q_{y_{t}}(\alpha \,\vert\,\mathcal{I}_{t-1})= & \,  \omega(\alpha)+\beta(\alpha) Q_{y_{t-1}}(\alpha \,\vert\,\mathcal{I}_{t-2})+\big( \gamma_{+}(\alpha)\mathds{1}\{y_{t-1} > 0\}+\gamma_{-}(\alpha)\mathds{1}\{y_{t-1} \leq 0\} \big) |y_{t-1}|,  
\end{aligned}
\label{AS} 
\end{equation}
where $\mathds{1}\{\cdot\}$ denotes the indicator function. 
This formulation allows the model to capture asymmetries in the conditional quantile dynamics by adjusting the slope based on whether the previous centered return was positive or negative, thus making it sensitive to directional changes in past returns.
The {CAViaR} models in (\ref{SAV}) and (\ref{AS}) allow the dynamics of $Q_{y_{t}}(\alpha \,\vert\,\mathcal{I}_{t-1})$ to vary across different probability levels $\alpha $ and have shown strong performance in empirical studies of VaR forecast accuracy \citep[see, e.g.,][]{Sener:2012}. However, while {CAViaR} models excel in predicting VaR, they do not inherently provide a mechanism for generating ES forecasts, as they focus solely on a particular quantile of the return distribution.

Recent advances in risk forecasting have focused on methods that jointly estimate VaR and ES, leveraging the fact that these two risk measures are jointly elicitable \citep{Fissler-Ziegel:2016}. A risk measure is said to be elicitable if it can be uniquely identified as the minimizer of a well-defined expected loss function, also known as a scoring function. In simpler terms, an elicitable risk measure can be evaluated and compared using a loss function that ranks different forecasting methods based on their accuracy. While ES is not elicitable on its own, the joint elicitability of VaR and ES allows for their simultaneous estimation and evaluation using a common loss function. This property has been pivotal in recent developments in financial risk forecasting.

\citet{Taylor:2019} proposes a joint VaR-ES model based on the asymmetric {Laplace} (AL) distribution, where the VaR component is modeled using a {CAViaR} specification. 
This approach uses the AL density to construct a likelihood function, whose maximization is equivalent to minimizing a strictly consistent \citet{Fissler-Ziegel:2016} scoring function.
Similarly, \citet{Patton-Ziegel-Chen:2019} introduce a semiparametric approach within the {Generalized Autoregressive Score} (GAS) framework of \citet{Creal-Koopman-Lucas:2013}, which also exploits the \citet{Fissler-Ziegel:2016} class of loss functions. Their GAS model dynamically adjusts both VaR and ES estimates based on past observations, specifically reacting to VaR violations while reverting to a long-run mean in the absence of breaches.

In this paper, we introduce a novel semiparametric {CAViaR}-based framework for forecasting VaR and ES, which we term the {Quantile-based Scale Dynamics} (QbSD) approach. 
The QbSD framework extends the widely used GARCH class by replacing the conditional standard deviation with a conditional scale---a measure analogous to the interquartile range (IQR), but more flexible as it is based on the difference between two specified quantiles of the return distribution, similar to \citet{Taylor:2005}.
To model the dynamics of the conditional scale, QbSD employs global {CAViaR} specifications in which persistence parameters are shared across quantiles to ensure parsimony, while local parameters account for level-specific adjustments. Parameter constraints enforce non-crossing quantile ordering and strictly positive scale values, ensuring internal consistency and robust tail risk estimation.

Beyond its theoretical advantages, the QbSD approach aligns with practical needs in financial risk management. Financial institutions rely on multiple model validations in modern risk management frameworks, particularly for stress-testing and regulatory reporting under Basel III guidelines. Given its flexibility and lack of reliance on specific distributional assumptions, QbSD can serve as a complementary tool for financial institutions in benchmarking VaR and ES estimates against conventional models. This is particularly relevant in periods of financial distress, where traditional parametric models may struggle to adapt to evolving risk dynamics. Furthermore, by providing more accurate tail risk estimates, QbSD can aid institutions in regulatory capital calculations and internal risk monitoring, supporting more resilient decision-making.

To achieve this, QbSD models the dynamics of conditional quantiles, enabling a structured estimation of both VaR and ES. 
Specifically, VaR is estimated from the left-tail quantile of the rescaled returns, while ES is approximated by averaging quantiles across probability levels below the VaR threshold. 
This approach leverages the representation of ES as the integral of conditional quantiles across the left tail of the distribution up to the VaR probability level \citep[see, e.g.,][]{Acerbi-Tasche:2002}. 
As a result, the method delivers valid, well-ordered risk estimates across all probability levels, including those in the extreme tails of the distribution.

This approach shares similarities with the Filtered Historical Simulation (FHS) method described by \citet[][Section 6.4]{Christoffersen:2012}, which first applies a volatility model---typically a GARCH specification with an assumed innovation distribution---to filter historical returns. FHS then standardizes the returns by dividing them by the conditional volatility estimates, and subsequently computes the quantiles of these residuals to estimate VaR. The average residual loss beyond the VaR estimate is used to calculate ES. In contrast, the semiparametric QbSD approach does not rely on a specific distributional assumption. Instead, it models the conditional scale dynamics using restricted {CAViaR} specifications, offering a robust and distribution-free framework for forecasting VaR and ES.

The remainder of the paper is organized as follows. Section 2 develops the QbSD approach for forecasting VaR and ES. Section 3 presents simulation experiments designed to evaluate the performance of the proposed method across a range of market conditions. In Section 4, we assess the QbSD method's effectiveness through an empirical application, using daily returns from major international stock indices, including the volatile period surrounding the COVID-19 pandemic. Finally, Section 5 concludes the paper. Additional numerical and empirical results are provided in the Supplementary material.

\section{Quantile-based modeling}

\subsection{Background}

Consider first the unconditional distribution of asset returns and suppose it belongs to the location-scale family, so that
\begin{equation} 
r_{t}=\mu+s\varepsilon_{t},\label{locationscale} 
\end{equation} 
where $\mu$ is the location parameter, $s>0$ is the scale parameter, and $\varepsilon_{t}$ is an i.i.d. innovation with distribution function $F_{\varepsilon}$ and zero location.

There are several ways to define the scale parameter. A commonly used one is the standard deviation, which is based on moments. Despite its widespread usage, the standard deviation is sensitive to the presence of outliers and may even be undefined or infinite for distributions with heavy tails. This sensitivity arises from the fact that the standard deviation has a zero breakdown point, meaning that even a single outlier can significantly distort the estimate \citep[see][Ch. 5]{Huber-Ronchetti:2009}. In contrast, scale estimators based on interquantile ranges have non-zero breakdown points, making them more robust to outliers and extreme values, which are common in financial return data.\footnote{\citet{Kim-White:2004} also advocate using quantiles to estimate characteristics of return distributions, focusing on robust measures for capturing skewness and kurtosis. Through simulations, they demonstrate that empirical moment-based estimators are significantly more sensitive to outliers. See also \citet{White-Kim-Manganelli:2010} and \citet{Ghysels:2014}. }

With the location-scale representation in (\ref{locationscale}), the $p$th quantile of the unconditional return distribution is $Q_{r_t}(p)=\mu+sF_{\varepsilon}^{^{-1}}(p)$. Let $0<p<0.5$ be a chosen quantile level. Following \citet{Taylor:2005}, consider the quantile at the $(1-p)$th level as a natural and compelling counterpart.
Even when $\mu$ is unknown, we can obtain $s$ as 
\begin{equation} s=\frac{Q_{r_t}(1-p)-Q_{r_t}(p)}{c_p},\label{s} 
\end{equation} 
where $c_p=F_{\varepsilon}^{-1}(1-p)-F_{\varepsilon}^{-1}(p)$ is a scaling term. Note that when $r_{t}\sim N(\mu,\sigma^{2})$, then (\ref{s}) yields $s=\sigma$. The quantity $\text{IQR}=Q_{r_t}(0.75)-Q_{r_t}(0.25)$ is the well-known \emph{interquartile range}, which takes the form in (\ref{s}) by setting $c_p=1$.
Another measure that fits this format is the \citet{PearsonTukey1965} scale measure that uses $p=0.05$ and $c_p=q_{0}(1-p)-q_{0}(p)$, where $q_{0}(p)$ is the $p$th quantile of the standard normal distribution. See \citet{Taylor:2005} and \citet[][Table 1]{KotzvanDorp2005} for other examples.

\subsection{Conditional dynamics}

We extend the basic specification in (\ref{locationscale}) to allow the location and  scale parameters to be conditionally dynamic so that
\begin{equation}
r_{t}=\mu_{t}+s_{t}\varepsilon_{t},\label{dynamiclocationscale}
\end{equation}
where $\mu_{t}$ and $s_{t}>0$ are functions of past information, $\mathcal{I}_{t-1}$. 
This structure mirrors the typical GARCH framework, where both the conditional mean and the  conditional standard deviation  evolve over time based on past information.
In the following, we focus primarily on first-order specifications when modeling the dynamics in $\mu_{t}$ and $s_{t}$.
While higher lag orders can be taken into account if required, such specifications are widely favored and commonly used in practice.

In some applications, it may be reasonable to set $\mu_{t}$ to zero or to assume that $\mu_{t} = \mu$, as done in \citet{Taylor:2019}.
However, autocorrelation in financial returns is often non-negligible \citep{Kuester-Mittnik-Paolella:2006}.
This property can be incorporated into our quantile-based modeling framework by allowing the returns to have a time-varying conditional median of the form
$\mu_{t}=Q_{r_{t}}(0.5\,\vert\,\mathcal{I}_{t-1})$, which may be captured by a quantile autoregression (QAR) model,  as proposed by \citet{KoenkerXiao2006}.
For example, a first-order QAR model is formulated as
\begin{equation}
Q_{r_{t}}(0.5\,\vert\,\mathcal{I}_{t-1})=\mu+\phi r_{t-1},
\label{QAR}
\end{equation}
where $\mu$ represents a baseline median, and $\phi$ captures the impact of $r_{t-1}$ on the current period's median.

Our analysis concentrates on downside market risk. Accordingly, we define $0 < \tau \leq 0.05$, where $\tau$
represents the typical left-tail region of interest in financial risk management, capturing extreme losses. 
We then set $\tau \leq p < 0.5$, where $p$ specifies the quantile levels used to construct the scale measure.
Consider the conditional version of (\ref{s}) written here as 
\begin{equation}
s_{t}=s_{t}(p)=\frac{Q_{r_{t}}(1-p\,\vert\,\mathcal{I}_{t-1})-Q_{r_{t}}(p\,\vert\,\mathcal{I}_{t-1})}{c_p}  ,
\label{scale}
\end{equation}
where $c_p=F_{\varepsilon}^{-1}(1-p)-F_{\varepsilon}^{-1}(p)$ is a time-invariant scaling term, as before.\footnote{As \citet{Taylor:2005} observes, estimating scale (or volatility) using the interval between tail quantiles shares similarities with range-based volatility estimation \citep[e.g.,][]{Parkinson:1980, Garman-Klass:1980, Alizadeh-Brandt-Diebold:2002}. This class of methods relies on the difference between the highest and lowest log prices, which essentially correspond to the quantiles at $1-p = 1$ and $p = 0$, respectively.  }  
It is interesting to note here again that $s_{t}$
is invariant to the specific process governing $\mu_{t}$ in (\ref{dynamiclocationscale}). 
We can then express the model specification as 
\begin{equation}
\begin{split}
 r_{t}      & =\mu_{t}+ s_{t}^{\ast}\varepsilon_{t}^{\ast},  \\
s_{t}^{\ast}& =Q_{y_{t}}(1-p\,\vert\,\mathcal{I}_{t-1})-Q_{y_{t}}(p\,\vert\,\mathcal{I}_{t-1}),
\end{split} \label{model}
\end{equation}
where $y_t = r_t - \mu_t$ and $\varepsilon_{t}^{\ast}= y_t/s_{t}^{\ast}=\varepsilon_{t}/c_p$ is simply a rescaled version of $\varepsilon_{t}$. 
The $\tau$th conditional quantile of $y_{t}= s_{t}^{\ast}\varepsilon_{t}^{\ast}$ is given by $Q_{y_{t}}(\tau \,|\, \mathcal{I}_{t-1}) = s_{t}^{\ast} F^{-1}_{\varepsilon^{\ast}}(\tau) $, where $F^{-1}_{\varepsilon^{\ast}}(\tau)$ is a constant.
Similar to the conditional standard deviation in a GARCH model, $s_{t}^{\ast}$ serves as a common factor across quantiles, ensuring coherence and building dependence between them in a parsimonious manner.

To model the dynamics of $s^{\ast}_t$, we draw a parallel with the absolute value GARCH model of \citet[][Section 3.6]{Taylor:1986} and \citet{Schwert:1990}, which captures the conditional standard deviation using past absolute deviations. We assume
\begin{equation}
s^{\ast}_t = \omega + \beta s^{\ast}_{t-1} + \gamma |y_{t-1}|, \label{st}
\end{equation}
so that the scale of returns adjusts dynamically in a manner similar to the absolute value GARCH framework. 
Observing that  $ s_{t-j}^{\ast} F^{-1}_{\varepsilon^{\ast}}(k) = Q_{y_{t-j}}(k \,|\, \mathcal{I}_{t-j-1})  $, we adopt
 the approach of \citet{Xiao-Koenker:2009}, who use an absolute value GARCH structure,  to express the two key quantiles in (\ref{model}) with the following global SAV {CAViaR} structure:
\begin{equation}
Q_{y_{t}}( k \,|\, \mathcal{I}_{t-1})   =  \omega(k) + \big( \beta + \gamma |\varepsilon_{t-1}^{\ast}| \big) Q_{y_{t-1}}(k \,|\, \mathcal{I}_{t-2}),
\label{globalSAV}
\end{equation}
for both $k=p$ and $k=1-p$, subject to $\omega(p) < \omega(1-p)$,  $\beta \ge 0$, and $\gamma \ge 0$. These restrictions ensure non-crossing quantiles and guarantee $s^{\ast}_t > 0 $ in (\ref{model}), provided the recursion begins with $Q_{y_{1}}(p \,|\, \mathcal I_{0}) < Q_{y_{1}}(1-p \,|\, \mathcal I_{0})$.
Here,  the persistence parameters $\beta$ and $\gamma$ are global, meaning they do not vary with $k$, while $\omega(k)= \omega F^{-1}_{\varepsilon^{\ast}}(k)$ is local and dependent on the quantile level $k$.

As an alternative to (\ref{st}), we also consider a threshold-based specification, similar to the threshold GARCH model of \citet{Zakoian:1994}:
\[
s^{\ast}_t = \omega + \beta s^{\ast}_{t-1} + \big( \gamma_{+} \mathds{1}\{y_{t-1} > 0\} + \gamma_{-} \mathds{1}\{y_{t-1} \leq 0\} \big) |y_{t-1}|,
\]
which allows $s^{\ast}_t$ to respond differently to positive and negative values of $y_{t-1}$. 
This leads to the following global AS {CAViaR} structure:
\begin{equation}
Q_{y_{t}}(k \,|\, \mathcal{I}_{t-1}) =  \omega(k) + \Big( \beta + \big( \gamma_{+} \mathds{1}\{y_{t-1} > 0\} + \gamma_{-} \mathds{1}\{y_{t-1} \leq 0\} \big) |\varepsilon_{t-1}^{\ast}| \Big) Q_{y_{t-1}}(k \,|\, \mathcal{I}_{t-2}), 
\label{globalAS}
\end{equation}
for both $k=p$ and $k=1-p$, subject to $\omega(p) < \omega(1-p)$,  $\beta \ge 0$, $\gamma_+ \ge 0$, and $\gamma_- \ge 0$.
Here, $\beta$, $\gamma_{+}$, and $\gamma_{-}$ are global parameters.
This specification allows an asymmetric response in the conditional quantile process, capturing the differing impacts of positive and negative past shocks. 
Empirical studies by \citet{Nelson:1991}, \citet{Glosten-Jagannathan-Runkle:1993}, and \citet{Engle-Ng:1992}, among others, provide strong evidence that asymmetry is crucial for accurately modeling volatility dynamics, as negative stock returns are typically associated with greater increases in return volatility. This justifies the inclusion of an asymmetric term in financial time series models such as the AS {CAViaR}.
For simplicity, the notation in (\ref{SAV}), (\ref{AS}), (\ref{QAR}),  (\ref{globalSAV}), and (\ref{globalAS})  omits the dependence on the involved parameters.

For estimation purposes, we recommend the following sequential strategy: (i) estimate the parameters of (\ref{QAR}) and obtain the
fitted values $\hat{\mu}_{t}=\hat{Q}_{r_t}(0.5\,\vert\,\mathcal{I}_{t-1})$, and then (ii) estimate the parameters of the {CAViaR} specifications 
with $\mu_{t}$ replaced by $\hat{\mu}_{t}$. 
While this approach may not be as statistically efficient as joint estimation, it offers the advantage of being simpler and more numerically robust.
We begin with the estimation of the QAR model.
Let ${\boldsymbol{\theta}}_{k}$ represent the parameter vector for  (\ref{QAR}) with $k=0.5$.
The estimate $\hat{\boldsymbol{\theta}}_{k}$ can be obtained by solving the minimization problem 
\begin{equation}
\min_{\boldsymbol{\theta}_{k}}\sum_{t}\rho_{k}\big[r_{t}-Q_{r_{t}}(k\,\vert\,\mathcal{I}_{t-1})\big],\label{min}
\end{equation}
where $\rho_{k}[u]=u(k-\mathds{1}\{u \leq 0\})$ is the quantile regression ``check" loss function \citep{KoenkerBassett1978}.

The estimation of (\ref{globalSAV}) and (\ref{globalAS}) proceeds with restricted quantile regression. 
When fitting these {CAViaR} specifications to data, a choice needs to be made for the initial value  $Q_{y_{1}}(k \,|\, \mathcal I_{0}) $. 
A  simple and reasonable approach is to set  $Q_{y_{1}}(k \,|\, \mathcal I_{0}) = \hat Q_{y}(k)$, the corresponding sample quantile. 
This choice ensures that all subsequent conditional quantiles will satisfy the non-crossing condition, both in and out of sample.
For instance, in the case of (\ref{globalSAV}), the parameter estimates for $\boldsymbol{\theta} = \big( \omega(p), \omega(1-p),  \beta, \gamma \big)'$ are found by solving:
\begin{equation*}
\begin{split}
& \min_{ {\boldsymbol  \theta}   }  \sum_{k \in \{p, 1-p  \}} \sum_{t} \rho_{k} \left[  y_t  - Q_{y_t}(k \,|\, \mathcal I_{t-1})  \right]  \\
& \mbox{subject to } \left\{ \begin{array}{l} \omega(p) <   \omega(1-p), \\ 
                 \beta \ge 0, \, \gamma \ge 0, \\
        \end{array}           
        \right.             
\end{split} 
\end{equation*} 
with $Q_{y_{1}}(k \,|\, \mathcal I_{0}) = \hat Q_{y}(k )$, for $k \in \{p, 1-p  \}$. Note that the constraints apply only to the model parameters, without involving the data.
The estimation of (\ref{globalAS}) follows a similar procedure, with the constraint on $\gamma$ replaced by  $\gamma_+ \ge 0$ and $\gamma_- \ge 0$.

With the estimated parameters, the fitted value of ${s}_{t}^{\ast}= \hat{s}_{t}^{\ast}(p)$ in (\ref{model}) is calculated as  
\[
\hat{s}_{t}^{\ast}(p) = \hat{Q}_{y_{t}}(1-p\,|\,\mathcal{I}_{t-1}) - \hat{Q}_{y_{t}}(p\,|\,\mathcal{I}_{t-1}).
\]
Then, using the fitted location $\hat{\mu}_{t}$ and scale numerator $\hat{s}_{t}^{\ast}(p)$, it is natural to use the unconditional quantiles of 
$\hat{\varepsilon}_{t}^{\ast}(p) = (r_{t} - \hat{\mu}_{t}) / \hat{s}_{t}^{\ast}(p)$ as estimates for $F_{\varepsilon^{\ast}(p)}^{-1}(\tau)$, since we are not assuming any  particular distribution for $\varepsilon_{t}$ in (\ref{dynamiclocationscale}). 
The estimated left-tail $\tau$-quantile of the conditional return distribution is then given by 
\begin{equation}
\hat{Q}_{r_{t}}(\tau, p \,|\, \mathcal{I}_{t-1}) = \hat{\mu}_{t} + \hat{s}_{t}^{\ast}(p) \hat{F}_{\varepsilon^{\ast}(p)}^{-1}(\tau),
\label{Qhat}
\end{equation}
for $0 < \tau \leq 0.05$.

\subsection{Estimating VaR}

We estimate the VaR in (\ref{defVaR}) based on the left-tail quantile of the conditional return distribution. In our quantile-based framework, the VaR estimate for a given probability level $\alpha$ is defined as
\begin{equation}
\widehat{\text{VaR}}_{r_t}(\alpha, p) = \hat{Q}_{r_{t}}(\alpha, p \,|\, \mathcal{I}_{t-1}),
\label{VaRp}
\end{equation}
where $\hat{Q}_{r_{t}}(\alpha, p \,|\, \mathcal{I}_{t-1})$ is derived from the dynamic quantile estimate in (\ref{Qhat}), evaluated at $\tau=\alpha$.

While the VaR estimate in (\ref{VaRp}) can be computed for a given value of $p$, relying on a single scale-defining quantile level may lead to inefficiencies. Since the seminal work of \citet{PearsonTukey1965}, it has been well understood that using the interval between symmetric tail quantiles to estimate standard deviation is influenced by the skewness and kurtosis of the distribution. Furthermore, as early as \citet{Mosteller:1946}, it was suggested that interquantile range-based estimates of standard deviation could be improved by considering multiple quantile pairs. Building on these insights, we extend this principle to VaR estimation by computing estimates across multiple quantile levels and aggregating the results using an averaging scheme.

Specifically, we define a set of scale-defining quantile levels $\mathcal{P}$ and compute VaR for each $p \in \mathcal{P}$. The final VaR estimate is then given by 
\begin{equation} 
\widehat{\text{VaR}}_{r_t}(\alpha) = \frac{1}{|\mathcal{P}|} \sum_{p \in \mathcal{P}} \widehat{\text{VaR}}_{r_t}(\alpha, p), 
\end{equation} 
where $|\mathcal{P}|$ denotes the number of quantile levels used. For our implementation, we use the set $\mathcal{P} = \{ 0.05, 0.10, 0.15, 0.20, 0.25 \}$. While this choice is somewhat ad hoc, it corresponds to the relatively more robust values reported in \citet[][Exhibit 5.4]{Huber-Ronchetti:2009} for estimating scale using interquantile ranges.

We also considered alternative methods, such as using a single quantile level or aggregating via the median across $\mathcal{P}$. However, simulations indicate that averaging yields greater stability and lower overall estimation error. This approach mitigates fluctuations in tail behavior and enhances out-of-sample performance, particularly for heavy-tailed distributions. A detailed comparison of these methods is provided in Section~A of the Supplementary material.

\subsection{Approximating ES}

The ES in (\ref{defES}) represents the conditional expectation of loss, given that the loss has exceeded the VaR threshold at a specified probability level $\alpha$. 
As shown by \citet{Acerbi-Tasche:2002}, among others, ES can be expressed as
\begin{equation}
\text{ES}_{r_t}(\alpha) = \frac{1}{\alpha} \int_{0}^{\alpha} {Q}_{r_{t}}(\tau \,\vert\, \mathcal{I}_{t-1}) \, d \tau,
\label{ESintegral}
\end{equation}
where ${Q}_{r_{t}}(\tau \,\vert\, \mathcal{I}_{t-1})$ represents the conditional quantile for $\tau \leq \alpha$.

To compute this integral in practice, we approximate it using a sum over $N$ equally spaced subdivisions between 0 and $\alpha$, similar to a Riemann sum approximation:
\[
\widetilde{\text{ES}}_{r_t}(\alpha, p, N) = \frac{1}{N} \sum_{i=1}^{N} \hat{Q}_{r_{t}}(\tau_{i}, p \,\vert\, \mathcal{I}_{t-1}),
\]
where $\tau_{i} = i \alpha / N$ for $i = 1, \ldots, N$, and each $\hat{Q}_{r_{t}}(\tau_{i}, p \,\vert\, \mathcal{I}_{t-1})$ is the corresponding quantile estimate from (\ref{Qhat}).

As with VaR, we compute ES for several values of $p$ to obtain a more robust estimate. Specifically, for each fixed $N$, we take the average ES across different quantile levels $p$:
\[
\widehat{\text{ES}}_{r_t}(\alpha, N) = \frac{1}{|\mathcal{P}|} \sum_{p \in \mathcal{P}} \widetilde{\text{ES}}_{r_t}(\alpha, p, N).
\]
This averaging process reduces estimation errors associated with any single $p$, leading to a more efficient and reliable ES estimate.
The order of operations is important: by first approximating the integral over $\tau$ before averaging across $p$, the quantiles used in each $\widetilde{\text{ES}}_{r_t}(\alpha, p, N)$ share a consistent scale structure. Averaging across $p$ only afterward preserves the internal coherence of each quantile specification and avoids mixing quantiles constructed under differing scale definitions.

Rather than fixing $N$ a priori, we employ an adaptive selection method. We initialize with $N=4$ and iteratively increase it, computing ES at each step. The iteration stops when successive averaged ES estimates stabilize within a predefined tolerance level $\epsilon$:
\[
\big\vert \widehat{\text{ES}}_{r_t}(\alpha, N)  - \widehat{\text{ES}}_{r_t}(\alpha, N-1)  \big\vert < \epsilon,
\]
where we set $\epsilon = 0.0001$ in our implementation. This ensures sufficient numerical accuracy while avoiding unnecessary computations. The stabilized ES estimate is denoted $\widehat{\text{ES}}_{r_t}(\alpha)$.

An important advantage of the QbSD approach is its ability to prevent inconsistencies common in tail risk estimation. Specifically, it avoids: (i) quantile crossings, (ii) crossings among ES values at different probability levels, and (iii) crossings between quantiles and their corresponding ES values. 
This method ensures valid, well-ordered risk estimates across all probability levels $\alpha$, even in the extreme tails of the distribution.

\section{Simulation experiments}

In this section, we investigate the accuracy of the proposed QbSD approach with respect to the one-step-ahead VaR and ES forecasts at probability level  $\alpha$. 
The performance of the QbSD models is evaluated using  mean absolute error (MAE) and root mean squared error (RMSE), and compared to several benchmark models, including GARCH and joint VaR-ES models. The analysis also assesses the models' ability to capture skewness, heavy tails, and leverage effects in financial returns.

\subsection{Simulation design}

The data-generating process (DGP) is the asymmetric power ARCH (APARCH) model of \citet{Ding-Granger-Engle:1993}, specified as
\begingroup
\setlength{\abovedisplayskip}{4pt}
\setlength{\belowdisplayskip}{7pt}
\begin{equation}
\begin{split}
r_t & = \sigma_t \varepsilon_t,\\
\sigma_t^{\delta} & = \omega +  \beta \sigma_{t-1}^{\delta} + \gamma (\vert r_{t-1} \vert - \theta r_{t-1})^{\delta} ,
\end{split} \label{DGP}
\end{equation}
\endgroup
where $\omega, \beta, \gamma, \delta > 0$ and $-1 < \theta < 1$. The innovations $\varepsilon_t$ follow the skewed $t$-distribution of \citet{Hansen1994}.
This model nests several well-known specifications as special cases. In particular, when $\delta = 2$ and $\theta = 0$, the model reduces to a standard GARCH model. 
When $\delta = 1$, it becomes the absolute value GARCH model of \citet[][Section 3.6]{Taylor:1986} and \citet{Schwert:1990}, which implies direct {CAViaR} representations for the conditional quantiles \citep{Xiao-Koenker:2009}.
Furthermore, when $\theta \neq 0$, the model implies an asymmetric response of volatility to past returns. When $\theta > 0$, this asymmetry takes the form of a leverage effect, whereby negative returns have a larger impact on future volatility than positive returns of the same magnitude \citep{Black:1976, Christie:1982, French-Schwert-Stambaugh:1987}.

The density function of the innovation term $\varepsilon_t$ appearing in (\ref{DGP}) is defined as
\begin{equation}
f_{\varepsilon}(x; v, \lambda)= \left\{  \begin{array} {ll}  
                                        b c \left(  1 + \frac{1}{v-2}\left(  \frac{b x  + a}{ 1 - \lambda} \right)^2  \right)^{-(v+1)/2}, & \mbox{  if } x \leq -a/b \\[2.0ex]
                                        b c \left(  1 + \frac{1}{v-2}\left(  \frac{b x  + a}{ 1 + \lambda} \right)^2  \right)^{-(v+1)/2}, & \mbox{  if } x > -a/b            
     \end{array}  \right. \label{skt}
\end{equation}
where the constants $a$, $b$, and $c$ are themselves defined as
\[
a = 4 \lambda c \frac{v-2}{v-1}, \quad b = \sqrt{1 + 3 \lambda^2 - a^2}, \quad c=\frac{\Gamma(\frac{v+1}{2})}{\sqrt{\pi (v-2)} \Gamma(\frac{v}{2})},
\]
and the parameters $v > 2$ and $-1 < \lambda < 1$ represent the degrees of freedom and the asymmetry of the distribution, respectively. 
The skewed $t$-distribution has a mean of zero and unit variance.
When $ \lambda = 0 $, it reduces to a standardized version of the traditional Student-$t$ distribution. A positive $ \lambda $ implies right skewness. Financial returns typically have a higher probability of large negative returns than large positive ones, corresponding to left skewness ($\lambda < 0$).

The distribution in (\ref{skt}) is piecewise-defined around its mode at $x=-a/b$, which separates the left and right parts of the density and plays an important role in the derivation of VaR and ES. \citet{Jondeau-Rockinger2003} show that the associated cumulative distribution function (CDF)  is defined by
\[
F_{\varepsilon}(x;v,\lambda)=\left\{  \begin{array}{ll} 
                            (1 - \lambda) F_{\mathcal T} \left(  \frac{bx+a}{1-\lambda} \sqrt{\frac{v}{v-2}}; v \right), & \mbox{  if } x \leq -a/b \\[2.0ex]
                            (1 + \lambda) F_{\mathcal T} \left(  \frac{bx+a}{1+\lambda} \sqrt{\frac{v}{v-2}}; v \right) - \lambda, & \mbox{  if } x > -a/b \\[2.0ex]                            
\end{array}
 \right.
\]
where $F_{\mathcal T}(x;v)$ is the CDF of the Student-$t$ distribution with $v$ degrees of freedom, and the quantile function of the skewed $t$-distribution is given by
\[
F_{\varepsilon}^{-1}(u;v,\lambda)=\left\{  \begin{array}{ll} 
                          \frac{1}{b}  \left( (1 - \lambda) \sqrt{\frac{v-2}{v}} F_{\mathcal T}^{-1}\left(  \frac{u}{1-\lambda} ; v \right) - a\right), & \mbox{  if } u \leq \frac{1-\lambda}{2} \\[2.0ex]
                          \frac{1}{b}   \left(  (1 + \lambda) \sqrt{\frac{v-2}{v}} F_{\mathcal T}^{-1}\left(  \frac{u + \lambda}{1+\lambda}; v \right) -a \right), & \mbox{  if } u > \frac{1-\lambda}{2}. \\[2.0ex]                            
\end{array}
 \right.
\]

Under this DGP, both the VaR and ES of $r_t$ are proportional to $\sigma_{t}$.  The true one-step-ahead VaR forecast is 
$
\text{VaR}_{r_{t+1}}(\alpha; v, \lambda) =  \sigma_{t+1} F_{\varepsilon}^{-1}(\alpha;v,\lambda) 
$
and the true one-step-ahead ES forecast is 
$
\text{ES}_{r_{t+1}}(\alpha; v, \lambda) = \sigma_{t+1} \text{ES}_{\varepsilon}(\alpha; v, \lambda),
$
where the ES for  \possessivecite{Hansen1994} skewed $t$-distribution is given by \citet{Patton-Ziegel-Chen:2019} as\footnote{The expression in Appendix B.4 of \cite{Patton-Ziegel-Chen:2019} contains some typos. It can be verified, either analytically or by simulation, that \eqref{ESskewt} is the correct ES expression for Hansen’s skewed $t$-distribution. In the second branch of \eqref{ESskewt}, the parameters are flipped to $(v,-\lambda)$ (so the constants change to $a_2,b_2$ accordingly) when evaluating $\text{ES}^{\ast}_{\varepsilon}$. The prefactor $(1-\alpha)/\alpha$ converts that right-tail ES at level $1-\alpha$ back to the desired left-tail level $\alpha$.}
\begin{equation}
\text{ES}_{\varepsilon}(\alpha; v, \lambda) = \left\{ \begin{array}{ll}
                                                  \text{ES}^{\ast}_{\varepsilon}(\alpha; v, \lambda),                                     &  \mbox{  if } F_{\varepsilon}^{-1}(\alpha;v,\lambda) \leq -a/b \\
                                                  \frac{1-\alpha}{\alpha}\text{ES}^{\ast}_{\varepsilon}(1-\alpha; v, -\lambda),          &  \mbox{  if } F_{\varepsilon}^{-1}(\alpha;v,\lambda)  > -a/b 
                                                \end{array}
                                        \right. \label{ESskewt}
\end{equation}
with 
\[
\text{ES}^{\ast}_{\varepsilon}(\alpha; v, \lambda) = \frac{\tilde \alpha}{\alpha}(1-\lambda) \Big(-\frac{a}{b} + \frac{1-\lambda}{b} \text{ES}_{\mathcal T}(\tilde \alpha; v)   \Big)
\]
and
\[
\tilde \alpha = F_{\varepsilon} \bigg( \frac{b}{1-\lambda}\Big( F_{\varepsilon}^{-1}(\alpha;v,\lambda)  + \frac{a}{b} \Big)  ;v,0  \bigg).
\]
Here $\text{ES}_{\mathcal T}(\alpha; v)$ is the ES for a  Student-$t$ distribution, standardized  to have unit variance,   given by
\[
\text{ES}_{\mathcal T}(\alpha; v) =  -\frac{1}{\alpha}  f_{\mathcal T}( q_{\alpha}; v) \Big( \frac{v + q_{\alpha}^2}{v-1} \Big) \sqrt{\frac{v-2}{v} },
\]
where $f_{\mathcal T}( x; v)$ and $q_{\alpha} = F_{\mathcal T}^{-1}(\alpha; v)$ are, respectively, the density and $\alpha$-quantile of the Student-$t$ distribution with $v$ degrees of freedom \citep{BrodaPaolella2011, Dobrev-Nesmith-Oh:2017}.

In the DGP outlined in (\ref{DGP}), we set the parameters as follows: $\omega = 0.05$, $\beta = 0.85$, $\gamma = 0.10$, and $\delta = 1.5$, while allowing $\theta$ to take values in $\{0, 0.5\}$. For the skewed $t$-distribution in  (\ref{skt}), we select $v \in \{5, 20 \}$ to represent different tail thicknesses, with $v = 5$ corresponding to heavy tails and $v = 20$ representing thinner tails. We use $\lambda \in \{0, -0.5\}$ to capture different levels of skewness, with $\lambda = 0$ indicating  no skewness and $\lambda = -0.5$ implying left skewness.
The sample size is set to $T = 1250$, which corresponds to roughly five years of daily returns and matches the primary rolling-window length used in our empirical application.
The results presented in the main text focus on this baseline case; additional simulations for $T \in \{250, 2500\}$ are reported in Section~B of the Supplementary material. 
We consider three probability levels for the forecasts: $\alpha \in \{0.01, 0.025, 0.05\}$. The choice of $\delta = 1.5$ allows us to explore a setting where the power parameter lies between the standard GARCH model ($\delta = 2$) and the absolute value GARCH model ($\delta = 1$).\footnote{This value of $\delta$ is close to the estimate of $1.43$ reported by \citet{Ding-Granger-Engle:1993} in their study of S\&P 500 returns.}
We set $\theta \in \{0, 0.5 \}$ to explore the role of asymmetry in volatility dynamics. As noted earlier, $\theta = 0$ yields a symmetric response, while $\theta = 0.5$ introduces a leverage effect.

For each DGP configuration, we evaluate the forecasting accuracy by computing the MAE and the RMSE across all 1,000 generated return series. These metrics allow us to compare the performance of the models in capturing the true one-step-ahead VaR and ES forecasts.
In our simulations, we consider the QbSD model with the two global {CAViaR} specifications in (\ref{globalSAV}) and (\ref{globalAS}).
Consistent with the simulation DGP in (\ref{DGP}), all QbSD and benchmark models, presented next, assume a zero location for returns.
This assumption simplifies the modeling process, allowing us to focus on volatility and tail risk dynamics.

\subsection{Benchmark models}

To assess the performance of our QbSD models, we compare them with several established models commonly used in financial risk forecasting.
 The benchmarks include the VaR and ES forecasts from the standard GARCH model, the asymmetric GJR-GARCH model \citep{Glosten-Jagannathan-Runkle:1993}, and the EGARCH  model \citep{Nelson:1991}, each with normal innovations. Additionally, for the GARCH and GJR-GARCH models, we consider alternative distributional assumptions for the innovations, including the Student-$t$ distribution and \possessivecite{Hansen1994} more general 
 skewed $t$-distribution.\footnote{We consider the EGARCH model only with normal innovations, as it tends to exhibit instability when applied with $t$-distributed errors \citep[][p. 365]{Nelson:1991}.}

We also include the joint VaR-ES model proposed by \citet{Taylor:2019}, which employs a pseudo-maximum likelihood estimation approach based on the AL density function for the joint estimation of VaR and ES. The conditional AL density function is given by
\[
f(r_{t}) = \frac{(\alpha-1)}{\text{ES}_{t}}\exp\left(\frac{\left(r_{t}-\text{VaR}_{t}\right)\left(\alpha-   \mathds{1}\{ r_{t} \leq \text{VaR}_{t} \} \right)}{\alpha \text{ES}_{t}}\right),
\]
where $\text{VaR}_{t}$ is modeled using either the SAV or AS {CAViaR} specifications from expressions (\ref{SAV}) and (\ref{AS}), with $ y_{t} $  replaced by $ r_{t} $ under the mean zero assumption. The complete model includes two alternative specifications for the dynamics of the ES:
\begin{equation}
\text{ES}_{t} = \left(1 + \exp(\gamma_{0})\right) \text{VaR}_{t} \label{AL-Mult}
\end{equation}
or
\begin{equation}
\text{ES}_{t} = \text{VaR}_{t} - x_{t}, \label{AL-AR}
\end{equation}
where $ x_{t} $ follows a dynamic updating process:
\[
x_{t} = \begin{cases}
\gamma_{0} + \gamma_{1}(\text{VaR}_{t-1} - r_{t-1}) + \gamma_{2}x_{t-1}, & \text{if}\; r_{t-1} \leq \text{VaR}_{t-1} \\
x_{t-1}, & \text{otherwise}.
\end{cases}
\]
This formulation provides a smooth adjustment to the magnitude of exceedances beyond the VaR, dynamically updating based on past returns and exceedances.

To differentiate between the two AL-based specifications, we use the notation ``$\text{AL}_{\text{Mult.}}$'' for the multiplicative version in (\ref{AL-Mult}), where ES is modeled as a multiple of VaR, and ``$\text{AL}_{\text{AR}}$'' for the autoregressive version in (\ref{AL-AR}). For these AL-based models, we further indicate the choice of VaR specification using the suffixes ``-SAV'' for the SAV {CAViaR} model in (\ref{SAV}) and ``-AS''  for the AS {CAViaR} model in (\ref{AS}). For example, $\text{AL}_{\text{Mult.}}$-SAV refers to the multiplicative AL-based model with the SAV {CAViaR} specification.
The QbSD models, in contrast, rely on global {CAViaR} specifications, where the scale dynamics are shared across quantiles. Here, the suffix ``-gSAV'' denotes the global SAV {CAViaR} specification (\ref{globalSAV}), while ``-gAS'' refers to the global AS {CAViaR} specification (\ref{globalAS}). For example, QbSD-gSAV refers to the QbSD model with the global SAV {CAViaR} specification.

In addition to the quantile-based models, we also consider the one-factor GAS  model proposed by \citet{Patton-Ziegel-Chen:2019}, which provides an alternative framework for jointly modeling VaR and ES.
This model is based on the following member  of the \citet{Fissler-Ziegel:2016} class of loss functions:
\[
L_{\text{FZ0}} (r_{t}, \text{VaR}_{t}, \text{ES}_{t} ) = -\frac{1}{\alpha \text{ES}_{t}} \mathds{1}\{ r_{t} \leq \text{VaR}_{t}\} (\text{VaR}_{t} - r_{t}) + \frac{\text{VaR}_{t}}{\text{ES}_{t}} + \log(- \text{ES}_{t}) - 1.
\]
The $\text{FZ}_0$ loss function is strictly consistent for the pair (VaR, ES). Although the loss itself is not homogeneous of degree zero---due to the $\log(-\text{ES}_t)$ term---the differences in loss values between competing forecasts are. This ensures that forecast rankings based on average loss are invariant to positive scaling of both returns and risk measures. 

The joint specification for VaR and ES is
\[
\text{VaR}_{t} = \zeta \exp(\kappa_{t}), \quad \text{ES}_{t} = \xi \exp(\kappa_{t}), \quad \text{with} \ \xi < \zeta < 0,
\]
where the common factor $ \kappa_{t} $ evolves dynamically as
\[
\kappa_{t} = \omega + \beta \kappa_{t-1} + \gamma g_{t-1},
\]
and  the score function $ g_{t} $, derived from the FZ0 loss function, is given by
\[
g_{t} = -\frac{1}{\text{ES}_{t}} \left( \frac{1}{\alpha} \mathds{1}\{r_{t} \leq \text{VaR}_t \} r_{t} - \text{ES}_{t} \right).
\]
This approach dynamically updates both VaR and ES based on the information in past losses.\footnote{Since the Hessian of the FZ0 loss is constant, the scaling matrix in the GAS framework of \citet{Creal-Koopman-Lucas:2013} can be set to one without loss of generality. This simplifies estimation while preserving the model's essential dynamics.} Following \citet{Patton-Ziegel-Chen:2019}, we set $\omega = 0$ for identification and estimate the model by minimizing the average FZ0 loss.

These competing models offer a range of benchmarks for evaluating the forecasting performance of the QbSD models. In the next section, we present the simulation results and assess how well each model captures key features of financial returns across the various DGP configurations. 
It is interesting to note that none of the models exactly matches the DGP in (\ref{DGP}), which makes the forecasting task more challenging and offers a stricter test of adaptability.

\subsection{Numerical results}

This section presents the forecasting results for the QbSD and benchmark models across the simulation scenarios. Tables~1--4 report out-of-sample one-step-ahead VaR and ES accuracy (MAE and RMSE) under symmetric and asymmetric volatility dynamics. 

Table~1 reports VaR forecast accuracy with no leverage (symmetric volatility dynamics). Parametric GARCH-family models with correctly specified innovations (Student-$t$ under symmetry; skew-$t$ under left skew) lead across configurations. With symmetric innovations ($v=20,\lambda=0$), GARCH/GJR with Student-$t$ is best across $\alpha$ (e.g., MAE $0.071,0.054,0.044$; RMSE $0.095,0.073,0.059$), while QbSD-gSAV is competitive but not top. With left skewness ($v=20,\lambda=-0.5$), skew-$t$ dominates (e.g., GJR-skew-$t$ MAE $0.098,0.068,0.049$; RMSE $0.132,0.093,0.069$), with only minor differences between GARCH and GJR. Under heavier tails ($v=5$), the pattern persists: when $\lambda=0$, GARCH/GJR-$t$ remains best (MAE $0.103,0.068,0.050$; RMSE $0.143,0.097,0.072$); when $\lambda=-0.5$, GJR-GARCH with skew-$t$ leads (MAE $0.145,0.084,0.054$; RMSE $0.213,0.134,0.093$). Across all blocks, QbSD-gSAV improves on AL-based and GAS alternatives but trails the best-specified GARCH variants.

Turning to ES under symmetric volatility (Table~2), the same ordering emerges: matching the innovation distribution to the DGP is pivotal. In the symmetric case ($v=20,\lambda=0$), GARCH/GJR-$t$ attains the lowest errors (MAE $0.100,0.075,0.060$; RMSE $0.129,0.099,0.081$). Introducing left skewness ($v=20,\lambda=-0.5$) shifts the frontier to skew-$t$ (GJR-skew-$t$ MAE $0.146,0.105,0.080$; RMSE $0.191,0.140,0.108$). Under heavier tails ($v=5$), Student-$t$ remains best when $\lambda=0$ (MAE $0.180,0.118,0.086$; RMSE $0.241,0.162,0.120$), while skew-$t$ variants lead for $\lambda=-0.5$ (e.g., GJR-skew-$t$ MAE $0.117,0.173,0.117$; RMSE $0.174,0.247,0.383$). AL-based and GAS models are consistently less accurate.

Table~3 reports VaR accuracy with leverage ($\theta=0.5$). QbSD-gAS delivers the lowest errors in most configurations featuring skewness and/or heavy tails. For example, at $v=20,\lambda=-0.5$, it is best across $\alpha$ (MAE $0.172,0.125,0.096$; RMSE $0.238,0.183,0.138$). In the heavy-tailed but symmetric case ($v=5,\lambda=0$), QbSD-gAS is generally strongest, with EGARCH edging it at $\alpha=2.5\%$ (MAE $0.105$ vs.\ $0.110$). Under heavy tails with left skewness ($v=5,\lambda=-0.5$), GJR-skew-$t$ attains the lowest MAE ($0.233,0.152,0.108$), with QbSD-gAS close behind. A notable exception is the thin-tailed, symmetric DGP ($v=20,\lambda=0$), where EGARCH dominates (MAE $0.105,0.075,0.062$; RMSE $0.148,0.107,0.086$).

Table~4 presents ES accuracy with leverage. QbSD-gAS leads in most configurations. It is best across $\alpha$ at $v=20,\lambda=-0.5$ (MAE $0.235,0.174,0.140$; RMSE $0.314,0.238,0.195$) and at $v=5,\lambda=0$ (MAE $0.292,0.186,0.136$). In the heavy-tailed, left-skewed case $v=5,\lambda=-0.5$, GJR-skew-$t$ has the edge (MAE $0.388,0.263,0.191$; RMSE $0.526,0.352,0.256$). For the thin-tailed symmetric DGP $v=20,\lambda=0$, EGARCH is competitive and surpasses QbSD-gAS at higher $\alpha$ (e.g., MAE $0.113,0.086$ and RMSE $0.157,0.122$ at $2.5\%$ and $5\%$, vs.\ QbSD-gAS $0.125,0.103$ and $0.165,0.137$); QbSD-gAS is slightly better at $1\%$ (MAE $0.165$ vs.\ $0.167$).

Overall, the evidence suggests a clear division: when volatility responds symmetrically and the innovation distribution is correctly specified (e.g., appropriately heavy-tailed or skewed), parametric GARCH variants set the benchmark for VaR and ES; once leverage/asymmetry matters, allowing asymmetric-slope (gAS) scale dynamics is pivotal---the proposed QbSD-gAS specification is typically strongest, particularly for ES, with EGARCH competitive and occasionally best in thin-tailed symmetric settings. Across these scenarios, AL-based and GAS models remain less accurate. We next examine whether these patterns carry over to international stock index returns.

\section{Empirical Application}

We evaluate one-day-ahead VaR and ES forecasts for the daily log returns of the S\&P~500, DJIA, NASDAQ, EURO STOXX~50, FTSE~100, DAX, CAC~40, and TSX stock indices. The daily adjusted closing prices for these indices were obtained from Yahoo Finance, spanning October 4, 2002, to February 2, 2024, across indices. Due to differences in trading days and data availability, the number of observations initially varies across the series.

To ensure comparability across stock indices, we standardized the sample size to match the index with the shortest available data. Specifically, the EURO STOXX~50 has the shortest sample, starting on October 21, 2002, and containing 5{,}357 daily observations, ending on February 2, 2024. For consistency, we truncated the data for indices with longer samples by counting backward from February 2, 2024, to retain exactly 5{,}357 observations for each index. This adjustment implies the following starting dates: October 22, 2002, for the S\&P~500, DJIA, and NASDAQ; October 21, 2002, for the EURO STOXX~50; November 18, 2002, for the FTSE~100; January 3, 2003, for the DAX; February 27, 2003, for the CAC~40; and October 4, 2002, for the TSX. As returns are calculated as first differences of log prices, this sample size yields 5{,}356 daily return observations for all indices.

Figure~1 illustrates the time series of daily log returns (in percentages) for the S\&P~500, DJIA, NASDAQ, and EURO STOXX~50 indices, each exhibiting notable spikes in volatility during significant economic events such as the 2008 financial crisis and the COVID-19 pandemic. Figure~2 shows the daily log returns for the FTSE~100, DAX, CAC~40, and TSX indices, which similarly exhibit heightened volatility during periods of financial stress.

We use a rolling window of $R=1250$ estimation observations (approximately five trading years) to produce one-day-ahead forecasts for VaR and ES. This choice balances sampling variability against responsiveness to structural change.
Model parameters are estimated using the most recent 1{,}250 observations, and forecasts are updated daily as the window moves through the sample period. Given the common sample of 5{,}356 daily returns, this setup leaves an out-of-sample evaluation period of 4{,}106 observations. We forecast VaR and ES at probability levels of 1\%, 2.5\%, and 5\%. The 1\% and 5\% levels are widely used in the literature, whereas the 2.5\% level corresponds to the regulatory standard for ES adopted under Basel~III. We also examined alternative rolling-window sizes of $R=250$ and $R=2500$. To maintain focus in the main text, we present results for the 1{,}250-observation window, while these alternative specifications appear in Section~C of the Supplementary material.

We evaluate a total of twenty-eight competing risk models for forecasting VaR and ES, including: (i) fourteen GARCH-type models (GARCH and GJR-GARCH with normal, Student-$t$, and \citeauthor{Hansen1994} skew-$t$ innovations under zero or AR(1) conditional means, plus EGARCH with normal innovations under zero or AR(1) means), (ii) two GAS variants with zero and AR(1) means \citep{Patton-Ziegel-Chen:2019}, (iii) eight AL-based models following \citet{Taylor:2019} (SAV and AS CAViaR for VaR dynamics, multiplicative and autoregressive specifications for ES dynamics, each under zero or AR(1) means), and (iv) four QbSD models (gSAV and gAS scale dynamics under zero or QAR(1) conditional medians).

In the following tables, we use the prefixes ``AR-'' and ``QAR-'' to denote an AR(1) conditional mean and a QAR(1) conditional median, respectively; models without a prefix assume a zero conditional location.

\subsection{Evaluation of VaR forecasts}

We evaluate the accuracy of VaR forecasts using the model confidence set (MCS) procedure of \citet{Hansen-Lunde-Nason:2011}, which identifies a subset of models that are not significantly outperformed at a given confidence level. The MCS framework is particularly useful in a multi-model comparison setting, as it does not require selecting a single best model but instead acknowledges model uncertainty by retaining all models that cannot be statistically eliminated.

Following \possessivecite{Hansen-Lunde-Nason:2011} suggestion, we apply the range-based procedure using the maximum absolute $t$-statistic, which iteratively removes the model with the largest loss difference relative to the others until only models that cannot be statistically distinguished from the best-performing ones remain. The final MCS consists of models that survive this elimination process at the chosen confidence level.\footnote{Specifically, we base the MCS on the $T_{R, \mathcal{M}}$ test statistic developed by \citet[][Section 3.1.2]{Hansen-Lunde-Nason:2011} and compute it with 1{,}000 bootstrap iterations using the R package `MCS' \citep{Bernardi-Catania:2018}.}

The test statistic is computed using the quantile score loss, given by
\begin{equation}
S(r_{t+1}, \text{VaR}_{t+1}) = \left( r_{t+1} - \text{VaR}_{t+1} \right) \left( \alpha - \mathds{1}\left\{ r_{t+1} \leq \text{VaR}_{t+1} \right\} \right),
\label{QS}
\end{equation}
where $r_{t+1}$ is the realized return and $\text{VaR}_{t+1}$ is the predicted VaR at a given probability level $\alpha$. 
This loss function is asymmetric and penalizes underpredictions more heavily than overpredictions, making it well suited to the evaluation of downside risk.

Tables 5--7 present the MCS results for 1\%, 2.5\%, and 5\% VaR forecasts across the eight international stock indices. Models excluded from the 90\% MCS ($\widehat{\mathcal{M}}_{90\%}^{*}$) for a given index are left blank. 
The remaining models are ranked by their average quantile score $t$-statistics; the ``\#'' column counts the number of indices for which a model remains in the MCS, while the ``Avg. rank'' and ``Final rank'' summarize overall performance.

For the 1\% tail (Table~5), QbSD-gAS attains the best overall rank and appears in the MCS for all eight indices; its closest competitors are GARCH variants with skew-$t$ innovations (GJR-GARCH-skew-$t$, GARCH-skew-$t$), followed by GJR-GARCH-$t$, while AL-based specifications generally trail. 
At 2.5\% (Table~6), QbSD-gAS again ranks first overall (in the MCS for all eight indices), with AL-based models featuring AS dynamics (AL$_{\text{Mult.}}$-AS, AR-AL$_{\text{Mult.}}$-AS) close behind; among GARCH-type models, skew-$t$ remains the most competitive and EGARCH sits in the upper middle of the table. 
By 5\% (Table~7), QbSD-gAS still leads, EGARCH moves into second place, AL$_{\text{Mult.}}$-AS and the QAR-driven QbSD-gAS (QAR-QbSD-gAS) remain prominent, and skew-$t$ GARCH variants continue to perform well.

Across probability levels, the symmetric QbSD specification (QbSD-gSAV) underperforms its asymmetric counterpart, underscoring the value of allowing an asymmetric scale response in practice. Within the GARCH family, skew-$t$ innovations generally improve performance at tighter tails (1\% and 2.5\%), whereas the strong showing of EGARCH at 5\% suggests that volatility asymmetry can partly compensate when attention shifts away from the far tail.

While these results provide important insights into tail-risk quantile prediction, a fuller assessment requires joint evaluation of VaR and ES, which we consider next.

\subsection{Joint evaluation of VaR and ES forecasts}

To jointly evaluate VaR and ES forecasts, we use the AL log score proposed by \citet{Taylor:2019}. This scoring rule is specifically designed to assess both VaR and ES, reflecting their joint elicitability. The AL log score is given by
\begin{align}
S(r_{t+1}, \mu_{t+1}, \text{VaR}_{t+1}, \text{ES}_{t+1}) =
& -\log\left( \frac{\alpha - 1}{\text{ES}_{t+1} - \mu_{t+1}} \right) \nonumber \\[2.0ex]
& - \frac{(r_{t+1} - \text{VaR}_{t+1})(\alpha - \mathds{1}\{r_{t+1} \leq \text{VaR}_{t+1}\})}{\alpha (\text{ES}_{t+1} - \mu_{t+1})},
\label{ALS}
\end{align}
where $\mu_{t+1}$ is the model's conditional mean and $\text{ES}_{t+1} < \mu_{t+1}$ is required. 
This proper scoring rule penalizes both the incidence and severity of forecast errors in the left tail, encouraging accurate joint predictions of VaR and ES.

Tables~8--10 summarize the MCS rankings under AL log scores. At the 1\% tail (Table~8), the best overall performer is {AR-GJR-GARCH-skew-$t$}, with {QbSD-gAS} a close second; both appear in the MCS for all eight indices. Other skew-$t$ GARCH variants ({GARCH-skew-$t$}, {GJR-GARCH-skew-$t$}) are also highly competitive, whereas AL-based specifications sit further down the table.
At 2.5\% (Table~9), leadership shifts to {QbSD-gAS}, which ranks first overall and is included in the MCS for all eight indices. AL-based models with {asymmetric VaR dynamics} (AS CAViaR), namely {AL$_{\text{Mult.}}$-AS} and {AR-AL$_{\text{Mult.}}$-AS}, are close followers, with skew-$t$ GARCH entries forming the next tier.
By 5\% (Table~10), {QbSD-gAS} again takes the top spot (in the MCS for all indices), followed by {AL$_{\text{AR}}$-AS} and {AL$_{\text{Mult.}}$-AS}; the QAR-driven {QAR-QbSD-gAS} also ranks near the front. Skew-$t$ GARCH specifications remain competitive, while EGARCH occupies an upper-mid-table position.

Taken together, these results indicate that joint VaR-ES forecasting favors models that accommodate {asymmetric tail dynamics}. {QbSD-gAS} is consistently among the best across tail levels and leads at 2.5\% and 5\%, while skew-$t$ GARCH variants are particularly strong at the most extreme tail (1\%). AL-based models with AS CAViaR VaR dynamics provide robust competitors at 2.5\% and 5\%, reinforcing the importance of allowing asymmetry in the VaR/ES system.

\section{Conclusion}

This paper introduced QbSD for forecasting VaR and ES: a semiparametric, distribution-free approach that models conditional scale via restricted quantile regression and accommodates skewness, heavy tails, and leverage---features that intensify during periods of market stress.

Across simulations and international equity indices, a consistent picture emerges. In simulation designs without leverage, parametric GARCH models with heavy-tailed or skewed innovation distributions set a high benchmark for both VaR and ES; Student-$t$ and skew-$t$ specifications are especially effective. 
With leverage, allowing asymmetric-slope quantile dynamics is pivotal: QbSD-gAS is frequently among the best, particularly for ES, while for VaR, EGARCH and skew-$t$ GJR-GARCH can also be highly competitive depending on the tail level and distributional setting.

In the empirical evaluation with a five-year rolling window ($R=1250$), QbSD-gAS is the clearest overall performer for VaR under quantile-score MCS: it attains the best average rank at 1\%, 2.5\%, and 5\% across the eight indices, with skew-$t$ GARCH variants forming the next tier and EGARCH rising near the front at 5\%. When VaR and ES are assessed jointly via AL log scores, performance at the most extreme tail (1\%) tilts toward skew-$t$ GARCH (AR-GJR-GARCH-skew-$t$), with QbSD-gAS a close second; at 2.5\% and 5\%, QbSD-gAS leads, and AL-based specifications with AS CAViaR VaR dynamics (both multiplicative and autoregressive ES linkages) are robust followers.

Two practical lessons follow.
First, asymmetry helps---but not all asymmetries are equal: QbSD-gAS delivers the most consistent improvements; skew-$t$ innovations help when empirical skewness is material; and EGARCH’s volatility asymmetry tends to help when return innovations are near-symmetric with relatively light tails, whereas GJR’s benefits appear mainly when combined with skew-$t$ under skewed, heavy-tailed conditions.
Second, the preferred specification varies with the tail probability and the evaluation target: skew-$t$ GARCH may perform well at the far tail (1\%) under joint VaR-ES scoring, whereas QbSD-gAS provides the most consistent gains across tail levels for VaR and remains among the leaders at 2.5\% and 5\% in the joint evaluation.

Overall, QbSD---especially with asymmetric-slope (gAS) quantile dynamics---offers a robust, flexible addition to the risk-management toolkit. Its distribution-free construction, strong showing under leverage and during turbulent periods, and straightforward extensibility to log-location-scale settings (e.g., strictly positive losses such as credit losses and insurance claims) make it well suited for practical deployment.

\section*{Acknowledgments}

We thank the editor, {Dick van Dijk}, the associate editor, and an anonymous reviewer for their constructive feedback and insightful suggestions. 
{Richard Luger} is supported in part by funding from the {Social Sciences and Humanities Research Council of Canada}.

\section*{Data and code availability}
The data and code needed to reproduce the results presented in this paper are available at {https://github.com/richardluger/QbSD}.
The code is written in {R} with {C++} functions for computational efficiency.

\clearpage
\newpage

\renewcommand{\baselinestretch}{1.635}
\small\normalsize

\bibliographystyle{chicago}

\bibliography{QbSD_Refs}

\newpage

\begin{landscape}

\bigskip
\bigskip

\begin{table}[h]
\begin{center}
\begin{minipage}{8.8in}
\vspace{-1.4cm}
{\small {\bf Table 1.} VaR forecast accuracy when there are no leverage effects}
\footnotesize
\\[0.5ex]
\begin{tabular*}{\textwidth}{@{\extracolsep{\fill}}  lcrrrcrrrcrrrcrrr}
\toprule

                        &&   \multicolumn{3}{c}{$v=20$, $\lambda=0$}   && \multicolumn{3}{c}{$v=20$, $\lambda=-0.5$}   && \multicolumn{3}{c}{$v=5$, $\lambda=0$} && \multicolumn{3}{c}{$v=5$, $\lambda=-0.5$}    \\

                        \cline{3-5} \cline{7-9} \cline{11-13} \cline{15-17} \\[-2.0ex]
  
                                   &$\alpha=$ & 0.01  &   0.025    & 0.05       && 0.01  &   0.025    & 0.05    && 0.01  &   0.025    & 0.05   && 0.01  &   0.025    & 0.05  \\

\midrule

\multicolumn{17}{l}{Panel A: Mean absolute error} \\

QbSD approach                               &&          &             &               &&          &             &           &&          &             &           &&          &             &          \\
\quad QbSD-gSAV                             &&    0.105 &    0.079    &    0.061      &&    0.134 &    0.095    &    0.071  &&    0.148 &    0.094    &    0.066  &&    0.201 &    0.119    &    0.080  \\
\quad QbSD-gAS                              &&    0.118 &    0.088    &    0.070      &&    0.143 &    0.102    &    0.079  &&    0.157 &    0.102    &    0.075  &&    0.207 &    0.125    &    0.086  \\
AL approach                                 &&          &             &               &&          &             &           &&          &             &           &&          &             &          \\
\quad $\text{AL}_{\text{Mult.}}$-SAV        &&    0.153 &    0.101    &    0.073      &&    0.220 &    0.136    &    0.097  &&    0.226 &    0.129    &    0.083  &&    0.344 &    0.180    &    0.114  \\
\quad $\text{AL}_{\text{AR}}$-SAV           &&    0.154 &    0.102    &    0.075      &&    0.218 &    0.137    &    0.102  &&    0.229 &    0.132    &    0.085  &&    0.345 &    0.187    &    0.122  \\
\quad $\text{AL}_{\text{Mult.}}$-AS         &&    0.187 &    0.115    &    0.085      &&    0.258 &    0.157    &    0.112  &&    0.280 &    0.151    &    0.097  &&    0.390 &    0.214    &    0.133  \\
\quad $\text{AL}_{\text{AR}}$-AS            &&    0.187 &    0.118    &    0.089      &&    0.260 &    0.163    &    0.119  &&    0.280 &    0.156    &    0.103  &&    0.394 &    0.223    &    0.142  \\
GAS                                         &&    0.283 &    0.204    &    0.160      &&    0.338 &    0.236    &    0.165  &&    0.340 &    0.225    &    0.151  &&    0.429 &    0.252    &    0.163  \\
GARCH                                       &&          &             &               &&          &             &           &&          &             &           &&          &             &          \\
\quad Normal                                &&    0.086 &    0.056    &\bf 0.044      &&    0.477 &    0.303    &    0.180  &&    0.241 &    0.087    &    0.082  &&    0.766 &    0.367    &    0.153  \\
\quad Student-$t$                           &&\bf 0.071 &\bf 0.054    &\bf 0.044      &&    0.392 &    0.285    &    0.195  &&\bf 0.103 &\bf 0.068    &\bf 0.050  &&    0.602 &    0.370    &    0.218  \\
\quad Skew-$t$                              &&    0.080 &    0.060    &    0.047      &&    0.099 &\bf 0.068    &\bf 0.049  &&    0.111 &    0.073    &    0.053  &&    0.150 &    0.087    &    0.055  \\
GJR-GARCH                                   &&          &             &               &&          &             &           &&          &             &           &&          &             &          \\
\quad Normal                                &&    0.086 &    0.056    &\bf 0.044      &&    0.477 &    0.303    &    0.180  &&    0.241 &    0.087    &    0.082  &&    0.765 &    0.367    &    0.153  \\
\quad Student-$t$                           &&\bf 0.071 &\bf 0.054    &\bf 0.044      &&    0.392 &    0.285    &    0.195  &&\bf 0.103 &\bf 0.068    &\bf 0.050  &&    0.602 &    0.369    &    0.218  \\
\quad Skew-$t$                              &&    0.080 &    0.059    &    0.047      &&\bf 0.098 &\bf 0.068    &\bf 0.049  &&    0.111 &    0.073    &    0.053  &&\bf 0.145 &\bf 0.084    &\bf 0.054  \\
EGARCH                                      &&    0.093 &    0.066    &    0.054      &&    0.480 &    0.306    &    0.182  &&    0.245 &    0.099    &    0.093  &&    0.772 &    0.367    &    0.152  \\

\midrule                         
                         
\multicolumn{17}{l}{Panel B: Root mean squared error} \\

QbSD approach                               &&          &             &               &&          &             &           &&          &             &           &&          &             &          \\
\quad QbSD-gSAV                             &&    0.136 &    0.101    &    0.079      &&    0.173 &    0.123    &    0.094  &&    0.197 &    0.126    &    0.091  &&    0.266 &    0.161    &    0.112  \\
\quad QbSD-gAS                              &&    0.154 &    0.115    &    0.093      &&    0.186 &    0.134    &    0.105  &&    0.212 &    0.137    &    0.102  &&    0.274 &    0.170    &    0.119  \\
AL approach                                 &&          &             &               &&          &             &           &&          &             &           &&          &             &          \\
\quad $\text{AL}_{\text{Mult.}}$-SAV        &&    0.221 &    0.134    &    0.100      &&    0.311 &    0.187    &    0.133  &&    0.332 &    0.184    &    0.118  &&    0.524 &    0.237    &    0.168  \\
\quad $\text{AL}_{\text{AR}}$-SAV           &&    0.225 &    0.136    &    0.102      &&    0.313 &    0.186    &    0.140  &&    0.326 &    0.185    &    0.119  &&    0.516 &    0.267    &    0.173  \\
\quad $\text{AL}_{\text{Mult.}}$-AS         &&    0.259 &    0.153    &    0.114      &&    0.349 &    0.212    &    0.153  &&    0.396 &    0.209    &    0.133  &&    0.539 &    0.300    &    0.188  \\
\quad $\text{AL}_{\text{AR}}$-AS            &&    0.251 &    0.156    &    0.119      &&    0.359 &    0.219    &    0.160  &&    0.409 &    0.211    &    0.140  &&    0.546 &    0.318    &    0.197  \\
GAS                                         &&    0.380 &    0.270    &    0.214      &&    0.438 &    0.313    &    0.216  &&    0.493 &    0.320    &    0.206  &&    0.609 &    0.366    &    0.261  \\
GARCH                                       &&          &             &               &&          &             &           &&          &             &           &&          &             &          \\
\quad Normal                                &&    0.106 &    0.074    &    0.060      &&    0.493 &    0.317    &    0.194  &&    0.270 &    0.124    &    0.123  &&    0.802 &    0.400    &    0.193  \\
\quad Student-$t$                           &&\bf 0.095 &\bf 0.073    &\bf 0.059      &&    0.410 &    0.298    &    0.207  &&\bf 0.143 &\bf 0.097    &\bf 0.072  &&    0.625 &    0.386    &    0.231  \\
\quad Skew-$t$                              &&    0.106 &    0.080    &    0.063      &&    0.132 &    0.093    &    0.070  &&    0.153 &    0.102    &    0.075  &&    0.224 &    0.139    &    0.095  \\
GJR-GARCH                                   &&          &             &               &&          &             &           &&          &             &           &&          &             &          \\
\quad Normal                                &&    0.106 &    0.074    &    0.060      &&    0.493 &    0.317    &    0.194  &&    0.270 &    0.125    &    0.123  &&    0.801 &    0.399    &    0.193  \\
\quad Student-$t$                           &&\bf 0.095 &\bf 0.073    &\bf 0.059      &&    0.410 &    0.298    &    0.207  &&\bf 0.143 &\bf 0.097    &\bf 0.072  &&    0.625 &    0.386    &    0.231  \\
\quad Skew-$t$                              &&    0.106 &\bf 0.080    &    0.063      &&\bf 0.132 &\bf 0.093    &\bf 0.069  &&    0.153 &    0.103    &    0.075  &&\bf 0.213 &\bf 0.134    &\bf 0.093  \\
EGARCH                                      &&    0.121 &    0.087    &    0.070      &&    0.502 &    0.325    &    0.201  &&    0.292 &    0.144    &    0.131  &&    0.814 &    0.407    &    0.191  \\

\bottomrule

\end{tabular*}
\\[1.0ex]
\textit{Notes:} This table reports the MAE in Panel A and the RMSE in Panel B of various VaR forecasting models. The results are based on  1,000 replications for each configuration of the APARCH model in (\ref{DGP}) with innovations following the skewed $t$-distribution in (\ref{skt}).
There are no leverage effects ($ \theta=0$). Bolded values indicate the model with the lowest error.
\end{minipage}
\end{center}
\end{table}

\end{landscape}

\begin{landscape}

\bigskip
\bigskip

\begin{table}[h]
\begin{center}
\begin{minipage}{8.8in}
\vspace{-1.4cm}
{\small {\bf Table 2.} ES forecast accuracy when there are no leverage effects}
\footnotesize
\\[0.5ex]
\begin{tabular*}{\textwidth}{@{\extracolsep{\fill}}  lcrrrcrrrcrrrcrrr}
\toprule

                        &&   \multicolumn{3}{c}{$v=20$, $\lambda=0$}   && \multicolumn{3}{c}{$v=20$, $\lambda=-0.5$}   && \multicolumn{3}{c}{$v=5$, $\lambda=0$} && \multicolumn{3}{c}{$v=5$, $\lambda=-0.5$}    \\

                        \cline{3-5} \cline{7-9} \cline{11-13} \cline{15-17} \\[-2.0ex]
  
                              &$\alpha=$     & 0.01  &   0.025    & 0.05       && 0.01  &   0.025    & 0.05    && 0.01  &   0.025    & 0.05   && 0.01  &   0.025    & 0.05  \\

\midrule

\multicolumn{17}{l}{Panel A: Mean absolute error} \\

QbSD approach                               &&          &             &               &&          &             &           &&          &             &           &&          &             &          \\
\quad QbSD-gSAV                             &&    0.143 &    0.106    &    0.085      &&    0.189 &    0.134    &    0.107  &&    0.265 &    0.164    &    0.118  &&    0.370 &    0.222    &    0.156  \\
\quad QbSD-gAS                              &&    0.154 &    0.117    &    0.096      &&    0.196 &    0.142    &    0.114  &&    0.275 &    0.171    &    0.126  &&    0.376 &    0.228    &    0.161  \\
AL approach                                 &&          &             &               &&          &             &           &&          &             &           &&          &             &          \\
\quad $\text{AL}_{\text{Mult.}}$-SAV        &&    0.188 &    0.125    &    0.092      &&    0.271 &    0.173    &    0.126  &&    0.339 &    0.196    &    0.126  &&    0.527 &    0.281    &    0.182  \\
\quad $\text{AL}_{\text{AR}}$-SAV           &&    0.210 &    0.142    &    0.108      &&    0.292 &    0.196    &    0.142  &&    0.373 &    0.219    &    0.147  &&    0.551 &    0.315    &    0.211  \\
\quad $\text{AL}_{\text{Mult.}}$-AS         &&    0.228 &    0.146    &    0.108      &&    0.317 &    0.206    &    0.148  &&    0.392 &    0.225    &    0.145  &&    0.571 &    0.319    &    0.210  \\
\quad $\text{AL}_{\text{AR}}$-AS            &&    0.233 &    0.155    &    0.116      &&    0.327 &    0.222    &    0.158  &&    0.413 &    0.240    &    0.159  &&    0.583 &    0.345    &    0.227  \\
GAS                                         &&    0.341 &    0.258    &    0.203      &&    0.416 &    0.307    &    0.220  &&    0.474 &    0.323    &    0.221  &&    0.628 &    0.393    &    0.256  \\
GARCH                                       &&          &             &               &&          &             &           &&          &             &           &&          &             &          \\
\quad Normal                                &&    0.146 &    0.094    &    0.067      &&    0.689 &    0.501    &    0.367  &&    0.631 &    0.322    &    0.163  &&    1.463 &    0.898    &    0.563  \\
\quad Student-$t$                           &&\bf 0.100 &\bf 0.075    &\bf 0.060      &&    0.489 &    0.394    &    0.315  &&\bf 0.180 &\bf 0.118    &\bf 0.086  &&    0.945 &    0.656    &    0.470  \\
\quad Skew-$t$                              &&    0.110 &    0.083    &    0.067      &&    0.147 &    0.106    &\bf 0.080  &&    0.190 &    0.126    &    0.091  &&    0.122 &    0.179    &    0.122  \\
GJR-GARCH                                   &&          &             &               &&          &             &           &&          &             &           &&          &             &          \\
\quad Normal                                &&    0.146 &    0.095    &    0.067      &&    0.689 &    0.501    &    0.367  &&    0.631 &    0.322    &    0.163  &&    1.463 &    0.898    &    0.563  \\
\quad Student-$t$                           &&\bf 0.100 &\bf 0.075    &\bf 0.060      &&    0.489 &    0.394    &    0.315  &&\bf 0.180 &\bf 0.118    &\bf 0.086  &&    0.945 &    0.656    &    0.470  \\
\quad Skew-$t$                              &&    0.110 &    0.083    &    0.067      &&\bf 0.146 &\bf 0.105    &\bf 0.080  &&    0.190 &    0.127    &    0.092  &&\bf 0.117 &\bf 0.173    &\bf 0.117  \\
EGARCH                                      &&    0.150 &    0.100    &    0.075      &&    0.693 &    0.504    &    0.370  &&    0.637 &    0.327    &    0.169  &&    1.470 &    0.904    &    0.568  \\

\midrule                         
                         
\multicolumn{17}{l}{Panel B: Root mean squared error} \\

QbSD approach                               &&          &             &               &&          &             &           &&          &             &           &&          &             &          \\
\quad QbSD-gSAV                             &&    0.185 &    0.138    &    0.112      &&    0.240 &    0.172    &    0.138  &&    0.340 &    0.215    &    0.157  &&    0.463 &    0.285    &    0.205  \\
\quad QbSD-gAS                              &&    0.201 &    0.153    &    0.126      &&    0.254 &    0.185    &    0.149  &&    0.352 &    0.227    &    0.168  &&    0.473 &    0.294    &    0.211  \\
AL approach                                 &&          &             &               &&          &             &           &&          &             &           &&          &             &          \\
\quad $\text{AL}_{\text{Mult.}}$-SAV        &&    0.265 &    0.169    &    0.126      &&    0.382 &    0.237    &    0.173  &&    0.482 &    0.277    &    0.179  &&    0.770 &    0.401    &    0.262  \\
\quad $\text{AL}_{\text{AR}}$-SAV           &&    0.291 &    0.197    &    0.147      &&    0.408 &    0.267    &    0.195  &&    0.521 &    0.309    &    0.209  &&    0.763 &    0.444    &    0.304  \\
\quad $\text{AL}_{\text{Mult.}}$-AS         &&    0.313 &    0.193    &    0.143      &&    0.429 &    0.276    &    0.200  &&    0.536 &    0.306    &    0.198  &&    0.760 &    0.439    &    0.289  \\
\quad $\text{AL}_{\text{AR}}$-AS            &&    0.312 &    0.210    &    0.155      &&    0.439 &    0.300    &    0.219  &&    0.555 &    0.330    &    0.221  &&    0.792 &    0.480    &    0.319  \\
GAS                                         &&    0.452 &    0.345    &    0.275      &&    0.536 &    0.405    &    0.288  &&    0.716 &    0.459    &    0.304  &&    0.921 &    0.558    &    0.395  \\
GARCH                                       &&          &             &               &&          &             &           &&          &             &           &&          &             &          \\
\quad Normal                                &&    0.168 &    0.114    &    0.085      &&    0.707 &    0.517    &    0.382  &&    0.658 &    0.349    &    0.192  &&    1.507 &    0.934    &    0.596  \\
\quad Student-$t$                           &&\bf 0.129 &\bf 0.099    &\bf 0.081      &&    0.515 &    0.414    &    0.330  &&\bf 0.241 &\bf 0.162    &\bf 0.120  &&    0.982 &    0.682    &    0.489  \\
\quad Skew-$t$                              &&    0.145 &    0.111    &    0.090      &&    0.192 &\bf 0.140    &\bf 0.108  &&    0.255 &    0.172    &    0.127  &&    0.183 &    0.262    &    0.411  \\
GJR-GARCH                                   &&          &             &               &&          &             &           &&          &             &           &&          &             &          \\
\quad Normal                                &&    0.168 &    0.115    &    0.085      &&    0.707 &    0.517    &    0.382  &&    0.658 &    0.349    &    0.192  &&    1.506 &    0.933    &    0.595  \\
\quad Student-$t$                           &&\bf 0.129 &\bf 0.099    &\bf 0.081      &&    0.515 &    0.414    &    0.330  &&\bf 0.241 &\bf 0.162    &\bf 0.120  &&    0.982 &    0.682    &    0.489  \\
\quad Skew-$t$                              &&    0.145 &\bf 0.110    &\bf 0.089      &&\bf 0.191 &\bf 0.140    &\bf 0.108  &&    0.256 &    0.173    &    0.127  &&\bf 0.174 &\bf 0.247    &\bf 0.383  \\
EGARCH                                      &&    0.184 &    0.130    &    0.099      &&    0.717 &    0.526    &    0.390  &&    0.678 &    0.369    &    0.213  &&    1.524 &    0.948    &    0.606  \\

\bottomrule

\end{tabular*}
\\[1.0ex]
\textit{Notes:}  This table reports the MAE in Panel A and the RMSE in Panel B for various ES forecasting models. See Table 1 for a description of the simulation setup. 
The results are based on simulations without leverage effects ($\theta = 0$). Bolded values indicate the model with the lowest error.
\end{minipage}
\end{center}
\end{table}

\end{landscape}

\begin{landscape}

\bigskip
\bigskip

\begin{table}[h]
\begin{center}
\begin{minipage}{8.8in}
\vspace{-1.4cm}
{\small {\bf Table 3.} VaR forecast accuracy  in the presence of  leverage effects}
\footnotesize
\\[0.5ex]
\begin{tabular*}{\textwidth}{@{\extracolsep{\fill}}  lcrrrcrrrcrrrcrrr}
\toprule

                        &&   \multicolumn{3}{c}{$v=20$, $\lambda=0$}   && \multicolumn{3}{c}{$v=20$, $\lambda=-0.5$}   && \multicolumn{3}{c}{$v=5$, $\lambda=0$} && \multicolumn{3}{c}{$v=5$, $\lambda=-0.5$}    \\

                        \cline{3-5} \cline{7-9} \cline{11-13} \cline{15-17} \\[-2.0ex]
  
                                   &$\alpha=$ & 0.01  &   0.025    & 0.05       && 0.01  &   0.025    & 0.05    && 0.01  &   0.025    & 0.05   && 0.01  &   0.025    & 0.05  \\

\midrule

\multicolumn{17}{l}{Panel A: Mean absolute error} \\

QbSD approach                               &&          &        &        &&        &        &        &&        &        &        &&        &        &        \\
\quad QbSD-gSAV                             &&    0.214 &    0.170 &    0.139 &&    0.271 &    0.212 &    0.166 &&    0.240 &    0.170 &    0.130 &&    0.324 &    0.220 &    0.159 \\
\quad QbSD-gAS                              &&    0.124 &    0.093 &    0.075 &&\bf 0.172 &\bf 0.125 &\bf 0.096 &&\bf 0.170 &    0.110 &\bf 0.080 &&    0.255 &    0.159 &    0.109 \\
AL approach                                 &&          &        &        &&        &        &        &&        &        &        &&        &        &        \\
\quad $\text{AL}_{\text{Mult.}}$-SAV        &&    0.260 &    0.187 &    0.148 &&    0.347 &    0.242 &    0.184 &&    0.326 &    0.200 &    0.144 &&    0.473 &    0.273 &    0.188 \\
\quad $\text{AL}_{\text{AR}}$-SAV           &&    0.261 &    0.189 &    0.148 &&    0.351 &    0.244 &    0.187 &&    0.330 &    0.202 &    0.144 &&    0.480 &    0.277 &    0.199 \\
\quad $\text{AL}_{\text{Mult.}}$-AS         &&    0.198 &    0.127 &    0.093 &&    0.299 &    0.184 &    0.135 &&    0.312 &    0.164 &    0.106 &&    0.482 &    0.254 &    0.163 \\
\quad $\text{AL}_{\text{AR}}$-AS            &&    0.204 &    0.131 &    0.098 &&    0.305 &    0.193 &    0.142 &&    0.304 &    0.169 &    0.109 &&    0.480 &    0.267 &    0.176 \\
GAS                                         &&    0.391 &    0.264 &    0.174 &&    0.605 &    0.386 &    0.247 &&    0.447 &    0.254 &    0.164 &&    0.751 &    0.397 &    0.239 \\
GARCH                                       &&          &        &        &&        &        &        &&        &        &        &&        &        &        \\
\quad Normal                                &&    0.202 &    0.163 &    0.135 &&    0.539 &    0.346 &    0.215 &&    0.299 &    0.172 &    0.149 &&    0.879 &    0.413 &    0.180 \\
\quad Student-$t$                           &&    0.200 &    0.162 &    0.134 &&    0.440 &    0.326 &    0.231 &&    0.221 &    0.162 &    0.126 &&    0.682 &    0.420 &    0.249 \\
\quad Skew-$t$                              &&    0.205 &    0.165 &    0.135 &&    0.211 &    0.160 &    0.124 &&    0.226 &    0.166 &    0.128 &&    0.239 &    0.155 &    0.109 \\
GJR-GARCH                                   &&          &        &        &&        &        &        &&        &        &        &&        &        &        \\
\quad Normal                                &&    0.194 &    0.156 &    0.128 &&    0.538 &    0.345 &    0.214 &&    0.295 &    0.164 &    0.142 &&    0.879 &    0.413 &    0.180 \\
\quad Student-$t$                           &&    0.200 &    0.162 &    0.134 &&    0.440 &    0.326 &    0.231 &&    0.221 &    0.162 &    0.126 &&    0.682 &    0.420 &    0.249 \\
\quad Skew-$t$                              &&    0.205 &    0.165 &    0.135 &&    0.211 &    0.160 &    0.124 &&    0.227 &    0.166 &    0.128 &&\bf 0.233 &\bf 0.152 &\bf 0.108 \\
EGARCH                                      &&\bf 0.105 &\bf 0.075 &\bf 0.062 &&    0.558 &    0.356 &    0.213 &&    0.262 &\bf 0.105 &    0.099 &&    0.901 &    0.430 &    0.179 \\

\midrule                         
                         
\multicolumn{17}{l}{Panel B: Root mean squared error} \\

QbSD approach                               &&          &        &        &&        &        &        &&        &        &        &&        &        &        \\
\quad QbSD-gSAV                             &&    0.283 &    0.225 &    0.185 &&    0.363 &    0.283 &    0.224 &&    0.330 &    0.235 &    0.180 &&    0.451 &    0.326 &    0.230 \\
\quad QbSD-gAS                              &&    0.166 &    0.126 &    0.101 &&\bf 0.238 &\bf 0.183 &\bf 0.138 &&\bf 0.250 &    0.178 &\bf 0.130 &&    0.437 &    0.333 &    0.222 \\
AL approach                                 &&          &        &        &&        &        &        &&        &        &        &&        &        &        \\
\quad $\text{AL}_{\text{Mult.}}$-SAV        &&    0.346 &    0.252 &    0.198 &&    0.479 &    0.335 &    0.254 &&    0.465 &    0.282 &    0.204 &&    0.753 &    0.412 &    0.285 \\
\quad $\text{AL}_{\text{AR}}$-SAV           &&    0.361 &    0.255 &    0.197 &&    0.510 &    0.341 &    0.256 &&    0.463 &    0.287 &    0.205 &&    0.722 &    0.409 &    0.292 \\
\quad $\text{AL}_{\text{Mult.}}$-AS         &&    0.272 &    0.175 &    0.129 &&    0.419 &    0.267 &    0.199 &&    0.460 &    0.235 &    0.161 &&    0.738 &    0.397 &    0.272 \\
\quad $\text{AL}_{\text{AR}}$-AS            &&    0.303 &    0.181 &    0.136 &&    0.433 &    0.283 &    0.209 &&    0.436 &    0.244 &    0.162 &&    0.725 &    0.408 &    0.284 \\
GAS                                         &&    0.503 &    0.345 &    0.227 &&    0.800 &    0.547 &    0.356 &&    0.684 &    0.350 &    0.218 &&    1.307 &    0.627 &    0.397 \\
GARCH                                       &&          &        &        &&        &        &        &&        &        &        &&        &        &        \\
\quad Normal                                &&    0.264 &    0.215 &    0.180 &&    0.622 &    0.411 &    0.265 &&    0.377 &    0.239 &    0.213 &&    1.027 &    0.500 &    0.231 \\
\quad Student-$t$                           &&    0.268 &    0.216 &    0.177 &&    0.522 &    0.390 &    0.281 &&    0.312 &    0.229 &    0.177 &&    0.804 &    0.500 &    0.303 \\
\quad Skew-$t$                              &&    0.273 &    0.220 &    0.179 &&    0.278 &    0.211 &    0.164 &&    0.320 &    0.234 &    0.179 &&\bf 0.326 &\bf 0.212 &\bf 0.150 \\
GJR-GARCH                                   &&          &        &        &&        &        &        &&        &        &        &&        &        &        \\
\quad Normal                                &&    0.255 &    0.207 &    0.173 &&    0.620 &    0.409 &    0.263 &&    0.365 &    0.231 &    0.209 &&    1.027 &    0.500 &    0.231 \\
\quad Student-$t$                           &&    0.268 &    0.216 &    0.177 &&    0.522 &    0.390 &    0.281 &&    0.312 &    0.229 &    0.177 &&    0.804 &    0.500 &    0.303 \\
\quad Skew-$t$                              &&    0.274 &    0.220 &    0.179 &&    0.278 &    0.211 &    0.164 &&    0.319 &    0.234 &    0.179 &&    0.317 &    0.208 &    0.147 \\
EGARCH                                      &&\bf 0.148 &\bf 0.107 &\bf 0.086 &&    0.647 &    0.428 &    0.273 &&    0.334 &\bf 0.160 &    0.136 &&    1.132 &    0.584 &    0.287 \\

\bottomrule

\end{tabular*}
\\[1.0ex]
\textit{Notes:}  This table reports the MAE in Panel A and the RMSE in Panel B for various VaR forecasting models. See Table 1 for a description of the simulation setup. The results are based on simulations with leverage effects ($\theta = 0.5$).
Bolded values indicate the model with the lowest error.
\end{minipage}
\end{center}
\end{table}

\end{landscape}

\begin{landscape}

\bigskip
\bigskip

\begin{table}[h]
\begin{center}
\begin{minipage}{8.8in}
\vspace{-1.4cm}
{\small {\bf Table 4.} ES  forecast accuracy in the presence of  leverage effects}
\footnotesize
\\[0.5ex]
\begin{tabular*}{\textwidth}{@{\extracolsep{\fill}}  lcrrrcrrrcrrrcrrr}
\toprule

                        &&   \multicolumn{3}{c}{$v=20$, $\lambda=0$}   && \multicolumn{3}{c}{$v=20$, $\lambda=-0.5$}   && \multicolumn{3}{c}{$v=5$, $\lambda=0$} && \multicolumn{3}{c}{$v=5$, $\lambda=-0.5$}    \\

                        \cline{3-5} \cline{7-9} \cline{11-13} \cline{15-17} \\[-2.0ex]
  
                              &$\alpha=$     & 0.01  &   0.025    & 0.05       && 0.01  &   0.025    & 0.05    && 0.01  &   0.025    & 0.05   && 0.01  &   0.025    & 0.05  \\

\midrule

\multicolumn{17}{l}{Panel A: Mean absolute error} \\

QbSD approach                               &&          &             &               &&          &             &           &&          &             &           &&          &             &          \\
\quad QbSD-gSAV                             &&    0.253 &    0.208    &    0.179      &&    0.335 &    0.268    &    0.223  &&    0.354 &    0.248    &    0.193  &&    0.510 &    0.342    &    0.257  \\
\quad QbSD-gAS                              &&\bf 0.165 &    0.125    &    0.103      &&\bf 0.235 &\bf 0.174    &\bf 0.140  &&\bf 0.292 &\bf 0.186    &\bf 0.136  &&    0.445 &    0.279    &    0.198  \\
AL approach                                 &&          &             &               &&          &             &           &&          &             &           &&          &             &          \\
\quad $\text{AL}_{\text{Mult.}}$-SAV        &&    0.304 &    0.229    &    0.189      &&    0.414 &    0.303    &    0.244  &&    0.450 &    0.283    &    0.208  &&    0.680 &    0.403    &    0.287  \\
\quad $\text{AL}_{\text{AR}}$-SAV           &&    0.329 &    0.253    &    0.201      &&    0.449 &    0.335    &    0.266  &&    0.488 &    0.323    &    0.225  &&    0.733 &    0.457    &    0.341  \\
\quad $\text{AL}_{\text{Mult.}}$-AS         &&    0.245 &    0.158    &    0.116      &&    0.369 &    0.234    &    0.179  &&    0.442 &    0.241    &    0.157  &&    0.697 &    0.381    &    0.256  \\
\quad $\text{AL}_{\text{AR}}$-AS            &&    0.262 &    0.180    &    0.138      &&    0.403 &    0.275    &    0.217  &&    0.454 &    0.267    &    0.181  &&    0.752 &    0.442    &    0.316  \\
GAS                                         &&    0.475 &    0.336    &    0.232      &&    0.730 &    0.522    &    0.329  &&    0.598 &    0.369    &    0.235  &&    1.062 &    0.619    &    0.365  \\
GARCH                                       &&          &             &               &&          &             &           &&          &             &           &&          &             &          \\
\quad Normal                                &&    0.255 &    0.207    &    0.175      &&    0.780 &    0.566    &    0.416  &&    0.663 &    0.367    &    0.228  &&    1.687 &    1.033    &    0.644  \\
\quad Student-$t$                           &&    0.249 &    0.205    &    0.175      &&    0.539 &    0.441    &    0.357  &&    0.325 &    0.239    &    0.188  &&    1.063 &    0.743    &    0.532  \\
\quad Skew-$t$                              &&    0.253 &    0.210    &    0.178      &&    0.279 &    0.218    &    0.179  &&    0.335 &    0.245    &    0.192  &&    0.405 &    0.272    &    0.197  \\
GJR-GARCH                                   &&          &             &               &&          &             &           &&          &             &           &&          &             &          \\
\quad Normal                                &&    0.247 &    0.199    &    0.168      &&    0.779 &    0.565    &    0.415  &&    0.661 &    0.364    &    0.223  &&    1.687 &    1.033    &    0.644  \\
\quad Student-$t$                           &&    0.249 &    0.205    &    0.175      &&    0.539 &    0.441    &    0.357  &&    0.325 &    0.239    &    0.188  &&    1.063 &    0.743    &    0.532  \\
\quad Skew-$t$                              &&    0.254 &    0.210    &    0.178      &&    0.279 &    0.218    &    0.178  &&    0.334 &    0.245    &    0.193  &&\bf 0.388 &\bf 0.263    &\bf 0.191  \\
EGARCH                                      &&    0.167 &\bf 0.113    &\bf 0.086      &&    0.804 &    0.586    &    0.430  &&    0.682 &    0.349    &    0.180  &&    1.711 &    1.055    &    0.664  \\

\midrule                         
                         
\multicolumn{17}{l}{Panel B: Root mean squared error} \\

QbSD approach                               &&          &             &               &&          &             &           &&          &             &           &&          &             &          \\
\quad QbSD-gSAV                             &&    0.338 &    0.279    &    0.239      &&    0.450 &    0.359    &    0.299  &&    0.473 &    0.339    &    0.266  &&    0.675 &    0.464    &    0.356  \\
\quad QbSD-gAS                              &&\bf 0.215 &    0.165    &    0.137      &&\bf 0.314 &\bf 0.238    &\bf 0.195  &&\bf 0.392 &\bf 0.266    &\bf 0.204  &&    0.672 &    0.465    &    0.357  \\
AL approach                                 &&          &             &               &&          &             &           &&          &             &           &&          &             &          \\
\quad $\text{AL}_{\text{Mult.}}$-SAV        &&    0.404 &    0.312    &    0.252      &&    0.580 &    0.422    &    0.335  &&    0.641 &    0.403    &    0.294  &&    1.071 &    0.590    &    0.423  \\
\quad $\text{AL}_{\text{AR}}$-SAV           &&    0.443 &    0.342    &    0.270      &&    0.664 &    0.474    &    0.391  &&    0.682 &    0.447    &    0.334  &&    1.110 &    0.684    &    0.587  \\
\quad $\text{AL}_{\text{Mult.}}$-AS         &&    0.328 &    0.216    &    0.162      &&    0.513 &    0.336    &    0.262  &&    0.619 &    0.333    &    0.232  &&    1.015 &    0.563    &    0.404  \\
\quad $\text{AL}_{\text{AR}}$-AS            &&    0.376 &    0.250    &    0.192      &&    0.571 &    0.404    &    0.335  &&    0.611 &    0.380    &    0.260  &&    1.183 &    0.675    &    0.602  \\
GAS                                         &&    0.623 &    0.438    &    0.304      &&    0.991 &    0.724    &    0.467  &&    0.936 &    0.529    &    0.317  &&    1.809 &    0.950    &    0.578  \\
GARCH                                       &&          &             &               &&          &             &           &&          &             &           &&          &             &          \\
\quad Normal                                &&    0.330 &    0.270    &    0.230      &&    0.880 &    0.650    &    0.487  &&    0.772 &    0.449    &    0.295  &&    1.955 &    1.202    &    0.759  \\
\quad Student-$t$                           &&    0.331 &    0.274    &    0.233      &&    0.643 &    0.525    &    0.427  &&    0.456 &    0.337    &    0.266  &&    1.251 &    0.874    &    0.630  \\
\quad Skew-$t$                              &&    0.340 &    0.280    &    0.237      &&    0.367 &    0.288    &    0.235  &&    0.468 &    0.345    &    0.272  &&    0.562 &    0.372    &    0.269  \\
GJR-GARCH                                   &&          &             &               &&          &             &           &&          &             &           &&          &             &          \\
\quad Normal                                &&    0.319 &    0.260    &    0.222      &&    0.878 &    0.648    &    0.486  &&    0.759 &    0.437    &    0.285  &&    1.954 &    1.202    &    0.758  \\
\quad Student-$t$                           &&    0.331 &    0.274    &    0.233      &&    0.643 &    0.525    &    0.427  &&    0.456 &    0.337    &    0.266  &&    1.251 &    0.874    &    0.630  \\
\quad Skew-$t$                              &&    0.338 &    0.279    &    0.237      &&    0.367 &    0.288    &    0.261  &&    0.469 &    0.346    &    0.271  &&\bf 0.526 &\bf 0.352    &\bf 0.256  \\
EGARCH                                      &&    0.219 &\bf 0.157    &\bf 0.122      &&    0.912 &    0.676    &    0.507  &&    0.765 &    0.421    &    0.244  &&    2.077 &    1.309    &    0.851  \\

\bottomrule

\end{tabular*}
\\[1.0ex]
\textit{Notes:} This table reports the MAE in Panel A and the RMSE in Panel B for various ES forecasting models. See Table 1 for a description of the simulation setup. The results are based on simulations with leverage effects ($\theta = 0.5$).
Bolded values indicate the model with the lowest error.
\end{minipage}
\end{center}
\end{table}

\end{landscape}

\newpage
\clearpage

\begin{table}[!h]
\hspace*{-0.65cm}
\begin{minipage}{7.1in}
{\small {\bf Table 5.} Ranking of 1\% VaR forecasting models based on the MCS procedure using quantile scores}
\begin{center}
\footnotesize
\vspace{-0.25cm}
\begin{tabular*}{\textwidth}{@{\extracolsep{\fill}} lcccccccc|ccc}
\toprule
                     & \multirow{2}{*}{S\&P} & \multirow{2}{*}{DJIA} & \multirow{2}{*}{NASDAQ} & \multirow{2}{*}{STOXX} & \multirow{2}{*}{FTSE} & \multirow{2}{*}{DAX} & \multirow{2}{*}{CAC} & \multirow{2}{*}{TSX} & \multirow{2}{*}{\#} & Avg. & Final \\
                     &     &     &     &     &     &     &     &     &     & rank  & rank  \\
\midrule 
QbSD-gAS                         & 16 &  1 &  3 &  7 &  2 &  1 &  1 &  1 &  8 &  4.0 &  1 \\
GJR-GARCH-skew-$t$               & 15 &  6 & 12 &  4 &  7 &  8 &  3 &  4 &  8 &  7.4 &  2 \\
GARCH-skew-$t$                   & 20 &  3 &  2 &  2 &  9 &  6 & 16 &  6 &  8 &  8.0 &  3 \\
GJR-GARCH-$t$                    &  7 &  7 & 11 &  1 & 20 &  2 &  2 & 15 &  8 &  8.1 &  4 \\
AR-GJR-GARCH-skew-$t$            &  1 &  5 &  7 &  3 & 10 & 18 & 12 &  9 &  8 &  8.1 &  5 \\
AR-GARCH-skew-$t$                & 13 &  9 &  1 &  8 & 11 &  4 & 15 &  8 &  8 &  8.6 &  6 \\
AR-GJR-GARCH-$t$                 &  2 &  2 & 10 &  6 & 19 &  7 & 11 & 21 &  8 &  9.8 &  7 \\
AR-AL$_{\text{Mult.}}$-AS        &  4 & 10 & 16 & 20 &  4 & 12 & 14 &  2 &  8 & 10.2 &  8 \\
QAR-QbSD-gAS                     &  6 &  4 & 22 &  5 &  1 & 14 & 28 &  3 &  8 & 10.4 &  9 \\
AL$_{\text{Mult.}}$-AS           &  8 & 17 & 14 & 21 &  3 & 13 &  4 &  5 &  8 & 10.6 & 10 \\
AR-AL$_{\text{AR}}$-AS           & 18 &  8 & 13 & 14 &  5 & 11 &  9 & 11 &  8 & 11.1 & 11 \\
AL$_{\text{AR}}$-AS              & 14 & 18 & 15 & 19 &  6 & 10 &  5 &  7 &  8 & 11.8 & 12 \\
GARCH-$t$                        &  5 & 13 &  8 & 12 & 22 &  3 & 18 & 14 &  8 & 11.9 & 13 \\
AR-GARCH-$t$                     &  3 & 11 &  9 &  9 & 21 & 20 & 13 & 22 &  8 & 13.5 & 14 \\
AL$_{\text{Mult.}}$-SAV          & 12 & 19 &  5 & 16 & 16 & 19 &  6 & 20 &  8 & 14.1 & 15 \\
QAR-QbSD-gSAV                    & 27 & 21 & 25 & 10 &  8 &  5 &  7 & 10 &  8 & 14.1 & 16 \\
AR-AL$_{\text{Mult.}}$-SAV       &  9 & 14 &  6 & 17 & 18 & 22 & 10 & 19 &  8 & 14.4 & 17 \\
EGARCH                           & 21 & 15 & 18 & 11 & 14 & 16 & 19 & 12 &  8 & 15.8 & 18 \\
AL$_{\text{AR}}$-SAV             & 22 & 24 &  4 & 18 & 13 & 17 & 17 & 17 &  8 & 16.5 & 19 \\
AR-AL$_{\text{AR}}$-SAV          & 10 & 12 & 17 & 24 & 15 & 21 & 20 & 18 &  8 & 17.1 & 20 \\
QbSD-gSAV                        &    & 28 & 26 & 15 & 12 &  9 &  8 & 13 &  7 & 17.4 & 21 \\
AR-EGARCH                        & 19 & 16 & 19 & 13 & 17 & 26 & 21 & 16 &  8 & 18.4 & 22 \\
GAS                              & 11 & 22 & 20 & 27 & 23 & 27 & 24 & 23 &  8 & 22.1 & 23 \\
AR-GAS                           & 17 & 20 & 21 &    & 24 &    & 25 & 24 &  6 & 23.4 & 24 \\
GARCH-normal                     & 24 & 23 & 24 & 23 & 25 & 23 & 23 &    &  7 & 24.1 & 25 \\
GJR-GARCH-normal                 & 23 & 25 & 23 & 22 &    & 24 & 22 &    &  6 & 24.4 & 26 \\
AR-GJR-GARCH-normal              & 25 & 26 & 28 & 25 &    & 15 & 26 &    &  6 & 25.1 & 27 \\
AR-GARCH-normal                  & 26 & 27 & 27 & 26 &    & 25 & 27 &    &  6 & 26.8 & 28 \\
\bottomrule

\end{tabular*}
\vspace{-0.5cm}
\end{center}
{\footnotesize \textit{Notes:} A model that is excluded from $\widehat{\mathcal{M}}_{90\%}^{\ast}$ for a given stock index is left blank in the table. The remaining models in 
$\widehat{\mathcal{M}}_{90\%}^{\ast}$ are ranked based on their quantile score $t$-statistics, with lower values indicating better performance. The column ``\#'' represents the number of stock indices (out of 8) for which the model is included in $\widehat{\mathcal{M}}_{90\%}^{\ast}$. The ``Avg. rank'' column reports the average rank across all stock indices, where models eliminated from $\widehat{\mathcal{M}}_{90\%}^{\ast}$ were assigned a rank of 28. The ``Final rank'' column orders models from best to worst based on their average rank, summarizing their relative forecasting performance across stock indices. The rolling-window size is $R=1250$.}
\end{minipage}
\end{table}

\begin{table}[!h]
\hspace*{-0.65cm}
\begin{minipage}{7.1in}
{\small {\bf Table 6.} Ranking of 2.5\% VaR forecasting models based on the MCS procedure using quantile scores}
\begin{center}
\footnotesize
\vspace{-0.25cm}
\begin{tabular*}{\textwidth}{@{\extracolsep{\fill}} lcccccccc|ccc}
\toprule
                     & \multirow{2}{*}{S\&P} & \multirow{2}{*}{DJIA} & \multirow{2}{*}{NASDAQ} & \multirow{2}{*}{STOXX} & \multirow{2}{*}{FTSE} & \multirow{2}{*}{DAX} & \multirow{2}{*}{CAC} & \multirow{2}{*}{TSX} & \multirow{2}{*}{\#} & Avg. & Final \\
                     &     &     &     &     &     &     &     &     &     & rank  & rank  \\
\midrule 
QbSD-gAS                         &  2 &  1 &  1 & 16 &  1 &  1 &  1 &  3 &  8 &  3.2 &  1 \\
AL$_{\text{Mult.}}$-AS           &  9 &  3 &  2 &  9 &  2 &  2 &  2 &  2 &  8 &  3.9 &  2 \\
AR-AL$_{\text{Mult.}}$-AS        &  6 &  4 &  6 &  8 &  3 &  3 &  3 &  1 &  8 &  4.2 &  3 \\
AL$_{\text{AR}}$-AS              &  7 &  5 &  4 &  4 &  8 &  4 &  8 &  5 &  8 &  5.6 &  4 \\
GJR-GARCH-skew-$t$               &  4 &  8 &  5 &  6 & 12 &  7 &  7 &  8 &  8 &  7.1 &  5 \\
QAR-QbSD-gAS                     &  1 &  2 & 23 &  2 &  6 &  5 & 15 &  4 &  8 &  7.2 &  6 \\
GARCH-skew-$t$                   &  3 &  6 &  7 &  1 & 10 &  8 & 19 &  7 &  8 &  7.6 &  7 \\
EGARCH                           & 10 &  7 &  8 & 12 &  5 & 11 & 12 &  6 &  8 &  8.9 &  8 \\
AR-AL$_{\text{AR}}$-AS           & 12 & 12 &  3 & 11 &  4 & 10 & 26 & 13 &  8 & 11.4 &  9 \\
GJR-GARCH-$t$                    & 13 &  9 & 10 & 13 & 15 & 16 & 10 & 11 &  8 & 12.1 & 10 \\
AL$_{\text{Mult.}}$-SAV          & 18 & 16 & 17 &  5 & 11 & 20 &  4 &  9 &  8 & 12.5 & 11 \\
AR-GJR-GARCH-skew-$t$            &  5 & 13 & 11 & 19 & 14 &  9 & 17 & 15 &  8 & 12.9 & 12 \\
AR-EGARCH                        & 16 & 11 & 22 & 15 &  7 & 12 & 21 & 10 &  8 & 14.2 & 13 \\
GARCH-$t$                        & 11 & 10 & 12 & 26 & 16 & 17 & 11 & 12 &  8 & 14.4 & 14 \\
AR-GARCH-skew-$t$                &  8 & 19 &  9 & 22 & 17 &  6 & 20 & 14 &  8 & 14.4 & 15 \\
AL$_{\text{AR}}$-SAV             & 25 & 17 & 14 &  7 & 23 & 21 &  5 & 19 &  8 & 16.4 & 16 \\
AR-AL$_{\text{Mult.}}$-SAV       & 14 & 20 & 19 &  3 & 19 & 22 &  6 &    &  7 & 16.4 & 17 \\
GJR-GARCH-normal                 & 21 & 18 & 13 & 14 & 22 & 23 & 13 & 18 &  8 & 17.8 & 18 \\
QAR-QbSD-gSAV                    & 27 & 22 & 25 & 10 & 13 & 18 & 16 & 16 &  8 & 18.4 & 19 \\
GARCH-normal                     & 22 & 15 & 20 & 17 & 18 & 25 & 14 & 17 &  8 & 18.5 & 20 \\
AR-GJR-GARCH-$t$                 & 15 & 14 & 15 & 20 & 20 & 15 & 22 &    &  7 & 18.6 & 21 \\
QbSD-gSAV                        &    & 26 & 18 &  9 & 13 &  9 &    &    &  5 & 19.9 & 22 \\
AR-AL$_{\text{AR}}$-SAV          & 20 & 21 & 18 & 21 & 24 & 19 & 18 &    &  7 & 21.1 & 23 \\
AR-GARCH-$t$                     & 17 & 24 & 16 & 25 & 21 & 26 & 23 &    &  7 & 22.5 & 24 \\
AR-GJR-GARCH-normal              & 19 & 25 & 21 & 24 &    & 14 & 24 &    &  6 & 22.9 & 25 \\
AR-GARCH-normal                  & 23 & 26 & 24 & 23 & 25 & 24 & 25 &    &  7 & 24.8 & 26 \\
GAS                               & 24 & 23 &    &    &    &    &    &    &  2 & 26.9 & 27 \\
AR-GAS                            & 26 & 27 &    &    &    &    &    &    &  2 & 27.6 & 28 \\
\bottomrule
\end{tabular*}
\vspace{-0.5cm}
\end{center}
{\footnotesize \textit{Notes:} For explanations, see notes of Table 5.}
\end{minipage}
\end{table}

\begin{table}[!h]
\hspace*{-0.65cm}
\begin{minipage}{7.1in}
{\small {\bf Table 7.} Ranking of 5\% VaR forecasting models based on the MCS procedure using quantile scores}
\begin{center}
\footnotesize
\vspace{-0.25cm}
\begin{tabular*}{\textwidth}{@{\extracolsep{\fill}} lcccccccc|ccc}
\toprule
                     & \multirow{2}{*}{S\&P} & \multirow{2}{*}{DJIA} & \multirow{2}{*}{NASDAQ} & \multirow{2}{*}{STOXX} & \multirow{2}{*}{FTSE} & \multirow{2}{*}{DAX} & \multirow{2}{*}{CAC} & \multirow{2}{*}{TSX} & \multirow{2}{*}{\#} & Avg. & Final \\
                     &     &     &     &     &     &     &     &     &     & rank  & rank  \\
\midrule 
QbSD-gAS                    &  1 &  1 &  1 & 12 &  1 &  1 &  1 &  3 &  8 &  2.6 &  1 \\
EGARCH                      &  3 &  3 &  2 &  2 &  2 &  4 &  6 &  5 &  8 &  3.4 &  2 \\
AL$_{\text{Mult.}}$-AS      &  4 & 19 &  3 &  4 &  5 &  7 &  3 &  1 &  8 &  5.8 &  3 \\
QAR-QbSD-gAS                &  2 &  2 &  6 & 14 &  4 &  2 & 12 &  4 &  8 &  5.8 &  4 \\
AR-EGARCH                   & 11 &  4 & 14 &  7 &  3 &  5 &  7 & 12 &  8 &  7.9 &  5 \\
AR-AL$_{\text{Mult.}}$-AS   & 18 & 11 & 11 &  9 &  8 &  3 &  4 &  2 &  8 &  8.2 &  6 \\
GJR-GARCH-skew-$t$          & 10 &  8 &  4 &  6 & 12 & 10 & 13 &  8 &  8 &  8.9 &  7 \\
GARCH-skew-$t$              &  9 &  5 &  5 & 10 & 10 & 14 & 15 &  9 &  8 &  9.6 &  8 \\
AL$_{\text{AR}}$-AS         & 20 & 18 & 15 &  1 &  7 &  6 &  5 &  7 &  8 &  9.9 &  9 \\
GJR-GARCH-normal            &  7 & 10 & 10 &  3 & 21 & 11 & 10 & 10 &  8 & 10.2 & 10 \\
AL$_{\text{AR}}$-AS         & 17 & 20 & 16 & 17 &  6 &  8 &  2 &  6 &  8 & 11.5 & 11 \\
GJR-GARCH-$t$               &  6 &  6 & 12 & 22 &  9 & 16 &  8 & 14 &  8 & 11.6 & 12 \\
GARCH-$t$                   &  8 &  7 &  9 & 25 & 11 & 17 &  9 & 13 &  8 & 12.4 & 13 \\
GARCH-normal                & 12 &  9 & 13 & 16 & 16 & 13 & 11 & 11 &  8 & 12.6 & 14 \\
AR-GJR-GARCH-skew-$t$       &  5 & 15 &  7 & 21 & 18 & 18 & 16 &    &  7 & 16.0 & 15 \\
AR-GARCH-skew-$t$           & 13 & 16 &  8 & 24 & 17 & 15 & 22 &    &  7 & 17.9 & 16 \\
AR-GJR-GARCH-$t$            & 15 & 12 & 17 & 23 & 14 & 20 & 18 &    &  7 & 18.4 & 17 \\
AR-GARCH-normal             & 16 & 17 & 18 & 18 & 20 & 12 & 21 &    &  7 & 18.8 & 18 \\
AR-GJR-GARCH-normal         & 19 & 13 & 20 & 20 & 24 &  9 & 17 &    &  7 & 18.8 & 19 \\
AR-GARCH-$t$                & 14 & 14 & 19 & 26 & 15 & 22 & 20 &    &  7 & 19.8 & 20 \\
QbSD-gSAV                   & 25 & 21 & 22 & 19 & 13 & 25 & 14 & 19 &  8 & 19.8 & 21 \\
AL$_{\text{Mult.}}$-SAV     & 26 &    & 26 &  8 & 22 & 26 & 19 & 17 &  7 & 21.5 & 22 \\
QAR-QbSD-gSAV               & 22 &    & 24 & 15 & 19 & 21 &    & 18 &  6 & 21.9 & 23 \\
AL$_{\text{AR}}$-SAV        & 27 &    & 27 & 11 & 23 & 23 &    & 15 &  6 & 22.8 & 24 \\
AR-AL$_{\text{AR}}$-SAV     &    & 21 & 21 &  5 &    & 19 &    &    &  3 & 23.1 & 25 \\
AR-AL$_{\text{Mult.}}$-SAV  & 24 &    & 23 & 13 &    & 24 &    & 20 &  5 & 24.5 & 26 \\
AR-GAS                      & 23 &    & 25 &    &    & 27 &    & 16 &  4 & 25.4 & 27 \\
GAS                         & 21 &    & 28 & 27 &    &    &    &  3 &  3 & 27.0 & 28 \\
\bottomrule
\end{tabular*}
\vspace{-0.5cm}
\end{center}
{\footnotesize \textit{Notes:} For explanations, see notes of Table 5.}
\end{minipage}
\end{table}

\begin{table}[!h]
\hspace*{-0.65cm}
\begin{minipage}{7.1in}
{\small {\bf Table 8.} Ranking of 1\% VaR and ES forecasting models based on the MCS procedure using AL log scores}
\begin{center}
\footnotesize
\vspace{-0.25cm}
\begin{tabular*}{\textwidth}{@{\extracolsep{\fill}} lcccccccc|ccc}
\toprule
                           & \multirow{2}{*}{S\&P} & \multirow{2}{*}{DJIA} & \multirow{2}{*}{NASDAQ} & \multirow{2}{*}{STOXX} & \multirow{2}{*}{FTSE} & \multirow{2}{*}{DAX} & \multirow{2}{*}{CAC} & \multirow{2}{*}{TSX} & \multirow{2}{*}{\#} & Avg. & Final \\
                           &                        &                        &                          &                         &                        &                       &                       &                      &                      & rank  & rank  \\
\midrule
AR-GJR-GARCH-skew-$t$      &  1 &  5 &  3 &  1 &  3 &  3 &  8 & 11 &  8 &  4.4 &  1 \\
QbSD-gAS                   &  9 &  4 &  6 &  8 &  6 &  1 &  1 &  1 &  8 &  4.5 &  2 \\
GARCH-skew-$t$             &  3 &  2 &  1 &  3 &  4 &  5 & 14 & 10 &  8 &  5.2 &  3 \\
GJR-GARCH-skew-$t$         & 12 &  3 &  4 &  4 &  5 &  6 &  2 &  9 &  8 &  5.6 &  4 \\
QAR-QbSD-gAS               &  4 &  1 & 22 &  7 &  1 & 12 &  4 &  4 &  8 &  6.9 &  5 \\
AR-AL$_{\text{AR}}$-AS     &  5 &  8 & 11 & 11 & 12 &  7 &  7 &  5 &  8 &  8.2 &  6 \\
AR-GARCH-skew-$t$          &  2 & 16 &  2 & 10 &  8 &  2 & 16 & 12 &  8 &  8.5 &  7 \\
AR-AL$_{\text{Mult.}}$-AS  &  6 &  6 & 12 & 16 &  9 & 11 & 12 &  2 &  8 &  9.2 &  8 \\
QAR-QbSD-gSAV              & 21 & 15 & 18 &  2 &  2 &  4 &  6 &  6 &  8 &  9.2 &  9 \\
AL$_{\text{Mult.}}$-AS     &  7 & 18 & 10 & 17 & 10 & 14 &  5 &  3 &  8 & 10.5 & 10 \\
AL$_{\text{AR}}$-AS        & 11 & 14 & 13 & 12 &  7 & 13 &  9 &  8 &  8 & 10.9 & 11 \\
GJR-GARCH-$t$              & 17 &  9 & 15 &  6 & 21 &  8 &  3 & 18 &  8 & 12.1 & 12 \\
AR-AL$_{\text{Mult.}}$-SAV & 10 & 10 &  5 & 19 & 15 & 20 & 15 & 14 &  8 & 13.5 & 13 \\
AR-GJR-GARCH-$t$           & 14 &  7 & 16 &  5 & 19 & 10 & 10 &    &  7 & 13.6 & 14 \\
AR-AL$_{\text{AR}}$-SAV    &  8 & 11 &  7 & 20 & 13 & 16 & 22 & 15 &  8 & 14.0 & 15 \\
AL$_{\text{Mult.}}$-SAV    & 13 & 17 &  8 & 18 & 16 & 19 & 13 & 13 &  8 & 14.6 & 16 \\
GARCH-$t$                  & 18 & 12 & 14 & 14 & 22 &  9 & 18 & 17 &  8 & 15.5 & 17 \\
AL$_{\text{AR}}$-SAV       & 16 & 19 &  9 & 15 & 14 & 17 & 20 & 16 &  8 & 15.8 & 18 \\
QbSD-gSAV                  & 24 & 20 & 19 & 13 & 11 & 15 & 17 &  7 &  8 & 15.8 & 19 \\
AR-GARCH-$t$               & 15 & 13 & 17 &  9 & 20 & 18 & 11 &    &  7 & 16.4 & 20 \\
GAS                        & 19 &    & 21 &    & 17 & 21 & 19 & 20 &  6 & 21.6 & 21 \\
EGARCH                     & 23 & 23 & 20 & 21 & 23 & 22 &    & 19 &  7 & 22.4 & 22 \\
AR-GAS                     & 20 & 21 & 24 &    & 18 &    & 21 &    &  5 & 23.5 & 23 \\
AR-EGARCH                  & 22 & 22 & 23 & 22 & 24 &    &    &    &  5 & 24.6 & 24 \\
GARCH-normal               &    &    &    &    &    &    &    &    &  0 & 28.0 & 25 \\
GJR-GARCH-normal           &    &    &    &    &    &    &    &    &  0 & 28.0 & 26 \\
AR-GARCH-normal            &    &    &    &    &    &    &    &    &  0 & 28.0 & 27 \\
AR-GJR-GARCH-normal        &    &    &    &    &    &    &    &    &  0 & 28.0 & 28 \\
\bottomrule
\end{tabular*}
\vspace{-0.5cm}
\end{center}
{\footnotesize \textit{Notes:} 
A model that is excluded from $\widehat{\mathcal{M}}_{90\%}^{\ast}$ for a given stock index is left blank in the table. The remaining models in $\widehat{\mathcal{M}}_{90\%}^{\ast}$ are ranked based on their AL log score $t$-statistics, with lower values indicating better performance. The column ``\#'' represents the number of stock indices (out of 8) for which the model is included in $\widehat{\mathcal{M}}_{90\%}^{\ast}$. The ``Avg. rank'' column reports the average rank across all stock indices, where models eliminated from $\widehat{\mathcal{M}}_{90\%}^{\ast}$ were assigned a rank of 28. The ``Final rank'' column orders models from best to worst based on their average rank, summarizing their relative forecasting performance across stock indices. The rolling-window size is $R=1250$.}
\end{minipage}
\end{table}

\begin{table}[!h]
\hspace*{-0.65cm}
\begin{minipage}{7.1in}
{\small {\bf Table 9.} Ranking of 2.5\% VaR and ES forecasting models based on the MCS procedure using AL log scores}
\begin{center}
\footnotesize
\vspace{-0.25cm}
\begin{tabular*}{\textwidth}{@{\extracolsep{\fill}} lcccccccc|ccc}
\toprule
                           & \multirow{2}{*}{S\&P} & \multirow{2}{*}{DJIA} & \multirow{2}{*}{NASDAQ} & \multirow{2}{*}{STOXX} & \multirow{2}{*}{FTSE} & \multirow{2}{*}{DAX} & \multirow{2}{*}{CAC} & \multirow{2}{*}{TSX} & \multirow{2}{*}{\#} & Avg. & Final \\
                           &                        &                        &                          &                         &                        &                       &                       &                      &                      & rank  & rank  \\
\midrule
QbSD-gAS                   &  5 &  2 &  3 & 12 &  1 &  1 &  1 &  1 &  8 &  3.2 &  1 \\
AR-AL$_{\text{Mult.}}$-AS  &  7 &  8 &  8 &  4 &  4 &  2 &  3 &  4 &  8 &  5.0 &  2 \\
AL$_{\text{Mult.}}$-AS     & 11 &  7 &  6 &  8 &  2 &  3 &  2 &  2 &  8 &  5.1 &  3 \\
QAR-QbSD-gAS               &  1 &  1 & 18 &  2 &  3 &  8 &  4 &  5 &  8 &  5.2 &  4 \\
GJR-GARCH-skew-$t$         &  4 &  6 &  1 &  3 &  9 &  7 &  6 &    &  7 &  8.0 &  5 \\
AL$_{\text{AR}}$-AS        &  9 &  3 &  5 &  9 &    &  4 &  5 &  3 &  7 &  8.2 &  6 \\
GARCH-skew-$t$             &  2 &  5 &  4 &  5 &  8 &  6 & 14 &    &  7 &  9.0 &  7 \\
AR-AL$_{\text{AR}}$-AS     &  8 &  4 &  2 & 15 & 15 & 10 & 17 &  6 &  8 &  9.6 &  8 \\
AR-GJR-GARCH-skew-$t$      &  3 &  9 &  9 & 11 & 11 &  9 &  9 &    &  7 & 11.1 &  9 \\
QAR-QbSD-gSAV              & 14 & 13 & 14 &  1 &  5 & 11 &  8 &    &  7 & 11.8 & 10 \\
AR-GARCH-skew-$t$          &  6 & 12 &  7 & 13 & 12 &  5 & 15 &    &  7 & 12.2 & 11 \\
AL$_{\text{Mult.}}$-SAV    & 13 & 14 & 12 & 10 &  6 & 17 & 10 &    &  7 & 13.8 & 12 \\
AR-AL$_{\text{Mult.}}$-SAV & 10 & 18 & 11 &  6 & 10 & 15 & 12 &    &  7 & 13.8 & 13 \\
AL$_{\text{AR}}$-SAV       & 17 & 15 & 13 &  7 & 13 & 14 & 11 &    &  7 & 14.8 & 14 \\
AR-AL$_{\text{AR}}$-SAV    & 12 & 17 & 10 & 14 & 14 & 12 & 13 &    &  7 & 15.0 & 15 \\
GJR-GARCH-$t$              & 16 & 10 & 15 & 16 & 16 & 18 & 16 &    &  7 & 16.9 & 16 \\
QbSD-gSAV                  &    & 22 & 22 & 18 &  7 & 13 &  7 &    &  6 & 18.1 & 17 \\
AR-GJR-GARCH-$t$           & 18 & 11 & 19 & 17 & 19 & 16 & 19 &    &  7 & 18.4 & 18 \\
GARCH-$t$                  & 15 & 16 & 16 & 22 & 17 & 19 & 18 &    &  7 & 18.9 & 19 \\
EGARCH                     & 19 & 19 & 17 & 19 & 18 & 21 &    &    &  6 & 21.1 & 20 \\
AR-GARCH-$t$               & 21 & 20 & 20 & 20 &    & 20 & 20 &    &  6 & 22.1 & 21 \\
AR-EGARCH                  & 20 & 21 & 21 & 21 & 20 & 22 &    &    &  6 & 22.6 & 22 \\
GARCH-normal               &    &    &    &    &    &    &    &    &  0 & 28.0 & 23 \\
GJR-GARCH-normal           &    &    &    &    &    &    &    &    &  0 & 28.0 & 24 \\
AR-GARCH-normal            &    &    &    &    &    &    &    &    &  0 & 28.0 & 25 \\
AR-GJR-GARCH-normal        &    &    &    &    &    &    &    &    &  0 & 28.0 & 26 \\
GAS                        &    &    &    &    &    &    &    &    &  0 & 28.0 & 27 \\
AR-GAS                     &    &    &    &    &    &    &    &    &  0 & 28.0 & 28 \\
\bottomrule
\end{tabular*}
\vspace{-0.5cm}
\end{center}
{\footnotesize \textit{Notes:} For explanations, see notes of Table 8.}
\end{minipage}
\end{table}

\begin{table}[!h]
\hspace*{-0.65cm}
\begin{minipage}{7.1in}
{\small {\bf Table 10.} Ranking of 5\% VaR and ES forecasting models based on the MCS procedure using AL log scores}
\begin{center}
\footnotesize
\vspace{-0.25cm}
\begin{tabular*}{\textwidth}{@{\extracolsep{\fill}} lcccccccc|ccc}
\toprule
                           & \multirow{2}{*}{S\&P} & \multirow{2}{*}{DJIA} & \multirow{2}{*}{NASDAQ} & \multirow{2}{*}{STOXX} & \multirow{2}{*}{FTSE} & \multirow{2}{*}{DAX} & \multirow{2}{*}{CAC} & \multirow{2}{*}{TSX} & \multirow{2}{*}{\#} & Avg. & Final \\
                           &                        &                        &                          &                         &                        &                       &                       &                      &                      & rank  & rank  \\
\midrule
QbSD-gAS                   &  2 &  1 &  1 &  5 &  1 &  1 &  1 &  2 &  8 &  1.8 &  1 \\
AL$_{\text{AR}}$-AS        &  3 &  4 &  2 & 13 &  4 &  2 &  2 &  3 &  8 &  4.1 &  2 \\
AL$_{\text{Mult.}}$-AS     &  4 &  6 &  5 &  3 &  3 &  5 &  4 &  4 &  8 &  4.2 &  3 \\
QAR-QbSD-gAS               &  1 &  2 & 14 &  4 &  2 &  4 &  6 &  1 &  8 &  4.2 &  4 \\
AR-AL$_{\text{AR}}$-AS     & 10 &  9 &  7 &  1 &  5 &  6 &  5 &  5 &  8 &  6.0 &  5 \\
AR-AL$_{\text{Mult.}}$-AS  &  7 & 15 & 11 &  2 & 11 &  3 &  3 &  6 &  8 &  7.2 &  6 \\
GARCH-skew-$t$             &  6 &  5 &  4 & 10 &  7 &  8 &    &  9 &  7 &  9.6 &  7 \\
GJR-GARCH-skew-$t$         &  9 &  7 &  3 &  8 &  9 &  7 &    &  8 &  7 &  9.9 &  8 \\
EGARCH                     & 15 &  3 &  9 &  9 & 10 & 12 &    &  7 &  7 & 11.6 &  9 \\
AR-GJR-GARCH-skew-$t$      &  5 &  8 &  6 & 15 & 12 & 11 &    &    &  6 & 14.1 & 10 \\
QAR-QbSD-gSAV              & 11 & 16 & 10 &  6 &  6 &  9 &    &    &  6 & 14.2 & 11 \\
AR-GARCH-skew-$t$          &  8 & 10 &  8 & 16 & 13 & 10 &    &    &  6 & 15.1 & 12 \\
GJR-GARCH-$t$              & 12 & 12 & 13 & 18 & 17 & 16 &    &    &  6 & 18.0 & 13 \\
AR-EGARCH                  & 20 & 11 & 16 & 14 & 14 & 14 &    &    &  6 & 18.1 & 14 \\
AR-AL$_{\text{AR}}$-SAV    & 16 & 18 & 18 & 11 & 22 & 13 &    &    &  6 & 19.2 & 15 \\
GARCH-$t$                  & 13 & 13 & 12 & 26 & 18 & 17 &    &    &  6 & 19.4 & 16 \\
AL$_{\text{AR}}$-SAV       & 21 & 22 & 24 & 17 & 15 & 18 &    & 11 &  7 & 19.5 & 17 \\
AR-AL$_{\text{Mult.}}$-SAV & 17 &    & 15 &  7 & 21 & 15 &    &    &  5 & 19.9 & 18 \\
AR-GJR-GARCH-$t$           & 14 & 14 & 22 & 19 & 20 & 19 &    &    &  6 & 20.5 & 19 \\
AL$_{\text{Mult.}}$-SAV    &    & 20 & 11 & 16 & 22 &    & 10 &  5 &  5 & 20.5 & 20 \\
AR-GARCH-$t$               & 18 & 17 & 23 & 22 & 19 & 20 &    &    &  6 & 21.9 & 21 \\
QbSD-gSAV                  & 22 & 20 &    & 20 &  8 & 21 &    &    &  5 & 21.9 & 22 \\
GJR-GARCH-normal           & 19 & 21 & 19 & 21 &    & 25 &    &    &  5 & 23.6 & 23 \\
AR-GJR-GARCH-normal        &    & 19 &    & 24 &    & 23 &    &    &  3 & 25.8 & 24 \\
AR-GAS                     & 24 &    & 17 &    &    &    &    &    &  2 & 26.1 & 25 \\
GAS                        & 23 &    & 21 &    &    &    &    &    &  2 & 26.5 & 26 \\
GARCH-normal               &    &    & 22 &    & 26 &    &    &    &  2 & 27.1 & 27 \\
AR-GARCH-normal            &    &    & 25 & 25 &    & 25 &    &    &  2 & 27.1 & 28 \\
\bottomrule
\end{tabular*}
\vspace{-0.5cm}
\end{center}
{\footnotesize \textit{Notes:} For explanations, see notes of Table 8.}
\end{minipage}
\end{table}

\newpage

\begin{figure}[htbp]

\caption{Time series plots of the daily log returns for the S\&P 500, DJIA, NASDAQ, and EURO STOXX 50 indices over the sample period. Each plot represents the percentage daily returns computed from adjusted closing prices.}

    \begin{center}
        \subfloat[][S\&P 500]{
            \includegraphics[width=0.5\linewidth]{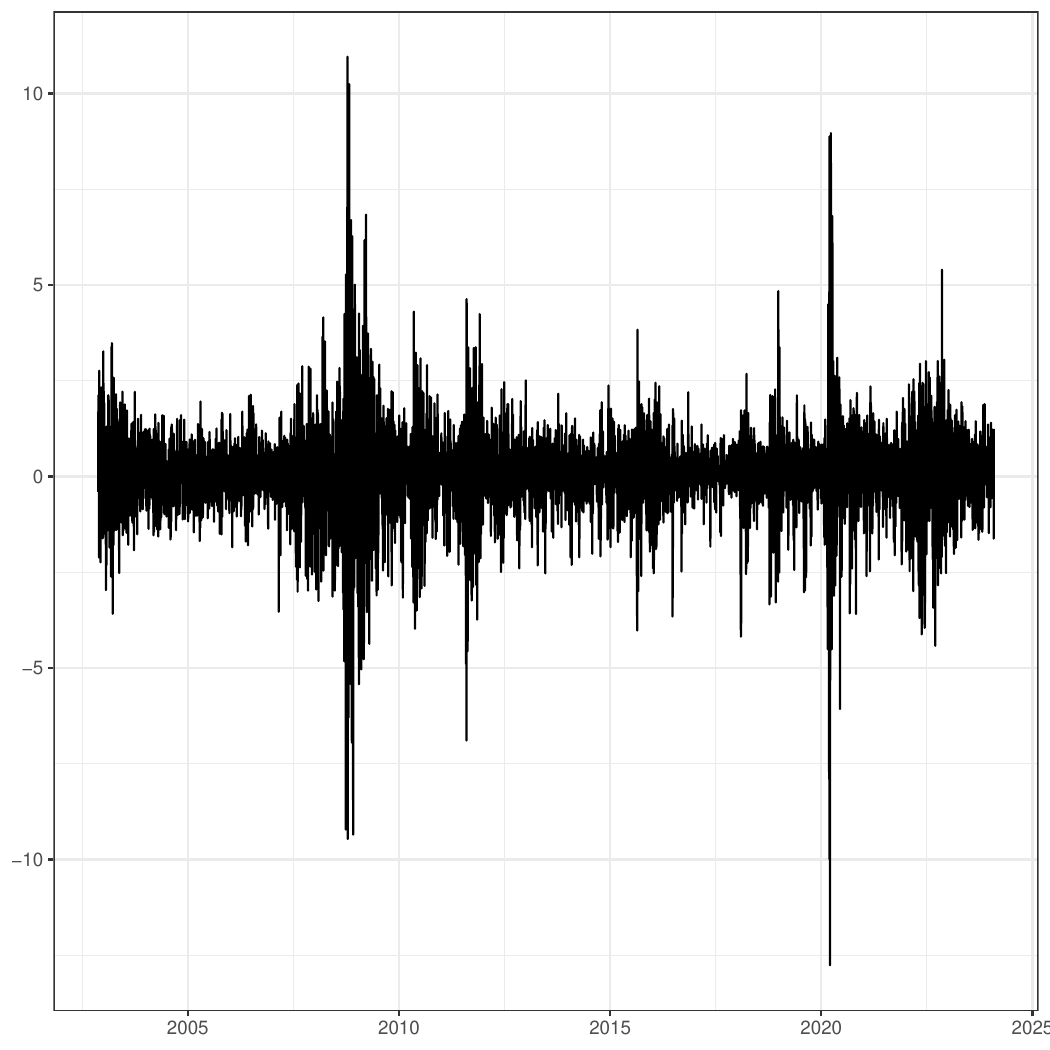}
        } 
        \subfloat[][DJIA]{

            \includegraphics[width=0.5\linewidth]{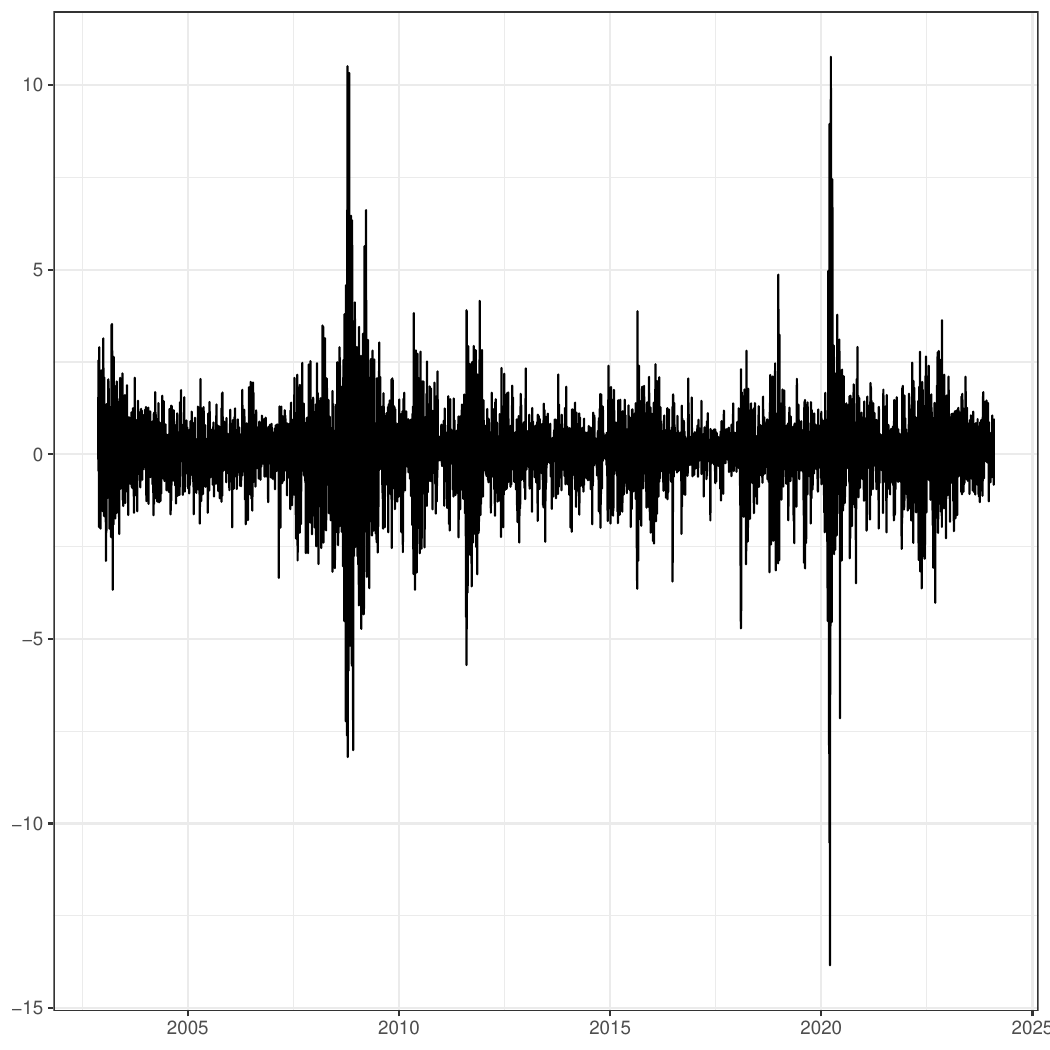}
        } \\ 
        \subfloat[][NASDAQ]{

            \includegraphics[width=0.5\linewidth]{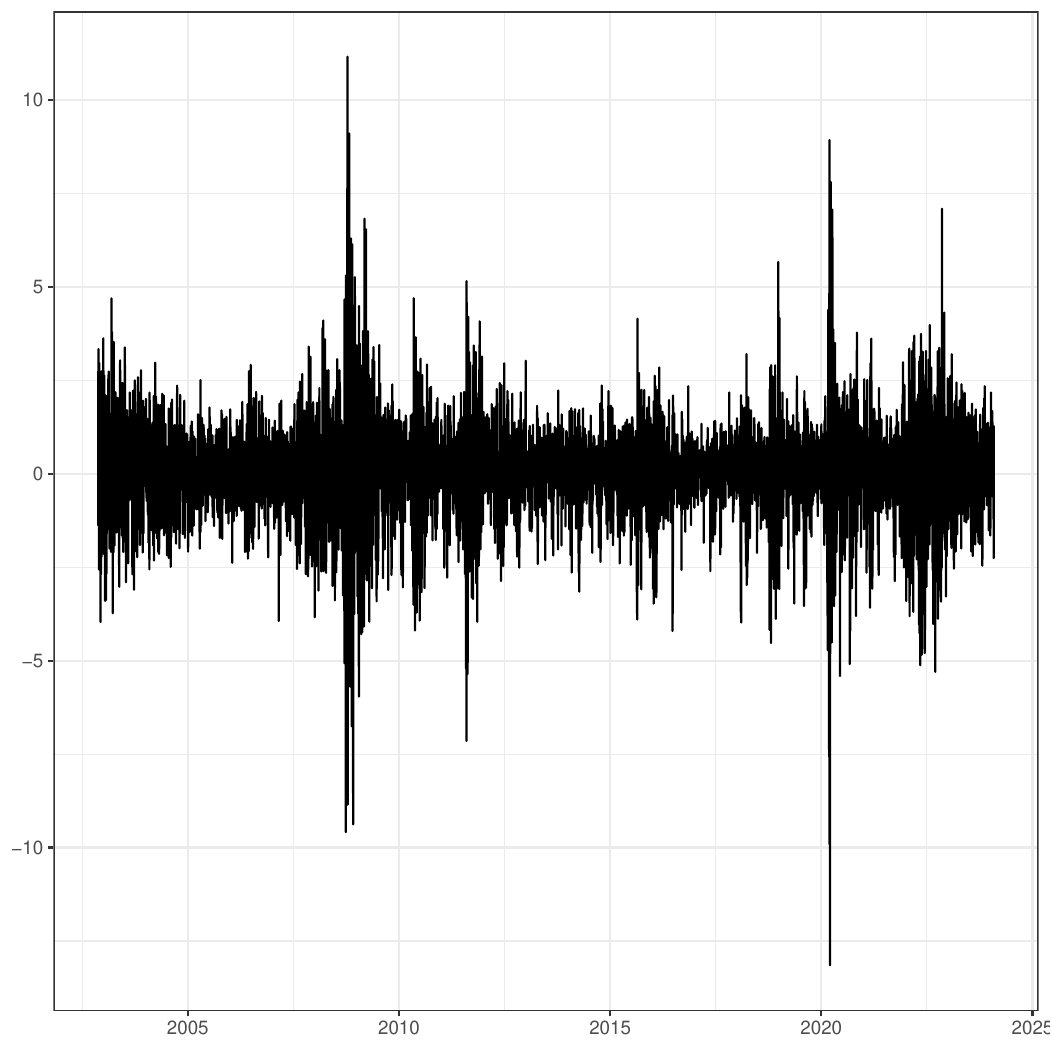}
        }
        \subfloat[][EURO STOXX 50]{

            \includegraphics[width=0.5\linewidth]{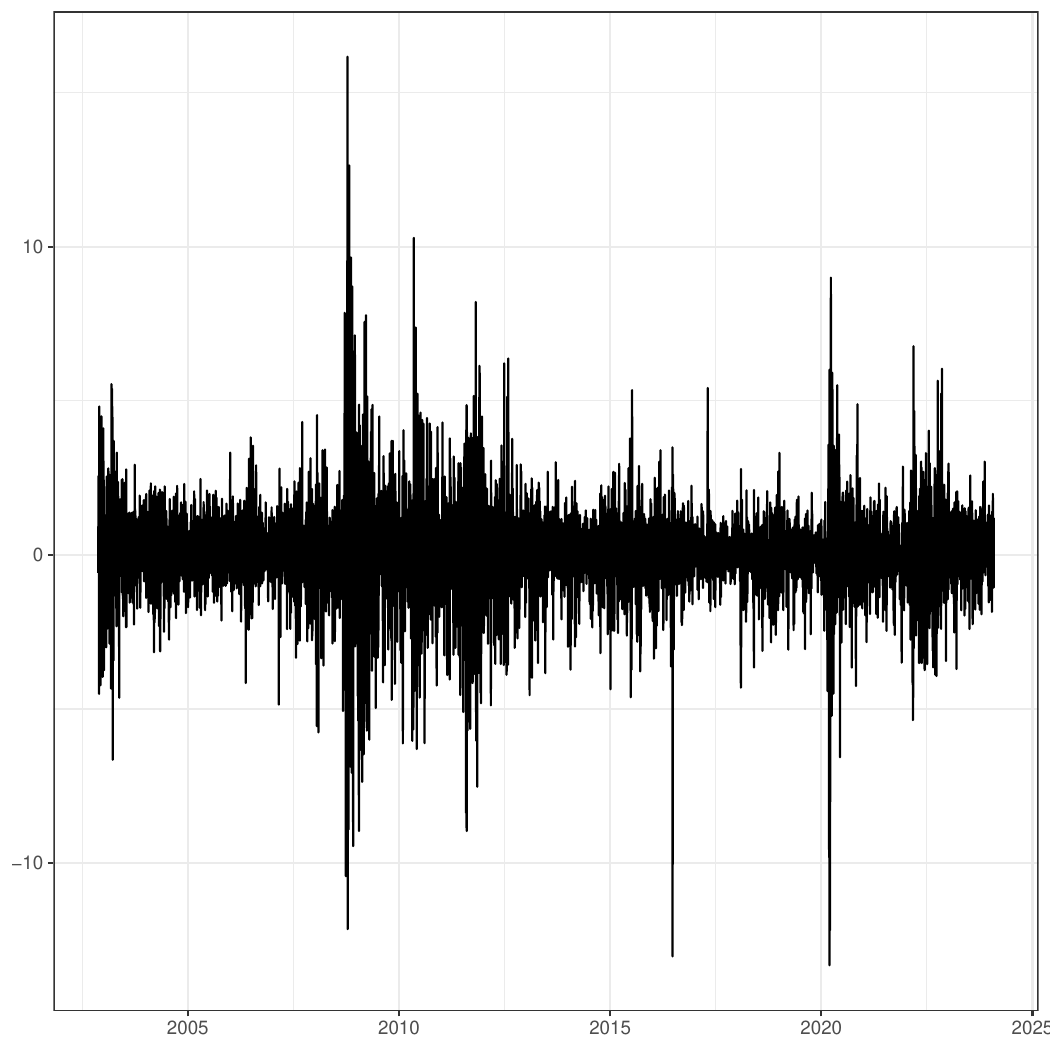}
        }

    \end{center}
\end{figure}

\begin{figure}[htbp]

    \caption{Time series plots of the daily log returns for the FTSE 100, DAX, CAC 40, and TSX indices over the sample period. Each plot displays percentage daily returns based on adjusted closing prices.}

    \begin{center}

        \subfloat[][FTSE 100]{

            \includegraphics[width=0.5\linewidth]{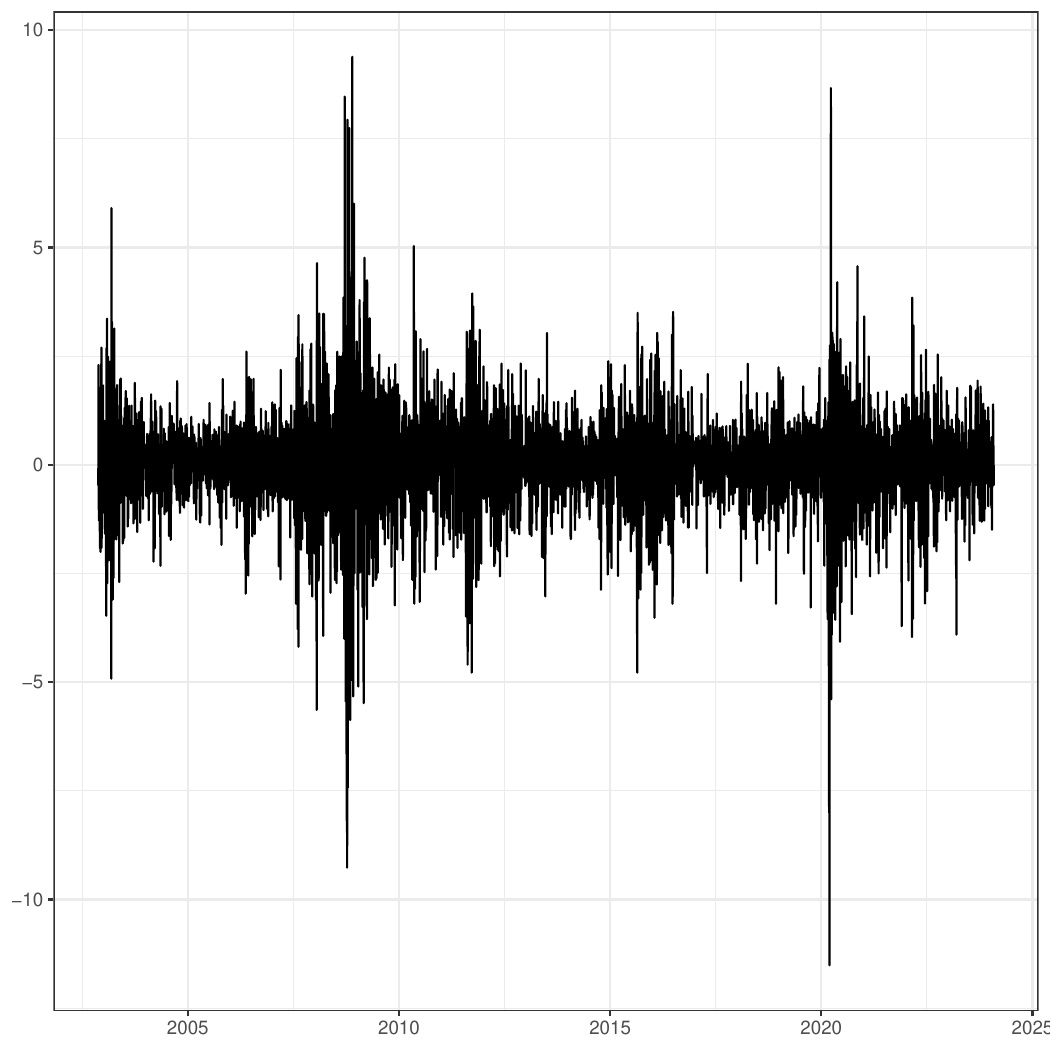}
        }       
        \subfloat[][DAX]{

            \includegraphics[width=0.5\linewidth]{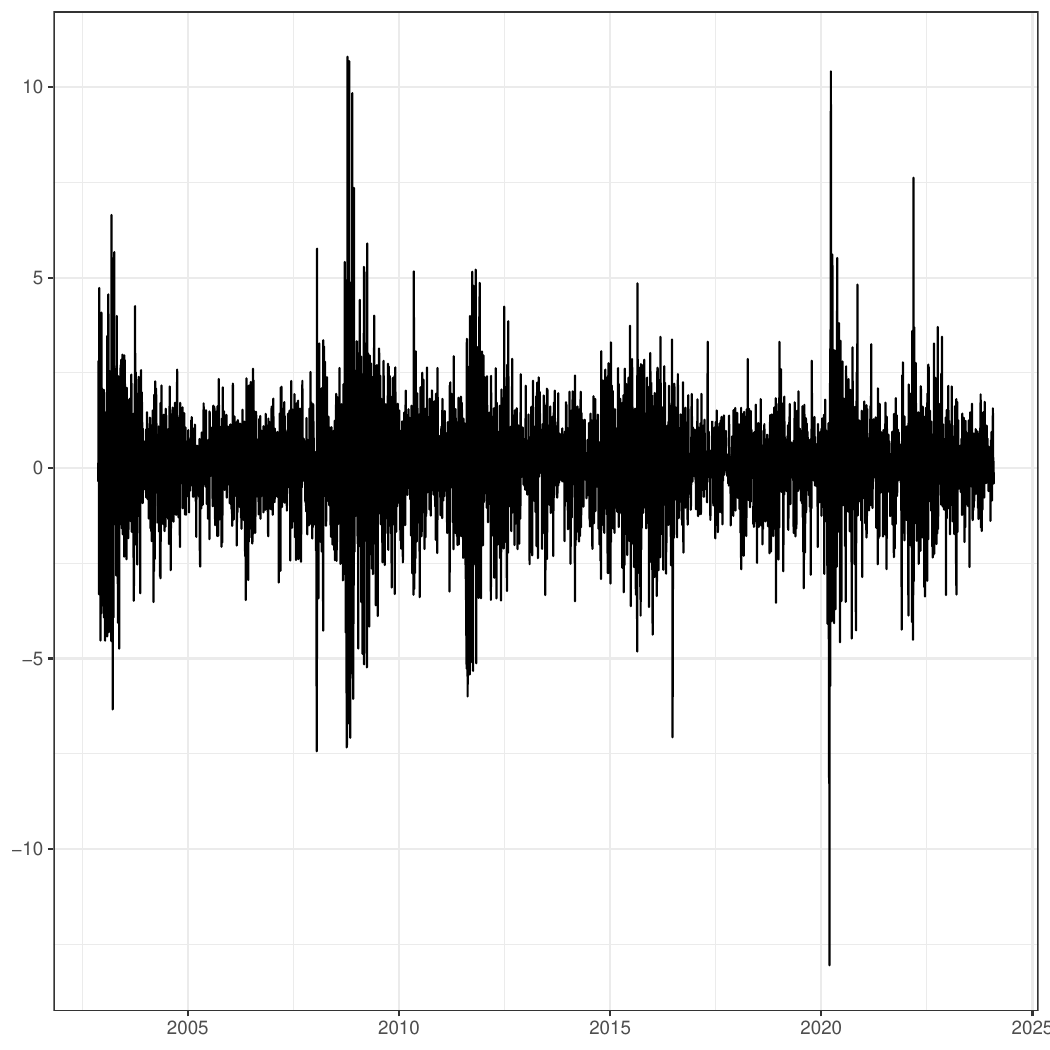}
        }    \\           

        \subfloat[][CAC 40 ]{

            \includegraphics[width=0.5\linewidth]{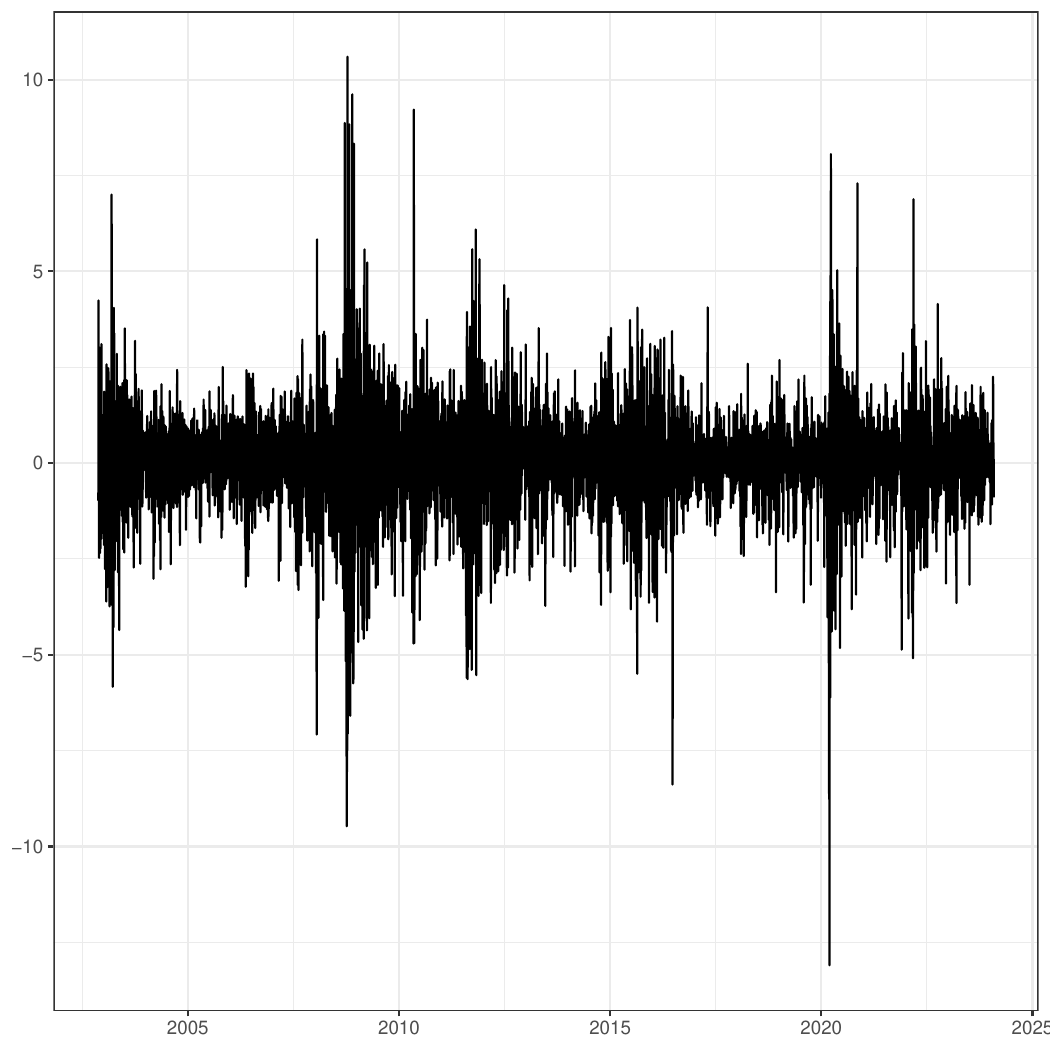}
        }       
        \subfloat[][TSX]{

            \includegraphics[width=0.5\linewidth]{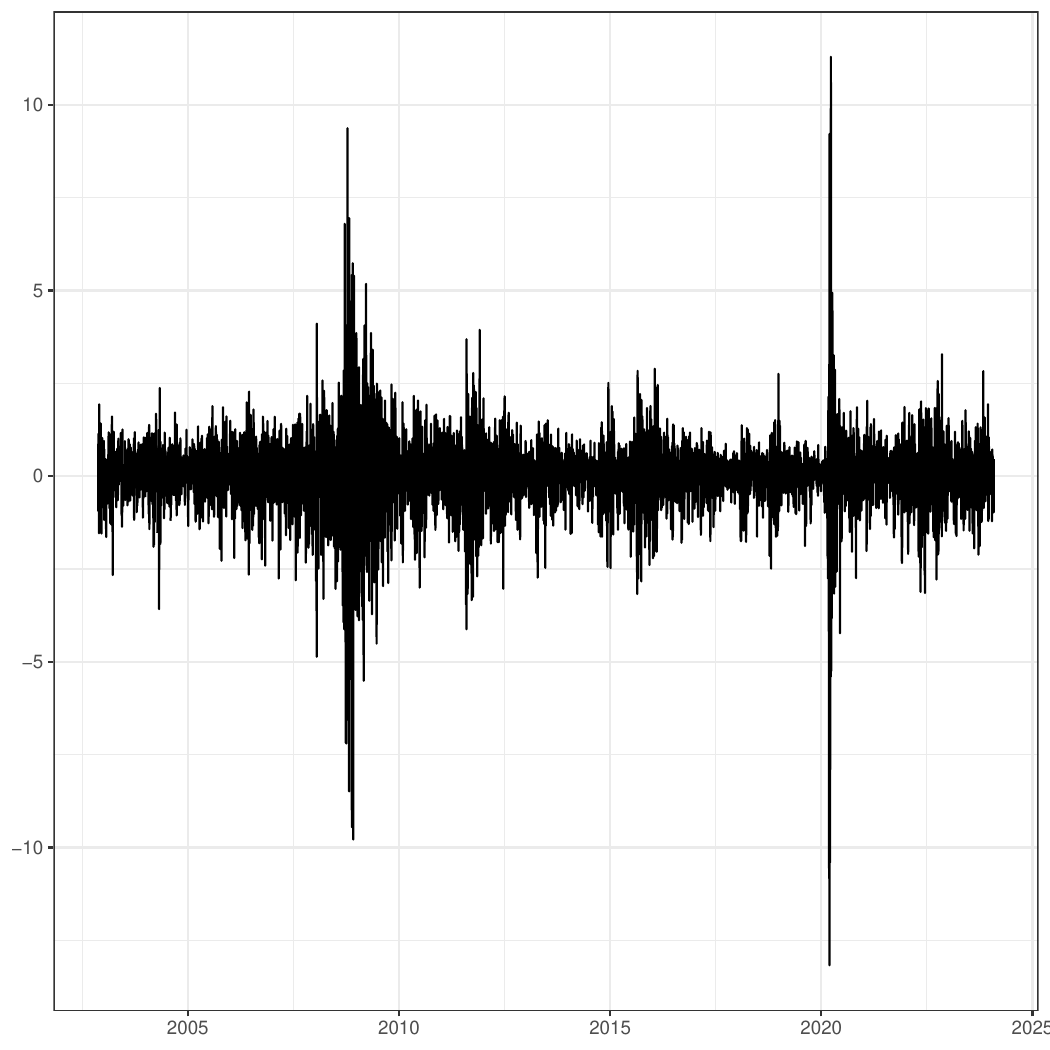}
        }

    \end{center}
\end{figure}

\clearpage
\setcounter{page}{1}

\renewcommand{\thefootnote}{\fnsymbol{footnote}}

\renewcommand{\baselinestretch}{1.75}
\small\normalsize

\begin{center}

Supplementary material for:

{\bf{\Large{Quantile-based modeling of scale dynamics in financial returns for Value-at-Risk and Expected Shortfall forecasting}}}

\bigskip

Xiaochun Liu and Richard Luger

\end{center}

\renewcommand{\baselinestretch}{1.5}
\small\normalsize

\section*{Section A}

Tables A1--A12 present comparisons of the mean absolute error (MAE) and root mean squared error (RMSE) for the QbSD forecasting approach, 
examining the impact of the scale-defining quantile level $p$ on the computation of VaR and ES. 
Specifically, we consider forecasts based on: (i) individual values of $p \in \{ 0.05, 0.10, 0.15, 0.20, 0.25 \}$; (ii) the mean over these $p$ values; and (iii) the median over them.

The results are obtained from 1,000 replications for each configuration of the asymmetric power ARCH model with innovations drawn from the skewed $t$-distribution, as specified in the main text.
The lowest MAE and RMSE values in each table are highlighted in bold.

Overall, the results in Tables A1--A12 suggest that combining forecasts over multiple quantile levels $p$---either by taking the mean or, typically to a slightly lesser degree, the median---tends to yield lower forecast errors than relying on a single $p$ in isolation. This advantage of averaging is most noticeable when $T$ is relatively small (e.g., $T=250$; see especially Tables A1--A4), where individual-$p$ estimates exhibit higher variability. As $T$ increases to 1250 (Tables A5--A8) or 2500 (Tables A9--A12), the performance gap narrows, yet the multi-$p$ strategies remain competitive or superior under most DGP configurations.

Although occasional scenarios arise where an individual $p$ may match or slightly outperform the combined forecasts (e.g., $p=0.10$ in Table A7), no single $p$ exhibits consistent dominance across the wide range of parameter settings (leverage $\theta$, tail thickness $v$, skewness $\lambda$) or sample sizes $T$. Consequently, our findings recommend the ``mean over $p$” method as a robust, practically simple strategy for balancing estimation accuracy and stability in quantile-based VaR and ES forecasting.

\begin{landscape}

\bigskip
\bigskip

\begin{table}[h]
\begin{center}
\begin{minipage}{8.8in}
\vspace{-1.0cm}
{\small {\bf Table A1.} VaR forecasting results with the QbSD approach when  $ \theta=0$ and $T=250$ }
\footnotesize
\\[0.5ex]
\begin{tabular*}{\textwidth}{@{\extracolsep{\fill}}  lcrrrcrrrcrrrcrrr}
\toprule

                        &&   \multicolumn{3}{c}{$v=20$, $\lambda=0$}   && \multicolumn{3}{c}{$v=20$, $\lambda=-0.5$}   && \multicolumn{3}{c}{$v=5$, $\lambda=0$} && \multicolumn{3}{c}{$v=5$, $\lambda=-0.5$}    \\

                        \cline{3-5} \cline{7-9} \cline{11-13} \cline{15-17} \\[-2.0ex]
  
                                  &$\alpha=$ & 0.01  &   0.025    & 0.05       && 0.01  &   0.025    & 0.05    && 0.01  &   0.025    & 0.05   && 0.01  &   0.025    & 0.05  \\

\midrule

\multicolumn{17}{l}{Panel A: Mean absolute error} \\

QbSD-gSAV model                            &&          &             &               &&          &             &        &&          &            &        &&       &            &       \\

\quad Individual values of $p$             &&          &             &               &&          &             &        &&          &            &        &&       &            &       \\

\quad \quad $p=0.05$                       &&    0.243 &    0.181    &    0.146      &&    0.304 &    0.222    &   0.173  &&    0.328 &   0.214    &   0.158    &&    0.436 &    0.269    & 0.189  \\
\quad \quad $p=0.10$                       &&    0.235 &    0.172    &    0.138      &&    0.281 &    0.204    &   0.157  &&    0.318 &   0.206    &   0.148    &&    0.402 &    0.251    & 0.173  \\
\bottomrule

\end{tabular*}
\end{minipage}
\end{center}
\end{table}

\end{landscape}

\begin{landscape}

\bigskip
\bigskip

\begin{table}[h]
\begin{center}
\begin{minipage}{8.8in}
\vspace{-1.0cm}
{\small {\bf Table A2.} ES forecasting results with the QbSD approach when $ \theta=0$ and $T=250$ }
\footnotesize
\\[0.5ex]
\begin{tabular*}{\textwidth}{@{\extracolsep{\fill}}  lcrrrcrrrcrrrcrrr}
\toprule

                        &&   \multicolumn{3}{c}{$v=20$, $\lambda=0$}   && \multicolumn{3}{c}{$v=20$, $\lambda=-0.5$}   && \multicolumn{3}{c}{$v=5$, $\lambda=0$} && \multicolumn{3}{c}{$v=5$, $\lambda=-0.5$}    \\

                        \cline{3-5} \cline{7-9} \cline{11-13} \cline{15-17} \\[-2.0ex]
  
                                  &$\alpha=$ & 0.01  &   0.025    & 0.05       && 0.01  &   0.025    & 0.05    && 0.01  &   0.025    & 0.05   && 0.01  &   0.025    & 0.05  \\

\midrule

\multicolumn{17}{l}{Panel A: Mean absolute error} \\

QbSD-gSAV model                            &&        &            &            &&       &            &        &&       &            &        &&       &            &       \\

\quad Individual values of $p$             &&        &            &            &&       &            &        &&       &            &        &&       &            &       \\

\quad \quad $p=0.05$                       &&    0.325 &     0.238    &    0.191      &&    0.424 &    0.299    &    0.232  &&    0.551 &    0.344    &    0.246  &&    0.762 &    0.458    &    0.315  \\
\quad \quad $p=0.10$                       &&    0.311 &     0.226    &    0.180      &&    0.395 &    0.275    &    0.212  &&    0.537 &    0.332    &    0.234  &&    0.732 &    0.429    &    0.290  \\
\bottomrule

\end{tabular*}
\end{minipage}
\end{center}
\end{table}

\end{landscape}

\begin{landscape}

\bigskip
\bigskip

\begin{table}[h]
\begin{center}
\begin{minipage}{8.8in}
\vspace{-1.0cm}
{\small {\bf Table A3.} VaR forecasting results with the QbSD approach when  $ \theta=0.5$  and $T=250$}
\footnotesize
\\[0.5ex]
\begin{tabular*}{\textwidth}{@{\extracolsep{\fill}}  lcrrrcrrrcrrrcrrr}
\toprule

                        &&   \multicolumn{3}{c}{$v=20$, $\lambda=0$}   && \multicolumn{3}{c}{$v=20$, $\lambda=-0.5$}   && \multicolumn{3}{c}{$v=5$, $\lambda=0$} && \multicolumn{3}{c}{$v=5$, $\lambda=-0.5$}    \\

                        \cline{3-5} \cline{7-9} \cline{11-13} \cline{15-17} \\[-2.0ex]
  
                                  &$\alpha=$ & 0.01  &   0.025    & 0.05       && 0.01  &   0.025    & 0.05    && 0.01  &   0.025    & 0.05   && 0.01  &   0.025    & 0.05  \\

\midrule

\multicolumn{17}{l}{Panel A: Mean absolute error} \\

QbSD-gSAV model                            &&          &             &               &&          &             &           &&          &             &           &&          &             &          \\
                           
\quad Individual values of $p$             &&          &             &               &&          &             &           &&          &             &           &&          &             &          \\
                           
\quad \quad $p=0.05$                       &&    0.327 &    0.252    &    0.206      &&    0.415 &    0.314    &    0.245  &&    0.401 &    0.277    &    0.206  &&    0.553 &    0.355    &    0.259  \\
\quad \quad $p=0.10$                       &&    0.316 &    0.241    &    0.195      &&    0.397 &    0.295    &    0.231  &&    0.380 &    0.257    &    0.190  &&    0.514 &    0.333    &    0.237  \\
\bottomrule

\end{tabular*}
\end{minipage}
\end{center}
\end{table}

\end{landscape}

\begin{landscape}

\bigskip
\bigskip

\begin{table}[h]
\begin{center}
\begin{minipage}{8.8in}
\vspace{-1.0cm}
{\small {\bf Table A4.} ES forecasting results with the QbSD approach when  $ \theta=0.5$  and $T=250$}
\footnotesize
\\[0.5ex]
\begin{tabular*}{\textwidth}{@{\extracolsep{\fill}}  lcrrrcrrrcrrrcrrr}
\toprule

                        &&   \multicolumn{3}{c}{$v=20$, $\lambda=0$}   && \multicolumn{3}{c}{$v=20$, $\lambda=-0.5$}   && \multicolumn{3}{c}{$v=5$, $\lambda=0$} && \multicolumn{3}{c}{$v=5$, $\lambda=-0.5$}    \\

                        \cline{3-5} \cline{7-9} \cline{11-13} \cline{15-17} \\[-2.0ex]
  
                                  &$\alpha=$ & 0.01  &   0.025    & 0.05       && 0.01  &   0.025    & 0.05    && 0.01  &   0.025    & 0.05   && 0.01  &   0.025    & 0.05  \\

\midrule

\multicolumn{17}{l}{Panel A: Mean absolute error} \\

QbSD-gSAV model                            &&          &             &               &&          &             &           &&          &             &           &&          &             &          \\
                           
\quad Individual values of $p$             &&          &             &               &&          &             &           &&          &             &           &&          &             &          \\
                           
\quad \quad $p=0.05$                       &&    0.420 &    0.324    &    0.266      &&    0.557 &    0.414    &    0.329  &&    0.654 &    0.425    &    0.314  &&    0.945 &    0.585    &    0.412  \\
\quad \quad $p=0.10$                       &&    0.401 &    0.309    &    0.254      &&    0.527 &    0.392    &    0.311  &&    0.619 &    0.398    &    0.291  &&    0.888 &    0.539    &    0.379  \\
\bottomrule

\end{tabular*}
\end{minipage}
\end{center}
\end{table}

\end{landscape}

\begin{landscape}

\bigskip
\bigskip

\begin{table}[h]
\begin{center}
\begin{minipage}{8.8in}
\vspace{-1.0cm}
{\small {\bf Table A5.} VaR forecasting results with the QbSD approach when  $ \theta=0$  and $T=1250$}
\footnotesize
\\[0.5ex]
\begin{tabular*}{\textwidth}{@{\extracolsep{\fill}}  lcrrrcrrrcrrrcrrr}
\toprule

                        &&   \multicolumn{3}{c}{$v=20$, $\lambda=0$}   && \multicolumn{3}{c}{$v=20$, $\lambda=-0.5$}   && \multicolumn{3}{c}{$v=5$, $\lambda=0$} && \multicolumn{3}{c}{$v=5$, $\lambda=-0.5$}    \\

                        \cline{3-5} \cline{7-9} \cline{11-13} \cline{15-17} \\[-2.0ex]
  
                                  &$\alpha=$ & 0.01  &   0.025    & 0.05       && 0.01  &   0.025    & 0.05    && 0.01  &   0.025    & 0.05   && 0.01  &   0.025    & 0.05  \\

\midrule

\multicolumn{17}{l}{Panel A: Mean absolute error} \\

QbSD-gSAV model                            &&          &             &               &&          &             &           &&          &             &           &&          &             &          \\
                           
\quad Individual values of $p$             &&          &             &               &&          &             &           &&          &             &           &&          &             &          \\
                           
\quad \quad $p=0.05$                       &&    0.106 &    0.080    &    0.064      &&    0.138 &    0.099    &    0.077  &&    0.152 &    0.099    &    0.073  &&    0.208 &    0.126    &    0.087  \\
\quad \quad $p=0.10$                       &&    0.106 &    0.078    &    0.061      &&    0.135 &    0.094    &    0.072  &&    0.148 &    0.092    &    0.065  &&    0.201 &    0.117    &    0.079  \\
\bottomrule

\end{tabular*}
\end{minipage}
\end{center}
\end{table}

\end{landscape}

\begin{landscape}

\bigskip
\bigskip

\begin{table}[h]
\begin{center}
\begin{minipage}{8.8in}
\vspace{-1.0cm}
{\small {\bf Table A6.} ES forecasting results with the QbSD approach when $ \theta=0$  and $T=1250$}
\footnotesize
\\[0.5ex]
\begin{tabular*}{\textwidth}{@{\extracolsep{\fill}}  lcrrrcrrrcrrrcrrr}
\toprule

                        &&   \multicolumn{3}{c}{$v=20$, $\lambda=0$}   && \multicolumn{3}{c}{$v=20$, $\lambda=-0.5$}   && \multicolumn{3}{c}{$v=5$, $\lambda=0$} && \multicolumn{3}{c}{$v=5$, $\lambda=-0.5$}    \\

                        \cline{3-5} \cline{7-9} \cline{11-13} \cline{15-17} \\[-2.0ex]
  
                                  &$\alpha=$ & 0.01  &   0.025    & 0.05       && 0.01  &   0.025    & 0.05    && 0.01  &   0.025    & 0.05   && 0.01  &   0.025    & 0.05  \\

\midrule
   
\multicolumn{17}{l}{Panel A: Mean absolute er   ror} \\
   
QbSD-gSAV model                            &&           &             &               &&          &             &           &&          &             &           &&          &              &          \\
                             
\quad Individual values of $p$             &&           &             &               &&          &             &           &&          &             &           &&          &              &          \\
                             
\quad \quad $p=0.05$                       &&     0.144 &    0.105    &    0.087      &&    0.191 &    0.137    &    0.109  &&    0.263 &    0.164    &    0.121  &&    0.367 &     0.222    &    0.160  \\
\quad \quad $p=0.10$                       &&     0.139 &    0.102    &    0.085      &&    0.184 &    0.131    &    0.105  &&    0.254 &    0.155    &    0.114  &&    0.357 &     0.213    &    0.151  \\
\bottomrule

\end{tabular*}
\end{minipage}
\end{center}
\end{table}

\end{landscape}

\begin{landscape}

\bigskip
\bigskip

\begin{table}[h]
\begin{center}
\begin{minipage}{8.8in}
\vspace{-1.0cm}
{\small {\bf Table A7.} VaR forecasting results with the QbSD approach when  $ \theta=0.5$ and $T=1250$ }
\footnotesize
\\[0.5ex]
\begin{tabular*}{\textwidth}{@{\extracolsep{\fill}}  lcrrrcrrrcrrrcrrr}
\toprule

                        &&   \multicolumn{3}{c}{$v=20$, $\lambda=0$}   && \multicolumn{3}{c}{$v=20$, $\lambda=-0.5$}   && \multicolumn{3}{c}{$v=5$, $\lambda=0$} && \multicolumn{3}{c}{$v=5$, $\lambda=-0.5$}    \\

                        \cline{3-5} \cline{7-9} \cline{11-13} \cline{15-17} \\[-2.0ex]
  
                                  &$\alpha=$ & 0.01  &   0.025    & 0.05       && 0.01  &   0.025    & 0.05    && 0.01  &   0.025    & 0.05   && 0.01  &   0.025    & 0.05  \\

\midrule

\multicolumn{17}{l}{Panel A: Mean absolute error} \\

QbSD-gSAV model                            &&          &               &               &&          &             &           &&          &             &           &&          &             &          \\
                             
\quad Individual values of $p$             &&          &               &               &&          &             &           &&          &             &           &&          &             &          \\
                             
\quad \quad $p=0.05$                       &&    0.226 &      0.180    &    0.147      &&    0.286 &    0.218    &    0.171  &&    0.252 &    0.180    &    0.138  &&    0.339 &    0.224    &    0.162  \\
\quad \quad $p=0.10$                       &&    0.225 &\bf   0.178    &    0.145      &&    0.286 &    0.218    &    0.170  &&    0.250 &    0.173    &    0.132  &&    0.339 &    0.222    &    0.157  \\
\bottomrule

\end{tabular*}
\end{minipage}
\end{center}
\end{table}

\end{landscape}

\begin{landscape}

\bigskip
\bigskip

\begin{table}[h]
\begin{center}
\begin{minipage}{8.8in}
\vspace{-1.0cm}
{\small {\bf Table A8.} ES forecasting results with the QbSD approach when  $ \theta=0.5$ and $T=1250$ }
\footnotesize
\\[0.5ex]
\begin{tabular*}{\textwidth}{@{\extracolsep{\fill}}  lcrrrcrrrcrrrcrrr}
\toprule

                        &&   \multicolumn{3}{c}{$v=20$, $\lambda=0$}   && \multicolumn{3}{c}{$v=20$, $\lambda=-0.5$}   && \multicolumn{3}{c}{$v=5$, $\lambda=0$} && \multicolumn{3}{c}{$v=5$, $\lambda=-0.5$}    \\

                        \cline{3-5} \cline{7-9} \cline{11-13} \cline{15-17} \\[-2.0ex]
  
                                  &$\alpha=$ & 0.01  &   0.025    & 0.05       && 0.01  &   0.025    & 0.05    && 0.01  &   0.025    & 0.05   && 0.01  &   0.025    & 0.05  \\

\midrule

\multicolumn{17}{l}{Panel A: Mean absolute error} \\

QbSD-gSAV model                            &&          &             &               &&          &             &           &&          &             &           &&          &              &       \\
                        
\quad Individual values of $p$             &&          &             &               &&          &             &           &&          &             &           &&          &              &       \\
                        
\quad \quad $p=0.05$                       &&    0.272 &    0.223    &    0.190      &&    0.360 &    0.284    &    0.237  &&    0.377 &    0.266    &    0.206  &&    0.539 &     0.357    &    0.268  \\
\quad \quad $p=0.10$                       &&\bf 0.270 &\bf 0.222    &    0.188      &&    0.357 &    0.283    &    0.236  &&    0.368 &    0.257    &    0.198  &&    0.534 &     0.354    &    0.265  \\
\bottomrule

\end{tabular*}
\end{minipage}
\end{center}
\end{table}

\end{landscape}

\begin{landscape}

\bigskip
\bigskip

\begin{table}[h]
\begin{center}
\begin{minipage}{8.8in}
\vspace{-1.0cm}
{\small {\bf Table A9.} VaR forecasting results with the QbSD approach when  $ \theta=0$  and $T=2500$}
\footnotesize
\\[0.5ex]
\begin{tabular*}{\textwidth}{@{\extracolsep{\fill}}  lcrrrcrrrcrrrcrrr}
\toprule

                        &&   \multicolumn{3}{c}{$v=20$, $\lambda=0$}   && \multicolumn{3}{c}{$v=20$, $\lambda=-0.5$}   && \multicolumn{3}{c}{$v=5$, $\lambda=0$} && \multicolumn{3}{c}{$v=5$, $\lambda=-0.5$}    \\

                        \cline{3-5} \cline{7-9} \cline{11-13} \cline{15-17} \\[-2.0ex]
  
                                  &$\alpha=$ & 0.01  &   0.025    & 0.05       && 0.01  &   0.025    & 0.05    && 0.01  &   0.025    & 0.05   && 0.01  &   0.025    & 0.05  \\

\midrule

\multicolumn{17}{l}{Panel A: Mean absolute error} \\
  
QbSD-gSAV model                            &&          &             &               &&          &             &           &&          &             &           &&          &               &          \\
                             
\quad Individual values of $p$             &&          &             &               &&          &             &           &&          &             &           &&          &               &          \\
                             
\quad \quad $p=0.05$                       &&    0.078 &    0.059    &    0.046      &&    0.100 &    0.073    &    0.054  &&    0.116 &    0.075    &    0.054  &&    0.150 &      0.095    &    0.063  \\
\quad \quad $p=0.10$                       &&    0.078 &    0.058    &    0.045      &&    0.098 &    0.072    &\bf 0.052  &&    0.112 &    0.072    &    0.050  &&    0.147 &      0.090    &    0.059  \\
\bottomrule

\end{tabular*}
\end{minipage}
\end{center}
\end{table}

\end{landscape}

\begin{landscape}

\bigskip
\bigskip

\begin{table}[h]
\begin{center}
\begin{minipage}{8.8in}
\vspace{-1.0cm}
{\small {\bf Table A10.} ES forecasting results with the QbSD approach when $ \theta=0$  and $T=2500$}
\footnotesize
\\[0.5ex]
\begin{tabular*}{\textwidth}{@{\extracolsep{\fill}}  lcrrrcrrrcrrrcrrr}
\toprule

                        &&   \multicolumn{3}{c}{$v=20$, $\lambda=0$}   && \multicolumn{3}{c}{$v=20$, $\lambda=-0.5$}   && \multicolumn{3}{c}{$v=5$, $\lambda=0$} && \multicolumn{3}{c}{$v=5$, $\lambda=-0.5$}    \\

                        \cline{3-5} \cline{7-9} \cline{11-13} \cline{15-17} \\[-2.0ex]
  
                                  &$\alpha=$ & 0.01  &   0.025    & 0.05       && 0.01  &   0.025    & 0.05    && 0.01  &   0.025    & 0.05   && 0.01  &   0.025    & 0.05  \\

\midrule

\multicolumn{17}{l}{Panel A: Mean absolute error} \\

QbSD-gSAV model                            &&          &              &               &&          &             &           &&          &             &           &&          &             &          \\
                            
\quad Individual values of $p$             &&          &              &               &&          &             &           &&          &             &           &&          &             &          \\
                            
\quad \quad $p=0.05$                       &&    0.106 &     0.080    &    0.066      &&    0.140 &    0.104    &    0.084  &&    0.203 &    0.131    &    0.097  &&    0.277 &    0.173    &    0.128  \\
\quad \quad $p=0.10$                       &&    0.104 &     0.078    &    0.064      &&    0.137 &\bf 0.101    &\bf 0.081  &&    0.197 &    0.126    &    0.092  &&\bf 0.271 &    0.167    &    0.122  \\
\bottomrule

\end{tabular*}
\end{minipage}
\end{center}
\end{table}

\end{landscape}

\begin{landscape}

\bigskip
\bigskip

\begin{table}[h]
\begin{center}
\begin{minipage}{8.8in}
\vspace{-1.0cm}
{\small {\bf Table A11.} VaR forecasting results with the QbSD approach when  $ \theta=0.5$ and $T=2500$ }
\footnotesize
\\[0.5ex]
\begin{tabular*}{\textwidth}{@{\extracolsep{\fill}}  lcrrrcrrrcrrrcrrr}
\toprule

                        &&   \multicolumn{3}{c}{$v=20$, $\lambda=0$}   && \multicolumn{3}{c}{$v=20$, $\lambda=-0.5$}   && \multicolumn{3}{c}{$v=5$, $\lambda=0$} && \multicolumn{3}{c}{$v=5$, $\lambda=-0.5$}    \\

                        \cline{3-5} \cline{7-9} \cline{11-13} \cline{15-17} \\[-2.0ex]
  
                                  &$\alpha=$ & 0.01  &   0.025    & 0.05       && 0.01  &   0.025    & 0.05    && 0.01  &   0.025    & 0.05   && 0.01  &   0.025    & 0.05  \\

\midrule

\multicolumn{17}{l}{Panel A: Mean absolute error} \\

QbSD-gSAV model                            &&          &             &               &&          &             &           &&          &             &           &&          &             &          \\
                           
\quad Individual values of $p$             &&          &             &               &&          &             &           &&          &             &           &&          &             &          \\
                           
\quad \quad $p=0.05$                       &&\bf 0.212 &\bf 0.174    &\bf 0.141      &&\bf 0.262 &    0.206    &\bf 0.162  &&\bf 0.229 &    0.171    &    0.130  &&\bf 0.293 &    0.203    &\bf 0.147  \\
\quad \quad $p=0.10$                       &&    0.214 &\bf 0.174    &\bf 0.141      &&\bf 0.262 &\bf 0.205    &\bf 0.162  &&    0.230 &    0.169    &    0.129  &&\bf 0.293 &\bf 0.202    &\bf 0.147  \\
\bottomrule

\end{tabular*}
\end{minipage}
\end{center}
\end{table}

\end{landscape}

\begin{landscape}

\begin{table}[h]
\begin{center}
\begin{minipage}{8.8in}
\vspace{-1.0cm}
{\small {\bf Table A12.} ES forecasting results with the QbSD approach when  $ \theta=0.5$ and $T=2500$ }
\footnotesize
\\[0.5ex]
\begin{tabular*}{\textwidth}{@{\extracolsep{\fill}}  lcrrrcrrrcrrrcrrr}
\toprule

                        &&   \multicolumn{3}{c}{$v=20$, $\lambda=0$}   && \multicolumn{3}{c}{$v=20$, $\lambda=-0.5$}   && \multicolumn{3}{c}{$v=5$, $\lambda=0$} && \multicolumn{3}{c}{$v=5$, $\lambda=-0.5$}    \\

                        \cline{3-5} \cline{7-9} \cline{11-13} \cline{15-17} \\[-2.0ex]
  
                                  &$\alpha=$ & 0.01  &   0.025    & 0.05       && 0.01  &   0.025    & 0.05    && 0.01  &   0.025    & 0.05   && 0.01  &   0.025    & 0.05  \\

\midrule

\multicolumn{17}{l}{Panel A: Mean absolute error} \\

QbSD-gSAV model                            &&          &              &               &&          &             &           &&          &             &           &&         &              &          \\
                            
\quad Individual values of $p$             &&          &              &               &&          &             &           &&          &             &           &&         &              &          \\
                            
\quad \quad $p=0.05$                       &&\bf 0.250 & \bf 0.212    &    0.183      &&\bf 0.320 &    0.263    &    0.223  &&    0.326 &    0.242    &    0.193  &&   0.449 &     0.317    &    0.246  \\
\quad \quad $p=0.10$                       &&    0.251 & \bf 0.212    &\bf 0.182      &&    0.319 &\bf 0.262    &\bf 0.222  &&\bf 0.323 &\bf 0.241    &\bf 0.191  &&\bf0.444 &\bf  0.314    &    0.243  \\
\bottomrule

\end{tabular*}
\end{minipage}
\end{center}
\end{table}

\end{landscape}

\section*{Section B}

Tables B1--B8 report one-step-ahead VaR/ES forecast errors (MAE, RMSE) for the proposed QbSD models (gSAV, gAS) and benchmarks (skew-$t$ GARCH/GJR/EGARCH, AL-based joint VaR-ES, and GAS) across data-generating designs varying by $T\in\{250,2500\}$, $v\in\{5,20\}$, $\lambda\in\{0,-0.5\}$, $\theta\in\{0,0.5\}$, and $\alpha\in\{0.01,0.025,0.05\}$. Minima are bolded.

For $T=250$ without leverage (Tables B1--B2), conventional GARCH-type models dominate. With symmetric innovations, Normal/Student-$t$ GARCH/GJR achieve most minima; under left-skew, their skew-$t$ counterparts lead. QbSD models are competitive but rarely best, while AL-based estimators and GAS trail.

Introducing leverage at $T=250$ (Tables B3--B4) shifts the lead within the GARCH family: EGARCH frequently attains the lowest errors in symmetric cases, and \sloppy skew-$t$ GARCH/GJR often prevail under skewed or heavy-tailed DGPs. QbSD-gAS improves relative to QbSD-gSAV but seldom displaces the top GARCH variants at this sample size.

With larger samples and no leverage (Tables B5--B6, $T=2500$, $\theta=0$), GARCH/GJR remain the primary winners. Student-$t$ specifications account for many minima under symmetry and heavy tails, whereas skew-$t$ versions excel when $\lambda=-0.5$. QbSD narrows the gap but generally does not surpass the best GARCH-type models here.

With larger samples under leverage (Tables B7--B8, $T=2500$, $\theta=0.5$), QbSD-gAS delivers many of the lowest ES MAE/RMSE values across $\alpha$ and $(v,\lambda)$ settings, outperforming AL and GAS and often surpassing standard GARCH models. Two caveats remain: (i) under symmetric innovations, EGARCH can match or exceed QbSD-gAS for ES RMSE at some $\alpha$ (e.g., $v=20$, $\lambda=0$ at $2.5\%$ and $5\%$); and (ii) for VaR the best model can shift with tail level---EGARCH under symmetry and skew-$t$ GJR-GARCH/GARCH under skewed, heavy-tailed designs---while QbSD-gAS remains the most robust ES performer under leverage.

Overall, while GARCH-type models (especially skew-$t$ and EGARCH) are very strong at $T=250$ and in several $T=2500$ designs without leverage, QbSD-gAS becomes particularly effective for ES once leverage is present and the sample is large, delivering broad gains except in the most extreme skew/heavy-tail scenario. AL-based estimators and GAS underperform across configurations.

\begin{landscape}

\bigskip
\bigskip

\begin{table}[h]
\begin{center}
\begin{minipage}{8.8in}
\vspace{-1.4cm}
{\small {\bf Table B1.} VaR forecasting results when $ \theta=0$  and $T=250$}
\footnotesize
\\[0.5ex]
\begin{tabular*}{\textwidth}{@{\extracolsep{\fill}}  lcrrrcrrrcrrrcrrr}
\toprule

                        &&   \multicolumn{3}{c}{$v=20$, $\lambda=0$}   && \multicolumn{3}{c}{$v=20$, $\lambda=-0.5$}   && \multicolumn{3}{c}{$v=5$, $\lambda=0$} && \multicolumn{3}{c}{$v=5$, $\lambda=-0.5$}    \\

                        \cline{3-5} \cline{7-9} \cline{11-13} \cline{15-17} \\[-2.0ex]
  
                                   &$\alpha=$ & 0.01  &   0.025    & 0.05       && 0.01  &   0.025    & 0.05    && 0.01  &   0.025    & 0.05   && 0.01  &   0.025    & 0.05  \\

\midrule

\multicolumn{17}{l}{Panel A: Mean absolute error} \\

QbSD approach                               &&        &       &        &&        &       &        &&        &       &        &&        &       &        \\
\quad QbSD-gSAV                              && 0.218  & 0.165 & 0.131  && 0.281  & 0.198 & 0.154  && 0.300  & 0.193 & 0.138  && 0.403  & 0.239 & 0.167  \\
\quad QbSD-gAS                               && 0.241  & 0.182 & 0.147  && 0.304  & 0.216 & 0.169  && 0.318  & 0.208 & 0.153  && 0.431  & 0.259 & 0.180  \\
AL approach                                  &&        &       &        &&        &       &        &&        &       &        &&        &       &        \\
\bottomrule

\end{tabular*}
\end{minipage}
\end{center}
\end{table}

\end{landscape}

\begin{landscape}

\bigskip
\bigskip

\begin{table}[h]
\begin{center}
\begin{minipage}{8.8in}
\vspace{-1.4cm}
{\small {\bf Table B2.} ES forecasting results when $ \theta=0$   and $T=250$}
\footnotesize
\\[0.5ex]
\begin{tabular*}{\textwidth}{@{\extracolsep{\fill}}  lcrrrcrrrcrrrcrrr}
\toprule

                        &&   \multicolumn{3}{c}{$v=20$, $\lambda=0$}   && \multicolumn{3}{c}{$v=20$, $\lambda=-0.5$}   && \multicolumn{3}{c}{$v=5$, $\lambda=0$} && \multicolumn{3}{c}{$v=5$, $\lambda=-0.5$}    \\

                        \cline{3-5} \cline{7-9} \cline{11-13} \cline{15-17} \\[-2.0ex]
  
                              &$\alpha=$ & 0.01  &   0.025    & 0.05       && 0.01  &   0.025    & 0.05    && 0.01  &   0.025    & 0.05   && 0.01  &   0.025    & 0.05  \\

\midrule

\multicolumn{17}{l}{Panel A: Mean absolute error} \\

QbSD approach                               &&        &       &       &&       &       &       &&       &       &       &&       &       &       \\
\quad QbSD-gSAV                              && 0.300 & 0.219 & 0.174 && 0.407 & 0.283 & 0.219 && 0.522 & 0.322 & 0.229 && 0.745 & 0.437 & 0.298 \\
\quad QbSD-gAS                               && 0.322 & 0.241 & 0.195 && 0.432 & 0.306 & 0.238 && 0.545 & 0.342 & 0.246 && 0.779 & 0.466 & 0.321 \\
AL approach                                  &&        &       &       &&       &       &       &&       &       &       &&       &       &       \\
\bottomrule
\end{tabular*}
\end{minipage}
\end{center}
\end{table}

\end{landscape}

\begin{landscape}

\bigskip
\bigskip

\begin{table}[h]
\begin{center}
\begin{minipage}{8.8in}
\vspace{-1.4cm}
{\small {\bf Table B3.} VaR forecasting results when $ \theta=0.5$  and $T=250$}
\footnotesize
\\[0.5ex]
\begin{tabular*}{\textwidth}{@{\extracolsep{\fill}}  lcrrrcrrrcrrrcrrr}
\toprule

                        &&   \multicolumn{3}{c}{$v=20$, $\lambda=0$}   && \multicolumn{3}{c}{$v=20$, $\lambda=-0.5$}   && \multicolumn{3}{c}{$v=5$, $\lambda=0$} && \multicolumn{3}{c}{$v=5$, $\lambda=-0.5$}    \\

                        \cline{3-5} \cline{7-9} \cline{11-13} \cline{15-17} \\[-2.0ex]
  
                                   &$\alpha=$ & 0.01  &   0.025    & 0.05       && 0.01  &   0.025    & 0.05    && 0.01  &   0.025    & 0.05   && 0.01  &   0.025    & 0.05  \\

\midrule

\multicolumn{17}{l}{Panel A: Mean absolute error} \\

QbSD approach                               &&          &             &               &&          &             &           &&          &             &           &&          &             &          \\
\quad QbSD-gSAV                             &&    0.302 &    0.226    &    0.185      &&    0.396 &    0.289    &    0.229  &&    0.373 &    0.243    &    0.181  &&    0.545 &    0.344    &    0.240  \\
\quad QbSD-gAS                              &&    0.262 &    0.196    &    0.159      &&    0.357 &    0.260    &    0.204  &&    0.356 &    0.232    &    0.172  &&    0.547 &    0.331    &    0.235  \\
AL approach                                 &&          &             &               &&          &             &           &&          &             &           &&          &             &          \\
\bottomrule

\end{tabular*}
\end{minipage}
\end{center}
\end{table}

\end{landscape}

\begin{landscape}

\bigskip
\bigskip

\begin{table}[h]
\begin{center}
\begin{minipage}{8.8in}
\vspace{-1.4cm}
{\small {\bf Table B4.} ES forecasting results when $ \theta=0.5$   and $T=250$}
\footnotesize
\\[0.5ex]
\begin{tabular*}{\textwidth}{@{\extracolsep{\fill}}  lcrrrcrrrcrrrcrrr}
\toprule

                        &&   \multicolumn{3}{c}{$v=20$, $\lambda=0$}   && \multicolumn{3}{c}{$v=20$, $\lambda=-0.5$}   && \multicolumn{3}{c}{$v=5$, $\lambda=0$} && \multicolumn{3}{c}{$v=5$, $\lambda=-0.5$}    \\

                        \cline{3-5} \cline{7-9} \cline{11-13} \cline{15-17} \\[-2.0ex]
  
                              &$\alpha=$ & 0.01  &   0.025    & 0.05       && 0.01  &   0.025    & 0.05    && 0.01  &   0.025    & 0.05   && 0.01  &   0.025    & 0.05  \\

\midrule

\multicolumn{17}{l}{Panel A: Mean absolute error} \\

QbSD approach                               &&          &             &               &&          &             &           &&          &             &           &&          &             &          \\
\quad QbSD-gSAV                             &&    0.396 &    0.300    &    0.243      &&    0.540 &    0.397    &    0.314  &&    0.633 &    0.402    &    0.289  &&    0.984 &    0.594    &    0.413  \\
\quad QbSD-gAS                              &&    0.348 &    0.258    &    0.209      &&    0.497 &    0.356    &    0.280  &&    0.595 &    0.376    &    0.273  &&    0.979 &    0.589    &    0.410  \\
AL approach                                 &&          &             &               &&          &             &           &&          &             &           &&          &             &          \\
\bottomrule
\end{tabular*}
\end{minipage}
\end{center}
\end{table}

\end{landscape}

\begin{landscape}

\bigskip
\bigskip

\begin{table}[h]
\begin{center}
\begin{minipage}{8.8in}
\vspace{-1.4cm}
{\small {\bf Table B5.} VaR forecasting results when $ \theta=0$  and $T=2500$}
\footnotesize
\\[0.5ex]
\begin{tabular*}{\textwidth}{@{\extracolsep{\fill}}  lcrrrcrrrcrrrcrrr}
\toprule

                        &&   \multicolumn{3}{c}{$v=20$, $\lambda=0$}   && \multicolumn{3}{c}{$v=20$, $\lambda=-0.5$}   && \multicolumn{3}{c}{$v=5$, $\lambda=0$} && \multicolumn{3}{c}{$v=5$, $\lambda=-0.5$}    \\

                        \cline{3-5} \cline{7-9} \cline{11-13} \cline{15-17} \\[-2.0ex]
  
                                   &$\alpha=$ & 0.01  &   0.025    & 0.05       && 0.01  &   0.025    & 0.05    && 0.01  &   0.025    & 0.05   && 0.01  &   0.025    & 0.05  \\

\midrule

\multicolumn{17}{l}{Panel A: Mean absolute error} \\

QbSD approach                               &&        &       &        &&        &       &        &&        &       &        &&        &       &        \\
\quad QbSD-gSAV                             && 0.078  & 0.057 & 0.046  && 0.099  & 0.070 & 0.054  && 0.112  & 0.071 & 0.052  && 0.150  & 0.090 & 0.062  \\
\quad QbSD-gAS                              && 0.088  & 0.066 & 0.053  && 0.107  & 0.077 & 0.059  && 0.121  & 0.079 & 0.058  && 0.154  & 0.095 & 0.065  \\
AL approach                                 &&        &       &        &&        &       &        &&        &       &        &&        &       &        \\
\bottomrule

\end{tabular*}
\end{minipage}
\end{center}
\end{table}

\end{landscape}

\begin{landscape}

\bigskip
\bigskip

\begin{table}[h]
\begin{center}
\begin{minipage}{8.8in}
\vspace{-1.4cm}
{\small {\bf Table B6.} ES forecasting results when $ \theta=0$   and $T=2500$}
\footnotesize
\\[0.5ex]
\begin{tabular*}{\textwidth}{@{\extracolsep{\fill}}  lcrrrcrrrcrrrcrrr}
\toprule

                        &&   \multicolumn{3}{c}{$v=20$, $\lambda=0$}   && \multicolumn{3}{c}{$v=20$, $\lambda=-0.5$}   && \multicolumn{3}{c}{$v=5$, $\lambda=0$} && \multicolumn{3}{c}{$v=5$, $\lambda=-0.5$}    \\

                        \cline{3-5} \cline{7-9} \cline{11-13} \cline{15-17} \\[-2.0ex]
  
                              &$\alpha=$     & 0.01  &   0.025    & 0.05       && 0.01  &   0.025    & 0.05    && 0.01  &   0.025    & 0.05   && 0.01  &   0.025    & 0.05  \\

\midrule

\multicolumn{17}{l}{Panel A: Mean absolute error} \\

QbSD approach                               &&          &       &       &&          &       &       &&          &       &       &&          &       &        \\
\quad QbSD-gSAV                             &&    0.104 & 0.079 & 0.064 &&    0.138 & 0.100 & 0.081 &&    0.199 & 0.126 & 0.092 &&    0.278 & 0.172 & 0.123 \\
\quad QbSD-gAS                              &&    0.113 & 0.086 & 0.072 &&    0.144 & 0.107 & 0.087 &&    0.202 & 0.131 & 0.097 &&    0.280 & 0.173 & 0.126 \\
AL approach                                 &&          &       &       &&          &       &       &&          &       &       &&          &       &        \\
\bottomrule
\end{tabular*}
\end{minipage}
\end{center}
\end{table}

\end{landscape}

\newpage

\begin{landscape}

\bigskip
\bigskip

\begin{table}[h]
\begin{center}
\begin{minipage}{8.8in}
\vspace{-1.4cm}
{\small {\bf Table B7.} VaR forecasting results when $ \theta=0.5$  and $T=2500$}
\footnotesize
\\[0.5ex]
\begin{tabular*}{\textwidth}{@{\extracolsep{\fill}}  lcrrrcrrrcrrrcrrr}
\toprule

                        &&   \multicolumn{3}{c}{$v=20$, $\lambda=0$}   && \multicolumn{3}{c}{$v=20$, $\lambda=-0.5$}   && \multicolumn{3}{c}{$v=5$, $\lambda=0$} && \multicolumn{3}{c}{$v=5$, $\lambda=-0.5$}    \\

                        \cline{3-5} \cline{7-9} \cline{11-13} \cline{15-17} \\[-2.0ex]
  
                                   &$\alpha=$ & 0.01  &   0.025    & 0.05       && 0.01  &   0.025    & 0.05    && 0.01  &   0.025    & 0.05   && 0.01  &   0.025    & 0.05  \\

\midrule

\multicolumn{17}{l}{Panel A: Mean absolute error} \\

QbSD approach                               &&        &       &        &&        &       &        &&        &       &        &&        &       &        \\
\quad QbSD-gSAV                              && 0.209  & 0.168 & 0.137  && 0.258  & 0.202 & 0.161  && 0.226  & 0.166 & 0.127  && 0.296  & 0.204 & 0.150  \\
\quad QbSD-gAS                               && 0.096  & 0.072 & 0.057  && \bf 0.135  & \bf 0.099 & \bf 0.077  && \bf 0.130  & 0.086 & \bf 0.063  && \bf 0.196  & \bf 0.126 & \bf 0.088  \\
AL approach                                 &&        &       &        &&        &       &        &&        &       &        &&        &       &        \\
\bottomrule

\end{tabular*}
\end{minipage}
\end{center}
\end{table}

\end{landscape}

\begin{landscape}

\bigskip
\bigskip

\begin{table}[h]
\begin{center}
\begin{minipage}{8.8in}
\vspace{-1.4cm}
{\small {\bf Table B8.} ES forecasting results when $ \theta=0.5$   and $T=2500$}
\footnotesize
\\[0.5ex]
\begin{tabular*}{\textwidth}{@{\extracolsep{\fill}}  lcrrrcrrrcrrrcrrr}
\toprule

                        &&   \multicolumn{3}{c}{$v=20$, $\lambda=0$}   && \multicolumn{3}{c}{$v=20$, $\lambda=-0.5$}   && \multicolumn{3}{c}{$v=5$, $\lambda=0$} && \multicolumn{3}{c}{$v=5$, $\lambda=-0.5$}    \\

                        \cline{3-5} \cline{7-9} \cline{11-13} \cline{15-17} \\[-2.0ex]
  
                              &$\alpha=$     & 0.01  &   0.025    & 0.05       && 0.01  &   0.025    & 0.05    && 0.01  &   0.025    & 0.05   && 0.01  &   0.025    & 0.05  \\

\midrule

\multicolumn{17}{l}{Panel A: Mean absolute error} \\

QbSD approach                               &&          &             &               &&          &             &           &&          &              &           &&         &             &          \\
\quad QbSD-gSAV                             &&    0.251 &    0.208    &    0.178      &&    0.319 &    0.259    &    0.217  &&    0.326 &    0.236     &    0.185  &&    0.459 &    0.319    &    0.243  \\
\quad QbSD-gAS                              &&\bf 0.123 &    0.095    &    0.078      &&\bf 0.178 &\bf 0.135    &\bf 0.109  &&\bf 0.216 &\bf 0.142     &\bf 0.106  &&\bf 0.335 &\bf 0.216    &\bf 0.157  \\
AL approach                                 &&          &             &               &&          &             &           &&          &              &           &&         &             &          \\
\bottomrule

\end{tabular*}
\end{minipage}
\end{center}
\end{table}

\end{landscape}

\section*{Section C}

This appendix reports additional rolling-window evaluations with $R=250$ and $R=2500$. Tables C1--C12 present 90\% Model Confidence Set (MCS) rankings for VaR and joint VaR-ES at the 1\%, 2.5\%, and 5\% levels.

Across windows, indices, and tails, the quantile-based scale-dynamics class performs very strongly overall. QbSD-gAS is the front-runner with $R=250$: it ranks first in most VaR and joint VaR-ES exercises and is almost always in the top two. Its QAR extension (QAR-QbSD-gAS) typically follows closely-often finishing in the top three---but its long-window 1\% VaR results slip on a few indices (see Table C7).

With the short window ($R=250$), patterns are clear. For VaR-only (quantile scores; Tables C1--C3), QbSD-gAS is the \emph{overall} top performer at 1\%, 2.5\%, and 5\%. For joint VaR-ES (AL log scores; Tables C4--C6), skewed-$t$ GARCH/GJR lead at 1\% (Table C4) with QbSD-gAS a close second, while QbSD-gAS returns to the top at 2.5\% and 5\% (Tables C5--C6).

With the long window ($R=2500$), leadership varies by tail and score. For VaR (Tables C7--C9), AL-based specifications dominate at 1\% (Table C7), QbSD-gAS ranks first at 2.5\% (Table C8), and at 5\% EGARCH narrowly takes the lead, with QbSD-gAS essentially tied on average but second in the final ranking (Table C9); note that both have the same average rank (1.9), with QbSD-gAS recording more \#1 finishes (4 of 8) while EGARCH is slightly more uniform across indices. For joint VaR-ES (Tables C10--C12), AL-based specifications lead at 1\% and 2.5\% (Tables C10--C11), whereas QbSD-gAS regains first place at 5\% (Table C12).

Among benchmarks, skewed-$t$ GARCH/GJR are the strongest non-QbSD competitors and are frequently retained in the MCS, especially for $R=250$ and in VaR-only comparisons. Normal-based GARCH variants generally underperform on average, with two caveats: (i) EGARCH is a strong contender at 5\% VaR for $R=2500$ (Table C9), and (ii) GJR-GARCH-normal is competitive at $R=250$ for 2.5\% and 5\% VaR (Tables C2--C3). Within the AL-based family, AS specifications consistently outperform their SAV counterparts. The gSAV variant of QbSD (QbSD-gSAV) typically trails QbSD-gAS, while QAR-QbSD-gSAV most often attains mid-table results.

Overall, these results align with the paper’s main finding: QbSD-gAS is frequently the strongest---especially with the short window---and remains consistently competitive across indices and tail levels under the long window; QAR-QbSD-gAS is often close behind. At the extreme tail and with long windows, AL-based variants tend to lead, and EGARCH is a notable runner-up at 5\% VaR.

\bigskip
\bigskip

\begin{table}[!h]
\hspace*{-0.65cm}
\begin{minipage}{7.1in}
{\small {\bf Table C1.} Ranking of 1\% VaR forecasting models using the MCS procedure with quantile scores, rolling window $R=250$}
\begin{center}
\footnotesize
\vspace{-0.25cm}
\begin{tabular*}{\textwidth}{@{\extracolsep{\fill}} lcccccccc|ccc}
\toprule
                     & \multirow{2}{*}{S\&P} & \multirow{2}{*}{DJIA} & \multirow{2}{*}{NASDAQ} & \multirow{2}{*}{STOXX} & \multirow{2}{*}{FTSE} & \multirow{2}{*}{DAX} & \multirow{2}{*}{CAC} & \multirow{2}{*}{TSX} & \multirow{2}{*}{\#} & Avg. & Final \\
                     &     &     &     &     &     &     &     &     &     & rank  & rank  \\
\midrule 
QbSD-gAS                    &  2 &  4 &  2 &  1 &  2 &  1 &  1 &  4 &  8 &  2.1 &  1 \\
QAR-QbSD-gAS                &  5 &  6 &  6 &  7 &  1 &  5 &  2 &  1 &  8 &  4.1 &  2 \\
GARCH-skew-$t$              &  1 &  1 &  1 & 10 &  3 & 11 &  4 &  3 &  8 &  4.2 &  3 \\
GJR-GARCH-skew-$t$          &  3 &  3 &  5 &  2 &  5 & 10 &  6 &  2 &  8 &  4.5 &  4 \\
GJR-GARCH-$t$               & 10 &  2 &  9 &  4 &  8 &  3 &  7 &  8 &  8 &  6.4 &  5 \\
AR-GARCH-skew-$t$           &  6 &  5 &  4 &  5 &  4 & 12 & 12 &  5 &  8 &  6.6 &  6 \\
\bottomrule
\end{tabular*}
\vspace{-0.5cm}
\end{center}
\end{minipage}
\end{table}

\begin{table}[!h]
\hspace*{-0.65cm}
\begin{minipage}{7.1in}
{\small {\bf Table C2.} Ranking of 2.5\% VaR forecasting models using the MCS procedure with quantile scores, rolling window $R=250$}
\begin{center}
\footnotesize
\vspace{-0.25cm}
\begin{tabular*}{\textwidth}{@{\extracolsep{\fill}} lcccccccc|ccc}
\toprule
                     & \multirow{2}{*}{S\&P} & \multirow{2}{*}{DJIA} & \multirow{2}{*}{NASDAQ} & \multirow{2}{*}{STOXX} & \multirow{2}{*}{FTSE} & \multirow{2}{*}{DAX} & \multirow{2}{*}{CAC} & \multirow{2}{*}{TSX} & \multirow{2}{*}{\#} & Avg. & Final \\
                     &     &     &     &     &     &     &     &     &     & rank  & rank  \\
\midrule
QbSD-gAS                    &  2 &  1 &  1 &  1 &  2 &  1 &  1 &  2 &  8 &  1.4 &  1 \\
GARCH-skew-$t$              &  3 &  2 &  2 &  4 &  4 &  6 &  8 &  3 &  8 &  4.0 &  2 \\
QAR-QbSD-gAS                &  1 &  5 & 18 &  3 &  1 &  4 &  2 &  1 &  8 &  4.4 &  3 \\
GJR-GARCH-normal            &  5 &  4 &  5 & 11 &  3 &  2 &  3 &  7 &  8 &  5.0 &  4 \\
GARCH-$t$                   &  8 &  7 &  7 &  9 &  9 &  5 &  4 &  9 &  8 &  7.2 &  5 \\
GJR-GARCH-$t$               & 11 &  6 &  9 & 10 &  7 &  3 &  5 &  8 &  8 &  7.4 &  6 \\
\bottomrule
\end{tabular*}
\vspace{-0.25cm}
\end{center}
\end{minipage}
\end{table}

\begin{table}[!h]
\hspace*{-0.65cm}
\begin{minipage}{7.1in}
{\small {\bf Table C3.} Ranking of 5\% VaR forecasting models using the MCS procedure with quantile scores, rolling window $R=250$}
\begin{center}
\footnotesize
\vspace{-0.25cm}
\begin{tabular*}{\textwidth}{@{\extracolsep{\fill}} lcccccccc|ccc}
\toprule
                     & \multirow{2}{*}{S\&P} & \multirow{2}{*}{DJIA} & \multirow{2}{*}{NASDAQ} & \multirow{2}{*}{STOXX} & \multirow{2}{*}{FTSE} & \multirow{2}{*}{DAX} & \multirow{2}{*}{CAC} & \multirow{2}{*}{TSX} & \multirow{2}{*}{\#} & Avg. & Final \\
                     &     &     &     &     &     &     &     &     &     & rank  & rank  \\
\midrule 
QbSD-gAS                     &  3 &  1 &  2 &  1 &  1 &  1 &  1 &  3 &  8 &  1.6 &  1 \\
QAR-QbSD-gAS                 &  1 &  2 &  6 &  3 &  2 &  2 &    &  1 &  7 &  5.6 &  2 \\
GARCH-skew-$t$               &  4 &  3 &  1 &  2 &  6 &  6 &    &  2 &  7 &  6.5 &  3 \\
GJR-GARCH-normal             &  2 &  4 &  4 &  5 &  4 &  3 &    &  7 &  7 &  7.1 &  4 \\
GJR-GARCH-skew-$t$           &  5 &  5 &  3 &  4 & 12 & 10 &    &  6 &  7 &  9.1 &  5 \\
EGARCH                       &  7 & 13 &  5 & 13 &  3 &  5 &    &  8 &  7 & 10.2 &  6 \\
\bottomrule
\end{tabular*}
\vspace{-0.25cm}
\end{center}
\end{minipage}
\end{table}

\begin{table}[!h]
\hspace*{-0.65cm}
\begin{minipage}{7.1in}
{\small {\bf Table C4.} Ranking of 1\% VaR and ES forecasting models based on the MCS procedure using AL log scores, rolling window $R=250$ }
\begin{center}
\footnotesize
\vspace{-0.25cm}
\begin{tabular*}{\textwidth}{@{\extracolsep{\fill}} lcccccccc|ccc}
\toprule
                     & \multirow{2}{*}{S\&P} & \multirow{2}{*}{DJIA} & \multirow{2}{*}{NASDAQ} & \multirow{2}{*}{STOXX} & \multirow{2}{*}{FTSE} & \multirow{2}{*}{DAX} & \multirow{2}{*}{CAC} & \multirow{2}{*}{TSX} & \multirow{2}{*}{\#} & Avg. & Final \\
                     &     &     &     &     &     &     &     &     &     & rank  & rank  \\
\midrule
GJR-GARCH-skew-$t$         &  2 &  4 &  2 &  1 & 13 & 11 &  3 &  2 &  8 &  4.8 &  1 \\
QbSD-gAS                   &  9 &  2 &  6 &  5 &  6 &  1 &  1 & 10 &  8 &  5.0 &  2 \\
GARCH-skew-$t$             &  1 &  1 &  1 &  3 &  9 & 13 &  4 &  9 &  8 &  5.1 &  3 \\
QAR-QbSD-gAS               &  8 &  7 &  7 & 16 &  1 &  5 &  2 &  1 &  8 &  5.9 &  4 \\
AR-AL$_{\text{Mult.}}$-AS  &  6 &  3 &  5 &  9 &  8 &  7 &  9 &  3 &  8 &  6.2 &  5 \\
AL$_{\text{Mult.}}$-AS     & 10 &  8 &  3 & 10 & 10 & 12 & 10 &  4 &  8 &  8.4 &  6 \\
\bottomrule
\end{tabular*}
\vspace{-0.5cm}
\end{center}
\end{minipage}
\end{table}

\begin{table}[!h]
\hspace*{-0.65cm}
\begin{minipage}{7.1in}
{\small {\bf Table C5.} Ranking of 2.5\% VaR and ES forecasting models based on the MCS procedure using AL log scores, rolling window $R=250$ }
\begin{center}
\footnotesize
\vspace{-0.25cm}
\begin{tabular*}{\textwidth}{@{\extracolsep{\fill}} lcccccccc|ccc}
\toprule
                     & \multirow{2}{*}{S\&P} & \multirow{2}{*}{DJIA} & \multirow{2}{*}{NASDAQ} & \multirow{2}{*}{STOXX} & \multirow{2}{*}{FTSE} & \multirow{2}{*}{DAX} & \multirow{2}{*}{CAC} & \multirow{2}{*}{TSX} & \multirow{2}{*}{\#} & Avg. & Final \\
                     &     &     &     &     &     &     &     &     &     & rank  & rank  \\
\midrule 
QbSD-gAS                   &  2 &  1 &  2 &  2 &  7 &  1 &  1 &  7 &  8 &  2.9 &  1 \\
QAR-QbSD-gAS               &  1 &  6 &  7 & 11 &  1 &  2 &  5 &  1 &  8 &  4.2 &  2 \\
GARCH-skew-$t$             &  3 &  2 &  1 &  6 &    &  7 &  9 &  4 &  7 &  7.5 &  3 \\
GJR-GARCH-skew-$t$         &  4 &  3 &  3 &  4 &    &  9 & 10 &  3 &  7 &  8.0 &  4 \\
AR-AL$_{\text{Mult.}}$-AS  &  6 &    &  5 &  9 &  6 &  5 &  7 &  2 &  7 &  8.5 &  5 \\
AL$_{\text{Mult.}}$-AS     &  8 &  7 &  6 &  7 &  3 &  4 &  8 &    &  7 &  8.9 &  6 \\
\bottomrule
\end{tabular*}
\vspace{-0.5cm}
\end{center}
\end{minipage}
\end{table}

\begin{table}[!h]
\hspace*{-0.65cm}
\begin{minipage}{7.1in}
{\small {\bf Table C6.} Ranking of 5\% VaR and ES forecasting models based on the MCS procedure using AL log scores, rolling window $R=250$ }
\begin{center}
\footnotesize
\vspace{-0.25cm}
\begin{tabular*}{\textwidth}{@{\extracolsep{\fill}} lcccccccc|ccc}
\toprule
                     & \multirow{2}{*}{S\&P} & \multirow{2}{*}{DJIA} & \multirow{2}{*}{NASDAQ} & \multirow{2}{*}{STOXX} & \multirow{2}{*}{FTSE} & \multirow{2}{*}{DAX} & \multirow{2}{*}{CAC} & \multirow{2}{*}{TSX} & \multirow{2}{*}{\#} & Avg. & Final \\
                     &     &     &     &     &     &     &     &     &     & rank  & rank  \\
\midrule 
QbSD-gAS                   &  2 &  1 &  1 &  1 &  2 &  1 &  1 & 10 &  8 &  2.4 &  1 \\
QAR-QbSD-gAS               &  1 &  4 &  6 &  6 &  1 &  5 &  8 &  1 &  8 &  4.0 &  2 \\
AL$_{\text{Mult.}}$-AS     &  5 &  5 &  5 &  8 &  5 &  9 &  5 &  4 &  8 &  5.8 &  3 \\
AR-AL$_{\text{Mult.}}$-AS  &  4 &     &  4 &  7 &  6 &  8 &  7 &  3 &  7 &  8.4 &  4 \\
GARCH-skew-$t$             &  6 &  3 &  2 &  3 &     &     &     &  6 &  5 & 13.0 &  5 \\
GJR-GARCH-skew-$t$         &  7 &  6 &  3 &  2 &     &     &     &  7 &  5 & 13.6 &  6 \\
\bottomrule
\end{tabular*}
\vspace{-0.25cm}
\end{center}
\end{minipage}
\end{table}

\begin{table}[!h]
\begin{minipage}{7.0in}
{\small {\bf Table C7.} Ranking of 1\% VaR forecasting models using the MCS procedure with quantile scores, rolling window $R=2500$}
\begin{center}
\footnotesize
\vspace{-0.25cm}
\begin{tabular*}{\textwidth}{@{\extracolsep{\fill}} lcccccccc|ccc}
\toprule
                     & \multirow{2}{*}{S\&P} & \multirow{2}{*}{DJIA} & \multirow{2}{*}{NASDAQ} & \multirow{2}{*}{STOXX} & \multirow{2}{*}{FTSE} & \multirow{2}{*}{DAX} & \multirow{2}{*}{CAC} & \multirow{2}{*}{TSX} & \multirow{2}{*}{\#} & Avg. & Final \\
                     &     &     &     &     &     &     &     &     &     & rank  & rank  \\
\midrule
AR-AL$_{\text{AR}}$-AS         & 1  & 1  & 1  & 9  & 3  & 4  & 6  & 5  & 8  & 3.8  & 1 \\
AR-AL$_{\text{Mult.}}$-AS      & 7  & 9  & 6  & 14 & 6  & 6  & 4  & 2  & 8  & 6.8  & 2 \\
AL$_{\text{Mult.}}$-AS         & 18 & 2  & 7  & 17 & 4  & 2  & 3  & 3  & 8  & 7.0  & 3 \\
AL$_{\text{AR}}$-AS            & 4  & 8  & 9  & 13 & 2  & 20 & 2  & 4  & 8  & 7.8  & 4 \\
QbSD-gAS                       &    & 6  & 25 & 1  & 1  & 1  & 1  & 6  & 7  & 8.6  & 5 \\
AR-AL$_{\text{Mult.}}$-SAV     & 2  & 3  & 3  & 16 & 16 & 18 & 7  & 8  & 8  & 9.1  & 6 \\
\bottomrule
\end{tabular*}
\vspace{-0.5cm}
\end{center}
\end{minipage}
\end{table}

\begin{table}[!h]
\begin{minipage}{7.0in}
{\small {\bf Table C8.} Ranking of 2.5\% VaR forecasting models using the MCS procedure with quantile scores, rolling window $R=2500$}
\begin{center}
\footnotesize
\vspace{-0.25cm}
\begin{tabular*}{\textwidth}{@{\extracolsep{\fill}} lcccccccc|ccc}
\toprule
                     & \multirow{2}{*}{S\&P} & \multirow{2}{*}{DJIA} & \multirow{2}{*}{NASDAQ} & \multirow{2}{*}{STOXX} & \multirow{2}{*}{FTSE} & \multirow{2}{*}{DAX} & \multirow{2}{*}{CAC} & \multirow{2}{*}{TSX} & \multirow{2}{*}{\#} & Avg. & Final \\
                     &     &     &     &     &     &     &     &     &     & rank  & rank  \\
\midrule 
QbSD-gAS                       & 1  & 2  & 1  & 2  & 1  & 2  & 2  & 2  & 8 & 1.6  & 1 \\
AL$_{\text{Mult.}}$-AS         & 3  & 4  & 3  & 6  & 7  & 5  & 1  & 1  & 8 & 3.8  & 2 \\
AR-AL$_{\text{Mult.}}$-AS      & 4  & 1  & 7  & 4  & 4  & 4  & 6  & 6  & 8 & 4.5  & 3 \\
AL$_{\text{AR}}$-AS            & 2  & 3  & 2  & 15 & 5  & 3  & 4  & 3  & 8 & 4.6  & 4 \\
EGARCH                         & 6  & 5  & 10 & 1  & 6  & 6  & 3  & 7  & 8 & 5.5  & 5 \\
AR-AL$_{\text{AR}}$-AS         & 5  & 22 & 4  & 5  & 3  & 1  & 5  & 4  & 8 & 6.1  & 6 \\
\bottomrule
\end{tabular*}
\vspace{-0.5cm}
\end{center}
\end{minipage}
\end{table}

\begin{table}[!h]
\begin{minipage}{7.0in}
{\small {\bf Table C9.} Ranking of 5\% VaR forecasting models using the MCS procedure with quantile scores, rolling window $R=2500$}
\begin{center}
\footnotesize
\vspace{-0.25cm}
\begin{tabular*}{\textwidth}{@{\extracolsep{\fill}} llccccccc|ccc}
\toprule
                     & \multirow{2}{*}{S\&P} & \multirow{2}{*}{DJIA} & \multirow{2}{*}{NASDAQ} & \multirow{2}{*}{STOXX} & \multirow{2}{*}{FTSE} & \multirow{2}{*}{DAX} & \multirow{2}{*}{CAC} & \multirow{2}{*}{TSX} & \multirow{2}{*}{\#} & Avg. & Final \\
                     &     &     &     &     &     &     &     &     &     & rank  & rank  \\
\midrule 
EGARCH                       &  2 &  2 &  2 &  2 &  2 &  1 &  2 &  2 &  8 &  1.9 &  1 \\
QbSD-gAS                     &  1 &  1 &  1 &  3 &  1 &  2 &  3 &  3 &  8 &  1.9 &  2 \\
AL$_{\text{Mult.}}$-AS       &  4 &  4 &  3 &  5 &  5 &  3 &  4 &  6 &  8 &  4.2 &  3 \\
AL$_{\text{AR}}$-AS          &  6 &  3 & 15 &  6 &  3 &  5 &  1 &  5 &  8 &  5.5 &  4 \\
QAR-QbSD-gAS                 &  7 &  7 &  6 &  4 &  6 &  8 & 11 &  1 &  8 &  6.2 &  5 \\
AR-EGARCH                    &  8 &  6 &  4 &  1 & 11 &  6 &  7 &  8 &  8 &  6.4 &  6 \\
\bottomrule
\end{tabular*}
\vspace{-0.5cm}
\end{center}
\end{minipage}
\end{table}

\begin{table}[!h]
\begin{minipage}{7.0in}
{\small {\bf Table C10.} Ranking of 1\% VaR and ES forecasting models based on the MCS procedure using AL log scores, rolling window $R=2500$ }
\begin{center}
\footnotesize
\vspace{-0.25cm}
\begin{tabular*}{\textwidth}{@{\extracolsep{\fill}} lcccccccc|ccc}
\toprule
                           & \multirow{2}{*}{S\&P} & \multirow{2}{*}{DJIA} & \multirow{2}{*}{NASDAQ} & \multirow{2}{*}{STOXX} & \multirow{2}{*}{FTSE} & \multirow{2}{*}{DAX} & \multirow{2}{*}{CAC} & \multirow{2}{*}{TSX} & \multirow{2}{*}{\#} & Avg. & Final \\
                           &                        &                        &                          &                         &                        &                       &                       &                      &                      & rank  & rank  \\
\midrule
AR-AL$_{\text{AR}}$-AS        &  1 &  6 &  1 & 10 &  1 &  4 &  3 &  5 &  8 &  3.9 &  1 \\
AR-AL$_{\text{Mult.}}$-AS     &  6 &  1 &  6 & 12 &  4 &  5 &  2 &  8 &  8 &  5.5 &  2 \\
QbSD-gAS                      & 15 &  3 & 14 &  4 &  2 &  3 &  4 &  1 &  8 &  5.8 &  3 \\
AL$_{\text{Mult.}}$-AS        & 12 &  2 & 11 & 14 &  3 &  1 &  6 &  4 &  8 &  6.6 &  4 \\
GJR-GARCH-skew-$t$            &  9 &  7 &  7 &  2 &  5 &  7 &  7 & 16 &  8 &  7.5 &  5 \\
AR-AL$_{\text{Mult.}}$-SAV    &  2 &  5 &  3 & 15 & 12 & 16 &  9 &  7 &  8 &  8.6 &  6 \\
\bottomrule
\end{tabular*}
\vspace{-0.5cm}
\end{center}
\end{minipage}
\end{table}

\begin{table}[!h]
\begin{minipage}{7.0in}
{\small {\bf Table C11.} Ranking of 2.5\% VaR and ES forecasting models based on the MCS procedure using AL log scores, rolling window $R=2500$ }
\begin{center}
\footnotesize
\vspace{-0.25cm}
\begin{tabular*}{\textwidth}{@{\extracolsep{\fill}} lcccccccc|ccc}
\toprule
                          & \multirow{2}{*}{S\&P} & \multirow{2}{*}{DJIA} & \multirow{2}{*}{NASDAQ} & \multirow{2}{*}{STOXX} & \multirow{2}{*}{FTSE} & \multirow{2}{*}{DAX} & \multirow{2}{*}{CAC} & \multirow{2}{*}{TSX} & \multirow{2}{*}{\#} & Avg. & Final \\
                          &                        &                        &                          &                         &                        &                       &                       &                      &                      & rank  & rank  \\
\midrule
AR-AL$_{\text{Mult.}}$-AS     &  1 &  2 &  6 &  1 &  1 &  1 &  4 &  5 &  8 &  2.6 &  1 \\
AL$_{\text{Mult.}}$-AS        &  6 &  3 &  1 &  3 &  6 &  4 &  2 &  2 &  8 &  3.4 &  2 \\
QbSD-gAS                      &  5 &  1 &  8 &  2 &  3 &  5 &  5 &  1 &  8 &  3.8 &  3 \\
QAR-QbSD-gAS                  &  3 &  4 & 12 &  5 &  5 &  6 &  6 &  3 &  8 &  5.5 &  4 \\
AL$_{\text{AR}}$-AS           &  2 &  5 &  4 & 21 &  4 &  2 &  3 &  6 &  8 &  5.9 &  5 \\
AR-AL$_{\text{AR}}$-AS        &  4 & 20 &  9 &  4 &  2 &  3 &  1 &  4 &  8 &  5.9 &  6 \\
\bottomrule
\end{tabular*}
\vspace{-0.5cm}
\end{center}
\end{minipage}
\end{table}

\begin{table}[!h]
\begin{minipage}{7.0in}
{\small {\bf Table C12.} Ranking of 5\% VaR and ES forecasting models based on the MCS procedure using AL log scores, rolling window $R=2500$ }
\begin{center}
\footnotesize
\vspace{-0.25cm}
\begin{tabular*}{\textwidth}{@{\extracolsep{\fill}} lcccccccc|ccc}
\toprule
                           & \multirow{2}{*}{S\&P} & \multirow{2}{*}{DJIA} & \multirow{2}{*}{NASDAQ} & \multirow{2}{*}{STOXX} & \multirow{2}{*}{FTSE} & \multirow{2}{*}{DAX} & \multirow{2}{*}{CAC} & \multirow{2}{*}{TSX} & \multirow{2}{*}{\#} & Avg. & Final \\
                           &                       &                        &                         &                        &                        &                       &                       &                       &                      & rank & rank  \\
\midrule
QbSD-gAS                    &  2 &  2 &  2 &  1 &  1 &  5 &  3 &  2 &  8 &  2.2 &  1 \\
AL$_{\text{Mult.}}$-AS      &  3 &  1 &  1 &  4 &  4 &  1 &  4 &  4 &  8 &  2.8 &  2 \\
AL$_{\text{AR}}$-AS         &  5 &  8 & 14 &  8 &  3 &  3 &  1 &  3 &  8 &  5.6 &  3 \\
QAR-QbSD-gAS                &  4 &  3 & 20 &  2 &  6 &  6 &  5 &  1 &  8 &  5.9 &  4 \\
AR-AL$_{\text{AR}}$-AS      & 11 &  7 &  5 & 20 &  2 &  4 &  2 &  6 &  8 &  7.1 &  5 \\
AR-AL$_{\text{Mult.}}$-AS   &  1 &  6 &  8 &  5 &  5 &  2 &    &  5 &  7 &  7.5 &  6 \\
\bottomrule
\end{tabular*}
\vspace{-0.5cm}
\end{center}
\end{minipage}
\end{table}

\end{document}